\documentclass[12pt,preprint]{aastex}
\shorttitle{Wolf-Rayet Stars in M33}

\shortauthors{Neugent \& Massey}

\begin{document}

\title{The Wolf-Rayet Content of M33\altaffilmark{1}}

\slugcomment{ApJ, in press}

\author{Kathryn F.\ Neugent and Philip Massey\altaffilmark{2}}
\affil{Lowell Observatory, 1400 W Mars Hill Road, Flagstaff, AZ 86001;\\ kneugent@lowell.edu; phil.massey@lowell.edu}

\altaffiltext{1}{Observations reported here were partially obtained at the MMT Observatory, a joint facility of the University of Arizona and the Smithsonian Institution. MMT telescope time was granted by NOAO, through the 
Telescope System Instrumentation Program (TSIP). TSIP is funded by the
National Science Foundation.}
\altaffiltext{2}{Visiting Astronomer, Kitt Peak National Observatory, National Optical Astronomy Observatory, which is operated by the Association of Universities for Research in Astronomy, Inc.\ (AURA) under cooperative agreement with the National Science Foundation.}

\begin{abstract}
Wolf-Rayet stars (WRs) are evolved massive stars, and the relative number of WC-type and WN-type WRs should vary with metallicity, providing a sensitive test of stellar evolutionary theory. The observed WC/WN ratio is much higher than that predicted by theory in some galaxies but this could be due to observational incompleteness for WN-types, which have weaker lines. Previous studies of M33's WR content show a galactocentric gradient in the relative numbers of WCs and WNs, but only small regions have been surveyed with sufficient sensitivity to detect all of the WNs. Here we present a sensitive survey for WRs covering all of M33, finding 55 new WRs, mostly of WN type. Our spectroscopy also improves the spectral types of many previously known WRs, establishing in one case that the star is actually a background quasar. The total number of spectroscopically confirmed WRs in M33 is 206, a number we argue is complete to $\sim$5\%, with most WRs residing in OB associations, although $\sim$2\% are truly isolated. The WC/WN ratio in the central regions ($<$2~kpc) of M33 is much higher than that predicted by the current Geneva evolutionary models, while the WC/WN ratios in the outer regions are in good accord, as are the values in the SMC and LMC. The WC/WN ratio and the WC subtype distribution both argue that the oxygen abundance gradient in M33 is significantly larger than found by some recent studies, but are consistent with the two-component model proposed by Magrini et al.
\end{abstract}


\section{Introduction}
\label{INTRO}

Wolf-Rayet (WR) stars are the final evolutionary stage of high mass stars before they erupt into spectacular Type Ibc supernovae (Meynet et al.\ 2011). While in this stage, stars can be thought of as ``bare stellar cores" characterized by highly evolved surface elements (Meynet \& Maeder 2005) and high mass-loss rates of $\sim$$10^{-5} M_\sun$ year$^{-1}$ (Abbot \& Conti 1987, Nugis et al.\ 1998, Crowther 2007). These strong stellar winds create the broad emission lines that characterize a WR star's spectrum (Cassinelli \& Hartmann 1975, Conti \& de Loore 1979). Depending on the emission lines present in the spectrum, a WR star can be one of two types: WN (nitrogen rich) or WC (carbon rich). Both types contain large amounts of helium with little to no hydrogen while WN stars contain nitrogen and WC stars contain both carbon and oxygen (Massey 2003 and references therein). 

According to modern stellar-evolutionary theory, the most massive stars (and thus, the stars with the highest mass-loss rates) will first lose enough of the outer layers to reveal the H-burning products, helium and nitrogen, (WN stage) before continuing to remove enough material to reveal the He-burning products, carbon and oxygen (WC stage). However, less massive stars will only reach the WN stage. This is known as the Conti Scenario (Conti 1976, Maeder \& Conti 1994). It follows that WC stars will be more common relative to WN stars in galaxies with high metallicities\footnote{Throughout this paper we use ``metallicity" to really mean the oxygen abundance. This shorthand is justified in studying the evolution of massive stars, as oxygen is one of the primary drivers of the stellar wind. See discussion in Massey et al.\  (2003).  Fortunately, too, the oxygen abundance is (in principle) straightforward to measure in extragalactic HII regions.}, as the higher metallicities will result in higher mass-loss rates. The WC phase will also last for a larger fraction of the He-burning phase (see, e.g., Meynet \& Maeder 2005).

Current evolutionary models do a good job at predicting the relative number of WCs to WNs at low metallicities (SMC, LMC), but seem to fail at high metallicities. This is well illustrated in Figure 11 of Meynet \& Maeder (2005) for the Geneva evolutionary models, where the data for the WC/WN ratios comes from Massey (2003) and references therein. Various efforts have been made to solve this problem, i.e., Eldridge \& Vink (2006) and Meynet et al.\ (2011). However, the fault may not be that of the models but of the observations themselves. Conti \& Massey (1989) showed that the strongest optical line in WCs (C III $\lambda 4650$) has a median line strength a factor of 4 higher than for the strongest optical line in WNs (He II $\lambda 4686$), and thus WNs are harder to detect. So, while a survey may be complete for WCs, it may not delve deep enough to detect all of the WNs. This would skew the WC to WN ratio and thus make the evolutionary models appear incorrect. At the same time, surveys of a few small regions within a galaxy may skew the results to either higher or lower WC/WN values as the population may not be well mixed in age. With this in mind, a galaxy-wide {\it complete} survey must be undertaken to measure the WC to WN ratio before the models can criticized and improved if needed.

In this paper, we present the first survey of the entire disk of M33 with sufficient sensitivity to detect even the weakest-lined WNs. M33 is particularly interesting to examine because of its uncertain metallicity. Studies of WR stars in other galaxies have shown that the WC to WN ratio is quite sensitive to metallicity. For example, compare the SMC ($\log(\rm O/H) + 12 = 8.13$) and the LMC ($\log(\rm O/H) + 12 = 8.37$) (Russell \& Dopita 1990). Massey et al.\ (2003) found the SMC ratio to be 0.09 while Breysacher et al.\ (1999) found the LMC ratio to be 0.22. Previous WR studies of M33, such as Massey \& Conti (1983) and Massey \& Johnson (1998) show a decrease in the ratio of WCs to WNs as the galactocentric distance increases (center ratio of 0.6, outer ratio of 0.2). Additionally, Massey \& Johnson (1998) found a change in the WC spectral properties; WCs in the outer regions showed broad lines while WCs in the inner regions showed narrow lines. These findings are consistent with a M33 having a significant metallicity gradient, such as that found by Garnett et al.\ (1997). However, recent studies have found that there is little or no metallicity gradient based upon HII regions with measured [OIII] $\lambda 4363$ electron temperatures (Crockett et al.\ 2006, Rosolowsky \& Simon 2008, Magrini et al.\ 2007, 2010, Bresolin 2011). Nevertheless, Magrini et al.\ (2007, 2010) argue that the metallicity in the central region of M33 is much higher based on the abundances in young massive stars. Thus, by presenting a complete WR survey of M33, we hope to further investigate the metallicity of M33 as well as compare our results to evolutionary theory. Subsequent studies will revisit the issue of the WC/WN ratio of the low-metallicity starburst galaxy IC10, and present the first galaxy-wide deep survey for WRs in the relatively metal-rich disk of M31.

In Section~\ref{ID} we describe how we identified candidate WR stars through the use of photometry and image subtraction techniques. In Section~\ref{SpecCon} we discuss the spectroscopic confirmation process. In Section~\ref{BigTable} we present an updated list of all of the currently known WR stars in M33. In Section~\ref{Res} we evaluate our completeness and discuss our results and finally, in Section~\ref{Con} we summarize our findings and list future goals.

\section{Identification of Candidates}
\label{ID}
We have begun a new survey for Wolf-Rayet stars among the star-forming galaxies of the Local Group (beyond the Magellanic Clouds), using the basic filter system introduced by Armandroff \& Massey (1985). The goal of the present survey is to remove the bias towards the detection of WC stars discussed extensively by Massey \& Holmes (2002).

Owing to their strong emission lines, Wolf-Rayet stars are relatively easy to detect. Although slitless spectroscopy (objective prism) surveys were effective at finding Wolf-Rayet stars in the Magellanic Clouds (Azzopardi \& Breysacher 1979a, 1979b), crowding and faintness (compared to the night sky) compromises their use for distant galaxies (e.g., Bohannan et al.\ 1985). However, the use of narrow-band interference filter imaging has proven highly effective.  This technique was first used by Wray \& Corso (1972) who surveyed two 8' diameter fields near the center of M33 using photography through an interference filter that included both the C III $\lambda 4650$ line and the He II $\lambda 4686$ emission lines. The same fields were photographed through a continuum filter, and 25 WR candidates were identified by visually blinking the on-band and off-band exposures, most of which were subsequently confirmed as WRs, predominately of WC type (see Table 9 of Massey \& Johnson 1998 and references there in). Armandroff \& Massey (1985) described a three-filter set that was designed to be more sensitive to finding WRs, and to distinguish between WNs and WCs: one filter (``WC") is centered on the C III $\lambda 4650$ line, another (``WN") is centered on He II $\lambda 4686$ (strong in both WNs and WCs), and a continuum filter (``CT") is centered at $\lambda 4750$. Each filter has a band-width of approximately 50 \AA.  Armandroff \& Massey (1985) used CCD imaging through this set to identify WR candidates in NGC 6822 and IC 1613, as well as two small fields in M33 that overlapped with those of Wray \& Corso (1972). Spectroscopy confirmed that the Wray \& Corso (1972) imaging had missed many WNs, not unexpected given the difference in line strengths between CIII $\lambda$4650 in WCs and He II $\lambda$4686 in WNs.  Massey \& Johnson (1998) used a version of this set to image eight $5\farcm2 \times 5\farcm2$ fields in M33, and Massey \& Duffy (2001) imagined the SMC through such a filter set and identified some additional WNs that had been missed by objective prism surveys.

For the work here on M33 (and our survey of other Local Group galaxies that will be described in subsequent papers) we used the Kitt Peak 4-meter Mayall telescope and the Mosaic CCD camera to image through a similar set of filters. The data on M33 were obtained on 2000 Sept 21, 23, and 24 (and allowed to properly age). Three overlapping $36'\times36'$ fields were used to cover a 0.32 deg$^2$ area, extending $\sim$8.5 kpc in radius from the center and covering all of the optical disk. Each field was imagined using three 300s exposures in each filter, with the telescope dithered slightly between the exposures to fill in the gaps between CCDs in the Mosaic camera. The seeing varied from $1\farcs1$ on the center and southern fields to $1\farcs5$ on the northern field.

Flat-fielding of the images was achieved by exposures of an illuminated uniform screen attached to the dome, and bias frames were obtained every night.  Reductions followed the same procedure as for the Local Group Galaxies Survey (LGGS) data described by Massey et al.\ (2006), and included corrections for ``cross-talk" between pairs of chips that shared the same electronics.  The only additional step was that since these filters were non-standard, higher-order spatial distortion terms had to be derived before transforming to a linear plate scale.  In analyzing the images, we treated each of the three dithers as separate exposures; i.e., we did not attempt to ``stack" the images as this would compromise their photometric integrity as emphasized by Massey et al.\ (2006).

As we now describe, we used two techniques with these images in order to identify stars that were ``significantly" brighter on the on-band rather than the continuum. The first involved photometry of individual stars while the second was based upon image subtraction techniques.

\subsection{Photometry}
\label{Phot}
Crowded field photometry was used to identify candidate WR stars by comparing the magnitude differences of WC -- CT and WN -- CT to the magnitude uncertainties based on poisson statistics and CCD read noise. Instrumental magnitudes of every detected star were obtained using IRAF's point-spread function (PSF) fitting routine ``DAOPHOT" (Stetson et al.\ 1990) and the magnitude differences were then computed. Next, the number of measured stars and their positions in clusters were matched between the WN, WC and CT frames. Stars were added or subtracted by eye to obtain matches between the frames, and the PSF-fitting routines were re-run. This allowed for the detection of stars in even the most crowded regions of M33. Stars with magnitude differences larger than -0.10 mag and a significance level greater than 3$\sigma$ were considered candidates. 

For the photometry candidates, there were 1693 WR candidates with counterparts in the LGGS found to a low significance level (3$\sigma$). If one were to use the absolute magnitude of M33 to scale from the LMC ($M_V=-18.9$ and $M_V=-18.5$, respectively; van den Bergh 2000), one might expect to find $\sim$190 WR stars in M33, or about 30\% more than the 141 WRs Massey \& Johnson (1998) list. Thus, while we do expect to find a few more WR stars, the number of discoveries should be on the order of tens, not hundreds. Therefore, this large number of candidates exceeds the expected number of WRs by about a factor of 100. This gave us some indication that we needed to be a little more selective with our list of candidates and that the instrumental errors were probably underestimated. Since we knew that the Massey \& Johnson (1998) survey was deep enough to detect even Of-type stars, we instead considered the known M33 WRs. On average, they were detected at a level of 25$\sigma$, and only six (out of 137 unambiguously identified) were found at a significance level less than 10$\sigma$.  The lowest significance one of these was found at 5.2$\sigma$, while the next least significance known WR was found at the 7.1$\sigma$ level. Thus we decided to keep only those photometric candidates with significance levels $\ge 5 \sigma$; there were 382 such objects including many of the known ones. 
 
\subsection{Image Subtraction}
We complemented the photometry with the use of image subtraction, a technique that has achieve much success in the detection of supernovae (see, for example, Yuan et al.\ 2010). Since WR stars will be brighter in either the WN or WC filters than in the CT images, simply subtracting the CT image from the WN image should reveal an image displaying only the WN stars. Similarly, subtracting the CT image from the WC image should reveal the WC stars. However, seeing variability often turns this simple idea into a much more complex problem. To solve this, we used the Astronomical Image Subtraction by Cross-Convolution program which employs point-spread fitting techniques and cross-convolution to help overcome these barriers (Yuan \& Akerlof 2008). 

\begin{figure}
\epsscale{1}
\plotone{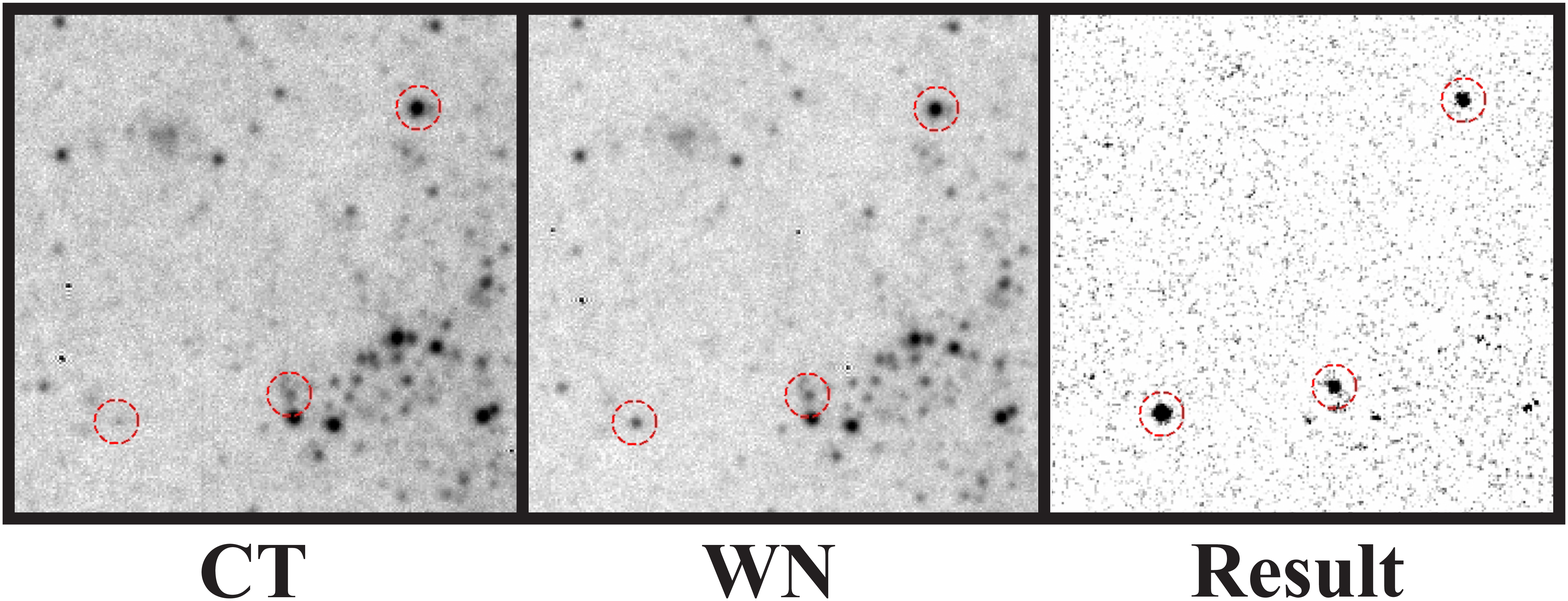}
\caption{\label{fig:imgSub} Demonstration of the image subtraction technique. The first image shows a region of M33 through the CT filter, the second through the WN filter and the third is the result of using the image subtraction program. Three confirmed WN stars are outlined in red dashed circles. While the WR star on the lower left is clearly more prominent through the WN filter than through the CT filter, the other two WR stars appear by eye to be the same brightness. However, the image subtraction program is able to detect the stars' small brightness differences through the two different filters. Notice that while there are many stars in the field, only three show up prominently in the result image.}
\end{figure}

After running the image subtraction program on each of the three dithered images in each field, we were able to proceed with identifying potential WR candidates. We first took the median of the images. This reduced the anomalies created by bad pixels as well as erroneous results due to poor image subtraction when dealing with bright foreground stars. By visually blinking through the images and comparing the stars on the WN or WC image and the result image to the original CT image, we were able to remove many false positives. As Figure~\ref{fig:imgSub} shows, the WR candidates we identified were of one of two types: either it was easy to visually identify the WR star since the magnitude differences were large, or it was virtually impossible to distinguish any magnitude difference between the two filters. Still, even stars with seemingly negligible magnitude differences showed up prominently as potential WR stars in the resulting image.

A study of all three M33 fields using image subtraction yielded 330 WR candidates, 225 (68\%) of which matched with candidates identified by photometry. Additionally, 129 of the 152 (85\%) known WR were identified using this method. Why the remaining 23 confirmed WR stars were not identified is discussed in Section~\ref{Complete}. 

\section{Spectroscopy}
\label{SpecCon}

In order to determine which of our WR candidates were real, we used the 6.5-m MMT with the 300 fiber-fed spectrometer Hectospec (Fabricant et al.\ 2005). The instrument covers a 1 deg diameter field of view, well matched to our survey area, and we prepared two configuration files with different field centers to cover the maximum number of new candidates. The spectra were taken on UT 2010 Oct 9 and 11 (program NOAO 2010B-0260) with the 270 line mm$^{-1}$ grating blazed at 5000 \AA, and covered the wavelength range 3700-9000 \AA. Beyond $\sim$7500 \AA\ the spectra may be contaminated by second-order blue light, but this has no impact on our project, which relied upon data in the 4000-6000 \AA\ range for identifying and classifying WR stars (i.e., from N IV $\lambda 4058$ through C IV $\lambda 5812$). The fibers have a diameter of 250 $\mu$m ($1\farcs5$ on the sky), and resulted in a spectral resolution of 6 \AA\ (5 pixels). Wavelength calibration was by means of a He-Ne-Ar comparison arc taken in the afternoon, with a wavelength shift (in pixel space) applied from the prominent $\lambda 8399$ OH night-sky line. Flat-fielding was accomplished by an exposure of an illumination calibration screen, with a through-put correction for each fiber determined by its location within the field.  Complete details of the reduction procedure can be found in Drout et al.\ (2009).  Fibers in each configuration were assigned to carefully selected blank sky positions.

We made use of the UCAC2 (Zacharias et al.\ 2004) to select suitable guide stars known to have negligible proper motions.  This catalog is on the global International Celestial Reference System (ICRS), while the LGGS coordinates are tied to the USNO-B1.0 catalog of Monet et al.\ (2003), a predecessor to the ICRS.  Using the UCAC2 catalog we applied the following transformations to the LGGS coordinates to make them match the ICRS: $\alpha_{\rm ucac2}=\alpha_{\rm LGGS}-0\fs030$ and $\delta_{\rm ucac2}=\delta_{\rm LGGS}-0\farcs14$.

In assigning priorities for the two field configurations, we concentrated first on the objects found both by the image-subtraction method {\it and} photometrically but were not previously known. The second priority targets were those found with a high significance level ($\ge6\sigma$) from the photometry comparison. Third priority were those stars found by image subtraction but {\it not} found by the photometry. Fourth priority were lower significance photometry candidates not found via the image subtraction. And the last priority were previously know WRs, as we did not want to waste fibers. (Not all of the known M33 WRs were included.) As we shall see, obtaining high quality spectra for some of the ``known" WRs proved illuminating.

\subsection{``Losers" and Non-WR Star ``Winners"}
In our first glance at the spectra, we were gratified to find a large number of newly discovered WR stars.  But even a cursory examination revealed that many of our WR candidates were in fact late-type stars. The reason was obvious from the spectra: there is a TiO molecular feature at $\lambda 4760$ within the CT bandpass that caused these stars to appear brighter in the emission line filters than in the continuum. This feature only becomes pronounced at spectral types of M3 and later (see Fig.~14.1 of Jaschek \& Jaschek 1990), and the effect was somehow overlooked when the second author originally designed the filter set, probably as there were only a few stars this late in the Jacoby et al.\ (1984) stellar library (see Armandroff \& Massey 1985). The problem equally affected the image subtraction and photometric candidates. Adding to our chagrin, of course, was the fact that such stars could have been readily removed from the sample simply by consulting the colors listed in the LGGS. Indeed, all of the stars for which we see TiO have $B-V>1.1$, with the exception of one star with unrealistic colors (i.e., $B-V\sim-1.3$) due to crowding issues.

We list in Table~\ref{tab:priorities} the total number of stars in each category and number observed, (more to the point) the number of non-red stars. We see that the vast majority of the stars detected with both methods (image subtraction and photometry) turn out to have emission, which we consider to then be a ``winner", even if the star is not actually a WR. The other 9 stars  (plus the 44 found only by image subtraction) invariably are bright stars, usually bright supergiants. These stars, although interesting in their own right, will be discussed elsewhere as part of our survey of the less evolved massive star population. Of the non-emission photometric candidates, the losers covered the entire range of possibilities, from O-type dwarfs (with He II $\lambda 4686$ in {\it absorption}) to mid F-type foreground dwarfs. The most likely explanation are mismatches in the number of stars identified on the continuum and emission-line images, despite our best efforts. If a tight grouping was photometered as 5 stars on the CT image but only as 3 or 4 stars on the WN or WC images, then there will be spurious high significance candidates.

\begin{figure}
\epsscale{1.0}
\plotone{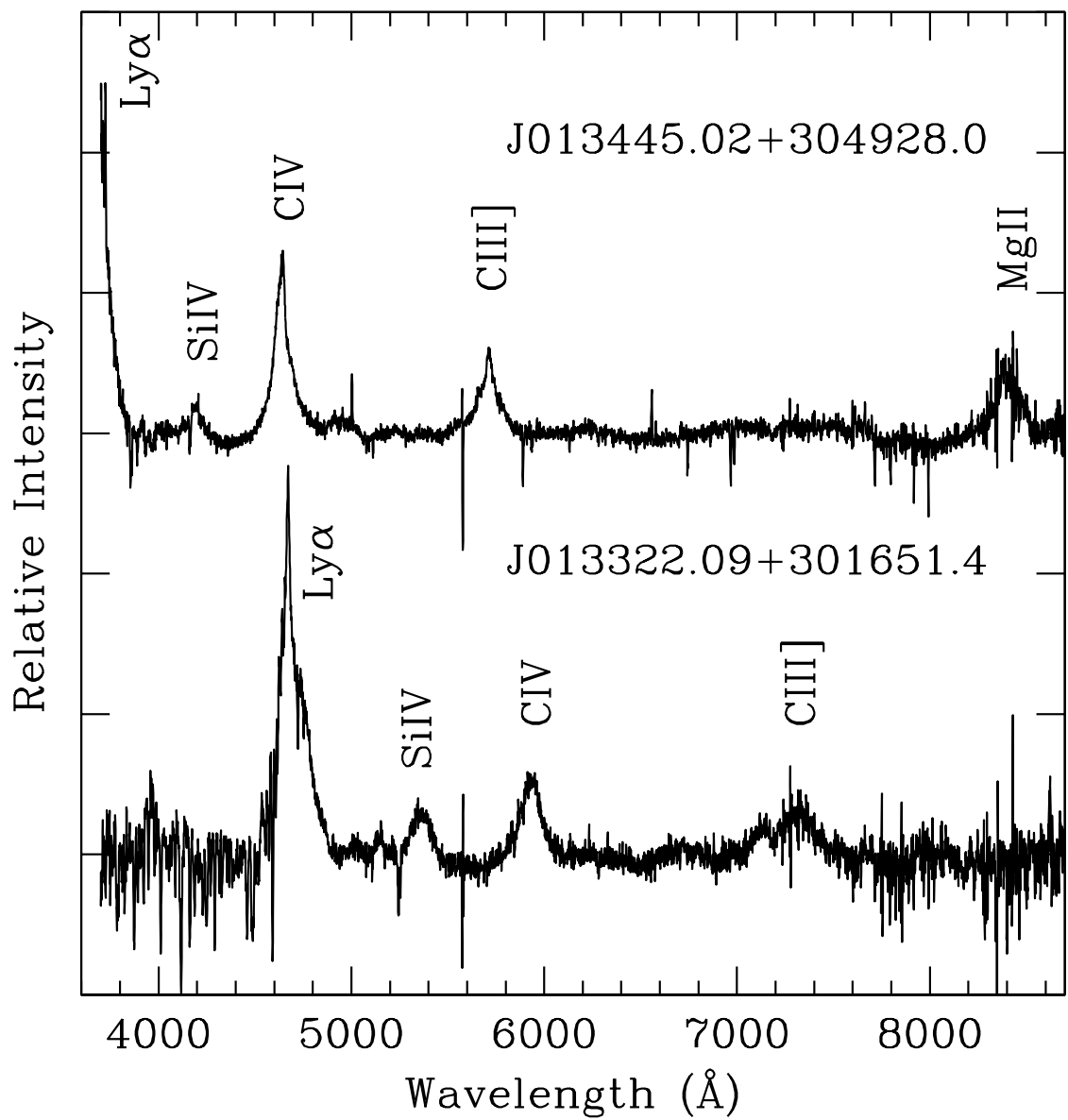}
\caption{\label{fig:QSO} The spectra of two newly discovered quasars. Upper: the star J0133445.02+304928.0 has a redshift of $z=1.99$, and its strong CIV $\lambda 1550$ line had been shifted to 4645 \AA, placing it near the center of our WC detection filter.  Lower: the star J013322.09+301651.4 has a redshift of $z=2.84$, and its Ly$\alpha$ line had been shifted to 4672 \AA, placing it in the WN filter bandpass.  That star had previously been called a WN-type Wolf-Rayet star (Massey \& Conti 1983), with the Ly$\alpha$ featured mistaken for a blend of NIII $\lambda 4634,42$ and He II $\lambda 4686$ by Massey \& Conti (1983).}
\end{figure}

Not all of our ``winners" were WRs or Of-type stars (Table~\ref{tab:nonWRs}). In fact our spectroscopy discovered two previously unrecognized quasars (QSOs), one of which had been previously classified as a Wolf-Rayet star! The spectra of the QSOs are shown in Figure~\ref{fig:QSO}. J013445.02+304928.0 has a redshift of $z=1.99$, while J013322.09+301651.4 had a redshift of $z=2.84$. The latter was classified as a WN star by Massey \& Conti (1983) based upon a much poorer signal-to-noise spectra, which showed only a single broad feature near $\lambda 4686$; see their Figure 2. The line is actually Lyman $\alpha$\footnote{``To a carpenter, everything looks like a nail." }. Note the rich Lyman $\alpha$ absorption forest short-wards of this line evident in our spectrum. Both objects are x-ray sources in the XMM-Newton Serendipitous Source Catalog\footnote{http://wmmssc-www.star.le.ac.uk/Catalogue} and also appear in the Atlas of Radio/X-ray Associations by Flesch (2010).  Perhaps some day these background objects can be used to probe the ISM of M33, although their faintness ($V=19.4$ and 20.1) would make such high dispersion studies difficult at present\footnote{We are indebted to David James for this suggestion.}. We also rediscovered the ``hot LBV" candidate J013459.47+303701.9 found by Massey et al.\ (2007b). Several of its relatively weak (-1 to -2 \AA\ equivalent width) FeII and [FeII] lines were in the WC and WN filters, but the CT bandpass happened to be free. One of our objects (J013331.23+303334.2) is an HII region with strong He II $\lambda 4686$ emission, apparently not previously known. The object was not considered an He II $\lambda 4686$ emission nebula by Kehrig et al.\ (2011) as it was too stellar (M. S. Oey, private communication 2010). 

\begin{figure}
\epsscale{0.48}
\plotone{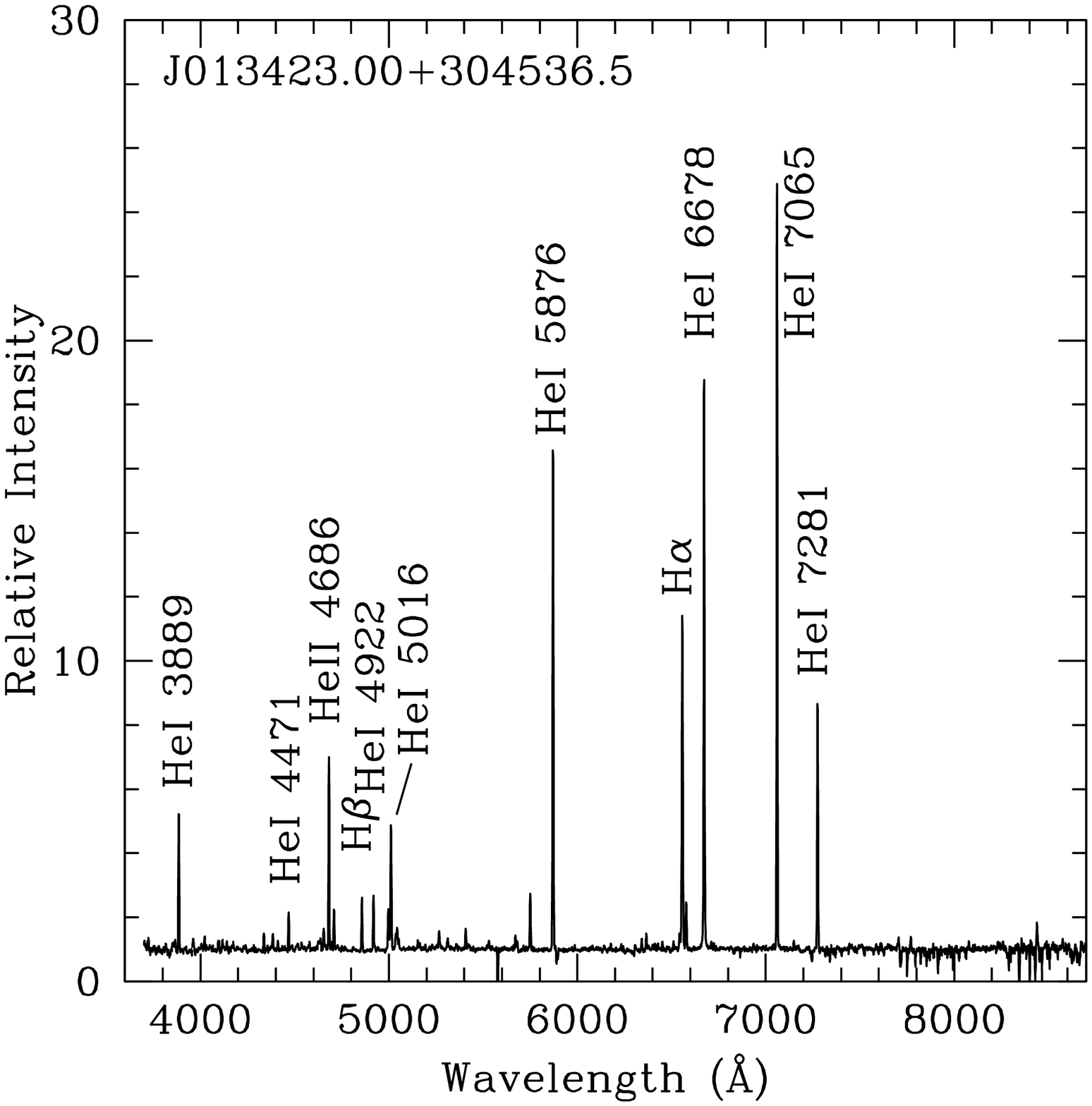}
\plotone{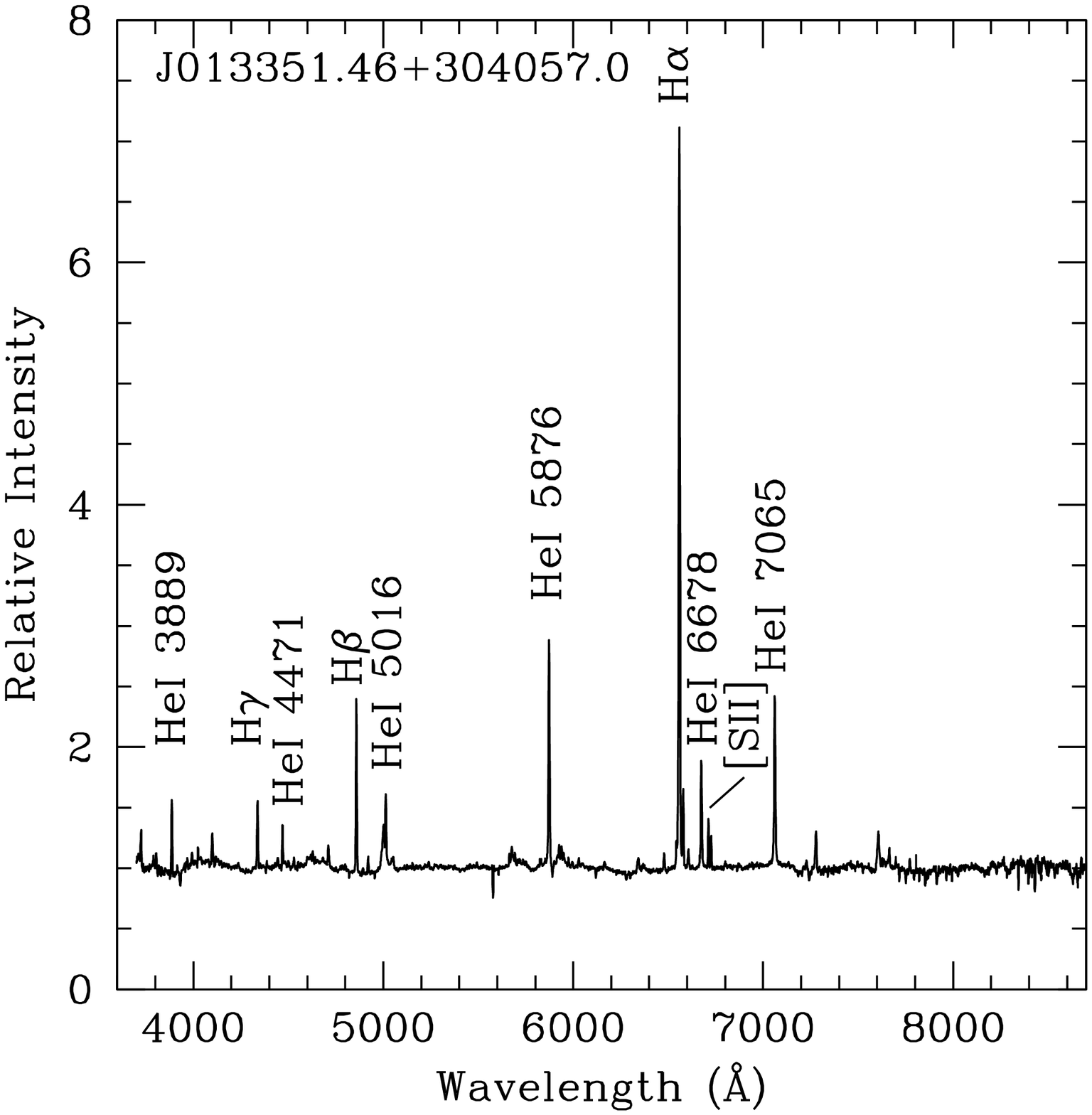}
\caption{\label{fig:wacky} Two non-WR emission-line stars.  Both stars show strong lines of H and He I, with weak [NII] $\lambda \lambda 6548, 6584$ nebular emission. Left: J013423.00+304536.5 additionally shows He~II emission lines, and at $M_V=-4.5$ may be a symbiotic star.  The counterargument is that there is no sign of a late-type companion visible in the far red. Right: J013351.46+304057.0 shows a broad feature that may include a He II $\lambda 4686$ component but not other He II lines.  Its spectrum shows [SII], originating in a low density region.  At $M_V=-7.2$ it is too luminous be a symbiotic star.} 
\end{figure}

Perhaps the most interesting two non-WR  ``winners" are shown in Figure~\ref{fig:wacky}.  The stars J013423.00+304536.5 and  J013351.46+304057.0 both show strong lines of H and He I, along with weak [NII] $\lambda \lambda 6548, 6584$ lines but no sign of the [OIII] emission that would characterize a normal HII region.   J013423.00+304536.5 shows weak lines of He II. The lines are all blue-shifted since the object has a velocity of -230 km s$^{-1}$.  Given the star's location within the plane of M33, this velocity is about as expected (Zaritsky et al.\  1989; Drout et al., in prep).  The emission-line spectrum resembles that of some symbiotic stars (Munari \& Zwitter 2002) although the late-type giant absorption features one expects in such objects in the far red are not present (Kenyon 1986). (The impression of absorption in the figure are due to slight over-subtraction of the night-sky spectrum.) The star is fairly bright ($V=20.47$, $B-V=0.27$, $U-B=-1.06$, all according to the LGGS) and if we adopt an average reddening of $E(B-V)=0.12$, then the absolute visual magnitude $M_V$ is $-4.5$, which is also consistent with a symbiotic.  J013351.46+304057.0, on the other hand, shows very broad, weak emission that extends from 4600 \AA\ to 4720 \AA, and appears to include some He II $\lambda 4686$ as well as He I $\lambda 4713$. This broad feature is why the star made it into our list, but the spectrum is not that of a WR star.  Its spectrum also includes the nebular [S II] $\lambda \lambda 6716, 6731$ (located in the figure just to the right of He I $\lambda 6678$). The flux ratio of [SII] $\lambda 6716$ to $\lambda 6731$ is about 1.4, requiring that these lines originate in a low density region (Osterbrock 1974). It is quite a bit brighter than that of J013423.00+304536.5, with $V=17.73$, $B-V=0.08$, and $U-V=-0.93$, according to the LGGS), and hence $M_V=-7.2$, which is too bright to be a symbiotic star. We leave the nature of both objects unresolved, a subject for future researchers.

\subsection{Newly Found and Previously Known Of-type and WR Stars}
In Table~\ref{tab:WRs} we list the Of-type and WRs that we observed spectroscopically, including previously known WRs. We measured the equivalent widths (EWs) and full widths at half maxima (FWHMs) of the ``discovery" lines,  He II $\lambda 4686$ for the WNs, and the C III $\lambda 4650$ and He II $\lambda 4686$ blend in WCs.  We also measure the EWs and FWHMs of C III $\lambda 4650$, and C IV $\lambda \lambda 5801, 5812$ feature (hereafter referred to as C IV $\lambda 5806$) if present.  For WCs, the very strong C III $\lambda 4650$ feature is blended with (and dominates over) He II $\lambda 4686$.  

Classification of these stars was, for the most part, straightforward. The primary classification scheme is derived from Smith (1968a, 1968b) for WN3--WN8 and WC5--WC9 stars, with the extension to WN2 and WN9 and WC4 by van der Hucht et al.\ (1981). See Conti (1988, p.\ 13) for a complete summary. In earlier investigations of the WRs stars located beyond the Magellanic Clouds, the signal to noise of the spectra was often too poor to make more than a crude division into ``early" or ``late" types (Massey et al.\ 1987).   The signal-to-noise we achieve here is modest by some standards (typically 40 per spectral resolution element), but the result is that our spectra often resemble textbook examples of their class.  We illustrate the spectra of a few representative examples in Figures~\ref{fig:wns} and \ref{fig:wcs} covering the range of subtypes we observed.

The higher signal-to-noise afforded by the 6.5-m aperture of the MMT and good throughput of Hectospec allowed us to detect the weak absorption lines in stars that might otherwise have been classified as a weak-lined late-type WN star.  For instance, J013333.58+304219.2 had been called a WN-type Wolf-Rayet star by Massey \& Johnson (1998), where it was listed as C22,  but our superior spectrum here shows that it is actually an O4 If star, with both He I $\lambda 4471$ and He II $\lambda 4200, 4542$ {\it absorption}. There is nebulosity present (as evident from [OIII] $\lambda \lambda$ 4959, 5007 and the [NII] lines flanking H$\alpha$), and the Balmer lines show a narrow emission component, further complicating the interpretation.  We show the spectrum in Figure~\ref{fig:o4}, as we believe this is the earliest supergiant yet found in M33 (i.e., based upon an inspection of Table 17 of Massey et al.\ 2007b and Table 9 of Massey et al.\ 2006). Although an equivalent width of He II $\lambda 4686$ of  -10 \AA\ is often taken as the dividing line between Of-type stars and WNs, there are stars whose EWs fall in the grey area, and J013333.58+304219.2 is one such star. Abbott et al.\ (2004) has similarly reclassified J013344.68+304436.7 (OB66-25) as an Of-type star rather than the WN8 classification offered by Massey et al.\ (1996), and we have reclassified J013432.68+304655.4 (N604-WR10 from Drissen et al.\ 2008) as an Of-type star rather than WN6 since we can see absorption. We list the stars that are no longer considered WRs in Table~\ref{tab:NoLonger}.  

The Ofpe/WN9 class (sometimes known as ``slash stars") was introduced by Walborn (1977) to describe four bright emission-line stars in the LMC with spectral properties intermediate between Of-type supergiants and WNs.  Spectral and photometric variability has linked these to the Luminous Blue Variables (LBVs); see Bohannan \& Walborn (1989).   The spectroscopic survey of UV-bright luminous stars in M33 by Massey et al.\ (1996) discovered a number of such objects.  Using  UV-spectroscopy, optical data and model fitting, Bianchi et al.\ (2004) showed that these M33 objects tend to have chemical abundances intermediate between O supergiants and WNs. Our survey here detected several of these objects, some previously known. These include the peculiar object found by Romano (1978) [J013509.73+304157.3] and classified by Massey et al.\ (2007b) as Ofpe/WN9, as well as J013416.07+303642.1 and J013418.37+303837.0, newly found here.  We illustrate the spectra of these three stars in Figure~\ref{fig:Ofpe}. Note the strong P Cygni features at He I $\lambda 4471$ and the broad sea of N II emission blended with He II $\lambda 4686$.

We also re-observed the Ofpe/WN9 star J013245.41+303858.3 (UIT 008), and were astonished to find that its characteristic P Cygni He I $\lambda 4471$ line has changed to a pure absorption profile. In Figure~\ref{fig:uit} we compare our 2010 spectrum with that obtained on 19 Nov 1995 and published in Massey et al.\ (1996). Both spectra were obtained with multi-fiber instruments, and hence the sky subtraction is not local; thus, nebulosity is not easily removed. The spectra are clearly contaminated by nebulosity, as evidenced by the strong [OIII] $\lambda \lambda$ 4959, 5007 lines in our 2010 spectrum. (The 1995 spectrum did not extend beyond 4900 \AA.) The fiber diameters used for our new observations have half the diameter (1\farcs5 vs.\ 3\farcs0) than those used in 1995, and indeed the Balmer lines are much stronger in the earlier spectrum.  While it is undecided whether these are primarily nebular in origin, there is no question that the He I $\lambda 4471$ profile in 1995 was P Cygni, not absorption contaminated by emission. Thus we might have naively classified this star as an O8If and attributed the remaining Balmer emission to nebulosity. We retain the Ofpe/WN9 designation here, although without He I $\lambda 4471$ in P Cygni this classification is somewhat a matter of curtesy\footnote{Note that Crowther et al.\ (1995)  prefer to call such stars  WN10 or  WN11 rather than maintaining the original ``slash" designation.}.

Another remarkable spectrum is shown in Figure~\ref{fig:ovia}.  The star J013510.27+304522 would normally be classified as a WN2, as its spectrum shows only lines of He~II, but no N~III, N~IV, or N~V.  What makes this spectrum so peculiar is the presence of extremely strong O VI $\lambda 3811$. Similar WC stars are known, and such objects have given rise to the (seldom-used) ``WO" classification (Barlow \& Hummer 1982).  Nugis \& Niedzielski (1990) argue that this line is weakly present in some WNs.  But, it is clear from Figure~\ref{fig:ovia} that the O~VI line is exceptionally strong in this star.  Oxygen is enriched as a result of He burning in massive stars, and so we would naively expect to see strong carbon lines, while only very weak C~IV $\lambda 5806$ is present. Such a star would offer a strong challenge to single-star evolutionary models if it signals high oxygen abundance. 

\section{The Current Census of Wolf-Rayet Stars in M33}
\label{BigTable}
We list in Table~\ref{BigTable} all of the 206 currently known Wolf-Rayet stars including the 55 new WR stars discovered in this survey.  We have removed the 3 stars from the 141 ``confirmed" WRs of Massey \& Johnson (1998) that we now consider to be non-WRs (Table~\ref{tab:nonWRs}), and added in the 4 newly found WRs by Massey et al.\ (2007b), the 7 by Drissen et al.\ (2008) and the 2 by Kehrig et al.\ (2011). Although Massey \& Johnson (1998) attempted to introduce a WR numbering scheme for M33, we adopted a coordinate based system here, as we were pleased to find that all of the known WRs have a counterpart in the LGGS survey of M33, even in the crowded HII regions studied by Drissen et al.\ (2008) with {\it HST}.  We acknowledge that the designations are cumbersome, but at least it should remove any ambiguity about which star is which.  We include cross references to the Massey \& Johnson (1998) numbers, as well as some previous IDs, in Table~\ref{BigTable}. Additionally, we give the designations from Drissen et al.\ (2008) even though in many cases their new designations referred to previously known WRs.

Included in the table are what we consider to be the  best spectral types; for instance, Abbott et al.\ (2004) obtained better spectra for a number of the Massey \& Conti (1983) stars and thus we've adopted Abbott et al.'s spectral types, unless our current spectroscopy improves upon it.

There are admittedly some arguable inclusions in the list.  For instance, J013332.64+304127.2 currently shows a spectrum resembling that of a hot LBV star, but we have included it as it was spectroscopically WN for many years, as discussed in Massey et al.\ (2006).  Yet we do not include J013416.07+303642.1, which Crowther et al.\ (1997) consider to be a ``WN11h", as discussed below.

We include in the table the value $\rho$ ($\rho$ of 1 = 7.53 kpc after adopting a distance of 840 kpc to M33, following Freedman et al.\ 1991), the de-projected distance from the center of M33 within its plane.  These values differ substantially from the older ones we used, primarily as the result of using the de Vaucouleurs et al.\ (1976) Second Reference Catalogue $\mu_{25}$ isophotoal radius of $30\farcm8$ rather than the often-quoted Holmberg radius of $25\arcmin$ as the normalization factor. We also adopted more modern values for the inclination ($56^\circ$) and position angle ($23^\circ$), following Zaritsky et al.\ (1989), although these are quite similar to what we had previously used, following Kwitter \& Aller (1981).  The center of M33 is taken to be $\alpha_{\rm 2000}$=1$^h$33$^m$50\fs89 and $\delta_{\rm 2000}$=30$^\circ$39$^\prime$36\farcs8.  

\subsection{Cross-identifications}
In creating a single list containing all of the known Wolf-Rayet stars in M33 we were occasionally stymied by some of the old identifications.  For instance, Massey \& Conti (1983) identify two WN stars (MC8 and MC9) near the HII region MA1 but provided only ``approximate" coordinates taken from instrument header.  Massey \& Johnson (1998) retained these two stars as ``MC8" and ``MC9" and repeat the approximate coordinates. In Table 9 of Massey et al.\ (2006), both (!) J013303.19+301124.2 {\it and} J013302.55+301135.8 are identified as MC 9 while J013302.94+301122.8 is identified as MC 8. Compounding this, Kehrig et al.\ (2011) list both MC 8 and MC 9 as J013303.19+301124.2, although in point of fact the correct two stars are identified in their Figure 3 (as WR14 and WR15, respectively)\footnote{We are indebted to Carol Kehrig for useful correspondence on the matter.}. To resolve this, we successfully resurrected the original spectra of MC 8 and MC 9 taken by Massey \& Conti (1983). The authors had taken spectra of the two very bright knots in the MA 1 HII region, and MC 8 corresponds to J013302.94+301122.8 and MC 9 corresponds to J013303.19+301124.2. (The second star identified as MC9 in Table 9 of Massey et al.\ 2006, J013302.55+301135.8, is not a WR star.) Neither MC8 nor MC9 were identified in our current imaging study as WRs as the objects were too bright ($V=16.1$ and $V=17.2$) for the image subtraction to work reliably, and the region too crowded for the photometry to be believable. Given that the absolute magnitudes of early WNs tends to be $-3$ or $-4$, while these objects have $M_V$ of $-9$ and $-8$, we can safely conclude that these WRs are actually marginally unresolved clusters of early-type stars, perhaps scaled down versions of R136. See further discussion in Kehrig et al.\ (2011).

Another difficult case was that of J013416.07+303642.1. This star was ``discovered" by Massey et al.\ (2007b) to be an H$\alpha$ emission line star, and spectrally resembles that of P Cygni. It was included in Massey et al.\ (2007b) as an LBV-candidate, cross-referenced to H 108 of Corral (1996). This star turns out to be none other than Humphreys \& Sandage (1980) B517, the spectrum of which was classified by Crowther et al.\ (1997) as a Wolf-Rayet star of type ``WN11h". We would not ourselves consider P Cygni to be a WR star, nor do we consider this object to be one, and therefore we do not include it in our table here\footnote{The star had also been observed earlier as Esteban (1994), who had misidentified it as coincident with the WC-type WR MC70. Crowther et al.\ (1997) conclude that the two sources are not the same, and here we confirm that MC 70 is J013416.28+303646.4, located slightly a few arcseconds east and 4" north of B517.}.

Drissen et al.\ (2008) used a combination of interference filter imaging with {\it HST} and ground-based spectroscopy to identify WR stars (and candidates) in the giant HII regions of M33, i.e., NGC 592, NGC 595, and NGC 604.  The superior resolution of {\it HST} helps in such crowded fields (particularly NGC 604), but by carefully comparing their finding charts to the LGGS images, we found that all but two of their confirmed and candidate WRs were well resolved on our ground-based LGGS images and appear as unique objects in the LGGS catalog. The exceptions are J013333.75+304134.0 (MC 31) which their {\it HST} imaging resolves into  N595-WR2A and -WR2B, both of which may be WRs, and J013433.62+304704.6 (MC 76) which they call N604-WR6, which has an unresolved companion in our LGGS photometry. They list cross identifications of two of their stars each for MC 74 (CM 11)  and MC 75 (CM 12), but this must reflect their uncertainty in the cross-identifications rather than evidence of blending given the reasonable spatial separation between objects. We do not include their N604-WR2 or N604-WR3 stars in Table~\ref{tab:final}  as these have not yet been confirmed spectroscopically, and one of their WR candidates did prove to be an O6.5 Iaf star (N604-WR13).  In addition, our spectrum of  J013432.68+304655.4 (their NGC604-WR10) clearly shows He II $\lambda 4200$ and $\lambda 4542$ absorption lines, leading us to call this star an OIf rather than a WN. (The difference is presumably due to our higher signal to noise.)  Note that while their {\it HST}-derived coordinates were very precise they are not very accurate, as inconsistencies in the {\it HST} guide star reference system from one region to another results in errors as great as 1\farcs5 compared to the ICRS now used in astrometric catalogs\footnote{Recall that we needed to transform our own LGGS coordinates slightly to place them on the ICRS; i.e., Section~\ref{SpecCon}.}. 

Other cross-identification issues were extensively discussed in the appendix of Massey \& Johnson (1998). 

\subsection{OB Association Membership}
Massey \& Johnson (1998) list the membership in the OB associations identified by Humphreys \& Sandage (1980).  Here we have extended this to the newly found WRs, using their charts to assign the membership indicated in Table~\ref{tab:final}. Although {\it most} of the WRs in the tables are in or near OB associations, 42 of the 206 (20\%) are not\footnote{We include as members the stars in the MA1 (Mayall \& Aller 1942), even though this is not cataloged in Humphreys \& Sandage (1980) as it is outside their survey area.}. Most of these stars are nevertheless located along spiral arms, primarily near the center of M33 where the photographic work of Humphreys \& Sandage (1980) was compromised by saturation. But in a few cases (we estimate 2\%), these stars are truly isolated.  Examples of truly isolated WRs are listed in Table~\ref{tab:isolated}. At the opposite extreme, we find small OB associations that are chock-full of WRs.  A good example of this is the OB association 102, which has a diameter about $0\farcm$5 (about 120 pc) but contains 3 WRs.

\subsection{Absolute Visual Magnitudes}
It may be of interest to consider the absolute visual magnitudes ($M_V$) of M33 WRs. The values are poorly known for Milky Way WRs (due to uncertain distances and reddenings), and for the Magellanic Clouds, due to the smaller sample size and limited range of spectral subtypes. In Table~\ref{tab:Mvs} we list the average $M_V$ for each spectral subtype, where we have removed stars of uncertain type or whose spectrum shows an absorption-line signature. We compare these values to those for the  Magellanic Clouds. We have treated all of the data similarly, except that for the SMC nearly all of the stars have an absorption line signature and we have left these in.

There are three interesting things that we note from these data. First, the standard deviations ($\sigma$) are extremely large for all subtypes that have a reasonable number of stars ($N$) in the sample. The data here are for broad-band $V$, and hence can include emission lines, but similarly large scatter is seen in the narrow-band absolute magnitudes given in Table 3-2 of Conti (1988, p.\ 122). We believe that this reflects the fact that stars of a variety of masses and luminosities pass through each subtype; i.e., that WN3 stars (for instance) do not come from one mass progenitor, while WN5 stars come from some other. Second, it is interesting to note that the $M_V$'s of the M33 WRs are no higher or lower, in general, than that of the same subtypes in the LMC. For galaxies beyond the Milky Way, we expect that line-of-sight companions will increasingly contaminate the sample due to crowding, but we expect the presence of such companions bright enough to affect the $M_V$ significantly will manifest itself spectroscopically through the presence of absorption lines. The fact that the $M_V$'s are similar for LMC and M33 WRs without obvious absorption signatures confirms this.  Third, there are 3 WN2s in our ``clean" sample, and these have a considerably brighter $M_V$ than the (one) LMC WN2 star. Based upon very small number statistics one might otherwise infer that as a subclass these stars are particularly faint visually (Conti 1988, p.\ 122), but this may not be true in general.

\subsection{Additional Candidates}
There were only 7 ``rank 1" WR candidates that we didn't have a chance to observe spectroscopically; we expect 5 of these to actually turn out to be WR stars rather than Of-type stars, or (say) quasars.  We list these remaining stars in Table~\ref{tab:more}.

\section{Results}
\label{Res}
\subsection{Completeness}
\label{Complete}
To understand the completeness of our survey, we must examine any biases created during the candidate selection or spectroscopic confirmation processes. Luckily, we are able to ignore any bias caused by differential reddening in M33 because the range in E(B-V) is modest (0.09 -- 0.33, Massey et al.\ 1995). However, there are a few other aspects we must examine.

While the photometry method of candidate selection was able to identify all of the known WR stars, the image subtraction method only identified 129 (or 85\%) of the 153 previously known stars. However, this 15\% difference was not due to any shortcomings in image subtraction per se - instead, it was due to the Mosaic CCD images themselves. When identifying candidate WR stars using the image subtraction method, we required that the same star be seen on each of the three dithered images. Thus, only the area of M33 that was covered by three images was surveyed. This area works out to be $\sim$85\% of the area covered by photometry. All 23 known WR stars (except one which we just plain missed) that were not identified by the image subtraction technique were located in the ``gaps" caused by dithering the Mosaic CCD images. Therefore, we can safely assume that while image subtraction did a wonderful job of finding new WR stars, we missed finding $\sim$15\% of the yet-to-be-found WR stars. However, considering we found 55 new WR stars, and this method was complemented by the photometry, at the high-end this works out to only be 8 stars or $\sim$5\% of the WR stars in M33.

Crowding is often an issue when observing young massive stars because they haven't had time to drift away from their birthplaces in OB associations and tightly packed clusters (Massey \& Conti 1983). By using crowded field photometry, as described in Section~\ref{Phot}, we were able to partially circumvent this issue. However, even though a great deal of effort was spent on this, in some cases the process wasn't as successful as in others. Image subtraction appears to have done a better job identifying WR stars in clusters since the missed stars mostly fell in the Mosaic CCD gaps and not in dense clusters. Thus, the intersection of the two resulting sets forms our best list of candidates.

The most important question we must ask in terms of our completeness is: is our survey sensitive to the weakest-lined WNs? As can be inferred from Table~\ref{tab:WRs}, the identification lines in the weakest-lined WNs are more than $5\times$ weaker than the lines in the weakest-lined WCs. Thus, we will concern ourselves with our WN-completeness and ignore the WCs. It has long been known that the line strength of He II $\lambda$4686 of the most extreme Of-type stars is comparable to the that found in weakest WN stars (Conti \& Frost 1977). As our spectroscopy confirmed, we detected 14 Of-type stars due to their He II $\lambda$4686 emission, and so we might conclude from this alone that our survey was indeed sensitive enough to detect even the weakest lined of WR stars.  However, Massey \& Johnson (1998) argued that in a survey such as ours, what is primarily important are the emission-line {\it fluxes}.   It is clear that for two stars of equal brightness, but with different He II $\lambda 4686$ EWs, that it will be harder to detect the star with the weaker line.  However, it is also true that it will be easier to detect the brighter of two stars with the same EWs of He II $\lambda$ 4686, all other things being equal.  Massey \& Johnson (1998) used the V-band flux of a star (based upon its magnitude) times the EW to approximate the line flux, as absolute spectrophotometry was not available for their data, just as it is not for ours. (Fiber data is notoriously hard to flux-calibrate well.) In Figure~\ref{fig:ofwn} we show the data from the current study.  Among the M33 WNs (open circles) we see the same trend found by Massey \& Johnson (1998): the fainter stars also have the weaker line fluxes (i.e., that the EWs are very similar).  The solid points denote the Of-type stars, and they indeed have a weaker line-flux for He II $\lambda 4686$ than WRs of the same brightness, i.e., they have weaker equivalent widths. We include in this diagram line fluxes of the WNs in the LMC (blue $\times$'s, from Conti \& Massey 1989) as well as for the SMC (red asterisks, from Massey et al.\ 2003). We find that our new survey has detected WNs with lines as weak or weaker than those of WNs in the Magellanic Clouds, galaxies whose WR content is essentially completely known (see discussion in Massey 2003).

It is also important to point out the quality of our spectroscopy, which is better even than that of Massey \& Johnson (1998). With our spectra, we can easily distinguish between Of-type stars and WN+absorption (WN stars whose emission is diluted by the continuum of a companion). Massey \& Johnson (1998) were unable to do this reliably because their signal to noise was not alwasy able to detect the absorption. As an example, see J013419.58+303801.5, whose equivalent width is less than 10~\AA. Generally, we use -10~\AA\ as the division between an Of-type and a WN and so at first pass we may have not classified this star as a WN. However, the signal to noise in our new spectrum is 75 per 6~\AA\ spectral resolution element, and with that we can detect quite a few other He II lines in emission, thus making this star a WN as such emission is not found in Of-type stars. With such high signal to noise, we are confident that if we identified a WR candidate and took spectra of it, we were able to correctly classify the star. This was not always possible in previous studies.

\subsection{The Metallicity of M33}
\label{Sec-metallicity}
As discussed in Section~\ref{INTRO}, the relative number of WCs and WNs (WC/WN) is quite sensitive to metallicity. Massey \& Conti (1983) and Massey \& Johnson (1998) found that in M33 this ratio decreased with galactocentric distance consistent with what was believed to be a strong metallicity gradient at the time (Kwitter \& Aller 1981, Zaritsky et al.\ 1989, Garnett et al.\ 1997).  Recent studies have found a much shallower slope for the oxygen abundance (Crockett et al.\ 2006, Magrini et al.\ 2007, Rosolowsky \& Simon 2008, Bresolin 2011) based upon HII regions with [OIII] $\lambda 4363$ measurements, although Magrini et al.\ (2010) have emphasized that this selection may significantly bias the sample towards low metallicity, particularly in the inner regions.

Figure~\ref{fig:WNratio} shows the distribution of the 206 known WC (red +'s) and WN (blue $\times$'s) stars throughout M33. It is clear that, even with our more complete data, there is a significant difference in the relative numbers of these stars, with the WCs concentrated towards the center. The green ovals represent distances of $\rho=0.25$ (1.9 kpc) and $\rho=0.50$ (3.8 kpc) within the plane of M33. In Table~\ref{tab:WCWN} we list the number of WCs and WNs within the corresponding regions.  The stated errors on the ratios are based upon the assumption of 5\% incompleteness.  We plot these WC/WN values against $\rho$ in Figure~\ref{fig:wcwnrho}. 

How well do these values agree with the measured metallicities for M33? In Figure~\ref{fig:metal}(left) we show the medley of recent determinations of the oxygen abundance plotted against galactocentric distance.  (For easy comparison we have normalized the distance within the plane of M33 in the same way as for the WR data, i.e., $\rho=1$ corresponds to 7.53 kpc)  First, we see that there is good agreement for $\rho>0.4$ (3~kpc).  Inwards of this region, studies have been usually been compromised by the scant number of HII regions that have been analyzed (Crockett et al.\ 2006) or large scatter (Rosolowsky \& Simon 2008). Bresolin (2011) has demonstrated that this scatter is not intrinsic but is instead due to the measuring errors of the very weak [OIII] $\lambda 4363$ lines needed to derive the electron temperature in higher metallicity regions.  Bresolin (2011) was able to measure this line in 8 regions inwards of 3~kpc, and derives the oxygen abundances shown in Figure~\ref{fig:metal} by the dashed red line. He does show that if one ignores the direct electron temperatures, and instead uses the so-called $R_{23}$ method (Pagel et al.\ 1979) on a larger sample, one derives a much higher metallicity. However Bresolin (2011) emphasizes that the exact value depends upon which high-metallicity calibration of the $R_{23}$ index one adopts, although he does find a nearly identical slope as he does with the data for which there are direct temperature measurements. Magrini et al.\ (2010) argues that by restricting the analysis to HII regions with {\it measurable} [O III] $\lambda 4363$ lines, one is automatically biasing the result towards lower metallicities, a selection effect which Bresolin (2011) acknowledges.  Instead, Magrini et al.\ (2007) proposes adopting a two-component model (shown in Figure~\ref{fig:metal} by the solid green lines), where the inner oxygen abundances are based upon the oxygen abundances of young stars.  In Figure~\ref{fig:metal}(right) we emphasize just the Magrini et al.\ (2007)  and the Bresolin (2011) results, with vertical dashed lines to indicate the effective $\rho$ values for the three sample regions. 

We see from Figure~\ref{fig:wcwnrho} that the WC/WN ratio in the outer two bins ($\overline{\rho}=0.38$ and 0.69) are similar to that seen in the LMC.  Indeed the metallicities of these two bins are $\log (\rm O/H)+12=8.4$ and $8.3$, respectively, according to Figure~\ref{fig:metal}, perfectly consistent with the oxygen abundance of the LMC, 8.37 (Russell \& Dopita 1990). However, the innermost bin ($\overline{\rho}=0.16$) must have a metallicity that is {\it considerably} greater than that of the LMC based on its high WC to WN ratio. Were we to linearly extrapolate from the SMC to the LMC we would expect a $\log(\rm O/H)+12$=8.9-9.0.  This is clearly at variance with that of the Bresolin (2011) result, but is nicely in accord with the two-component gradient proposed by Magrini et al.\ (2007).

We can provide one other bit of information on the metallicity gradient of M33. It has long been known (Smith 1968c) that nearly all of the late-type WC stars are found inwards of the solar circle, while the early-type WCs are found outwards of the solar circle.  Smith (1968c) supposes that this is due to the metallicity gradient in the Milky Way, as the result is consistent with the fact that the SMC and LMC (which are metal poor compared to the solar neighborhood) have only early-type WCs. Massey \& Johnson (1998) note that it is only M31 (where the metallicity is higher) that contains an abundance of late-type WCs. The interpretation that the spectral subtype of the WC class is determined by initial metallicity has been given a theoretical underpinning by Crowther et al.\ (2002).  Furthermore, late-type WCs have CIV $\lambda 5806$ lines that are weaker (smaller EWs) and skinnier (smaller FWHMs) than in early-type WCs, so a change in average spectral type can be demonstrated using a plot of these two quantities whether or not the spectral subtype has been well determined.  Massey \& Johnson (1998) used such a diagram to demonstrate that the WCs in the inner region of the M33 were later in type than those in the outer regions. With the considerably more complete data here, we revisit this issue in Figure~\ref{fig:WCls}. In the three upper panels we see a very clear progression from the inner regions ($\rho<0.25$), where the WCs are dominated by stars with narrower CIV $\lambda 5806$ lines (i.e., mostly of later type), while the outer region ($\rho\ge 0.50$) are dominated by WCs with stronger and broader lines. The middle region ($0.25\le\rho<0.50$) is intermediate, but more similar to the outer region than the inner. This provides a largely {\it independent} confirmation that the metallicity difference from the inner region to the outer region differs by some significant amount.  It is hard to place an exact value on this, but we can compare the distributions to that of the Milky Way and LMC in the lower two panels.  Clearly the distribution of the WCs in the inner region (upper left) is more like that of the Milky Way (lower left) than like that of the LMC (lower right), while the outer region of M33 (upper right) is more like that of the LMC.  Thus, to the extent that the WC spectral properties are dominated by metallicity effects, the inner region of M33 ($\rho<0.25$) must have an oxygen abundance that is more similar that of the solar neighborhood ($\log(\rm O/H)+12 = 8.7$, Esteban \& Peimbert 1995) than that of the LMC ($8.4$).  This provides additional, independent support for the higher metallicity value for the center.

In the following section we will therefore adopt the two-component oxygen gradient of Magrini et al.\ (2007).

\subsection{Comparison with the Evolutionary Models}
We can now compare our observed WC to WN ratios with the predictions of the Geneva evolutionary models. In this paper we have chosen to use only the galaxies where the WR content is well-known (i.e., SMC, LMC and now M33) when comparing the ratios with the models. For other galaxies in the Local Group, the surveys are thought be selectively incomplete for WNs (Massey \& Holmes 2002). This is especially true for the Milky Way where the interstellar extinction further complicates the matter (Shara et al.\ 2009). Figure~\ref{fig:modelComp} shows both the expected ratio of WCs to WNs as computed by Meynet \& Maeder (2005) as well as what we observe, with the data taken from Table~\ref{tab:WCWN}. 

As described by Meynet \& Maeder (2005), the models fail for high WC to WN ratios (such as those believe to exist in M31 and the Milky Way), but there has always been the lingering doubt that incompleteness in the observational bias towards WCs might be possibly responsible (see discussion in, for example, Massey \& Holmes 2003 and Meynet \& Maeder 2005).  The result that the discrepancy exists in the inner part of M33 is new here. Note too that this result is robust to any uncertainty in the metallicity of the center of M33 (i.e., Section \ref{Sec-metallicity}) as the rotating models fail to predict this high a value of the WC/WN ratio at {\it any} metallicity.  As long as the WN content of the inner bin is relatively well known, as we argue above, then the problem exists.

\section{Summary and Conclusions}
\label{Con}
M33 contains 206 confirmed WR stars, 55 of which were discovered as part of this survey. Based on our remaining candidates, we expect that there may be another half-dozen WRs that have yet to be found. This number is in accord with the $\sim$190 WR stars we expect by scaling from the LMC based on luminosity. This suggests that the current massive star formation rates in M33 are comparable with those in the LMC per unit luminosity. Overall, the dominant WC subclass is WC4 while there are a few later type WCs (WC7s) located in the central region of M33. The WNs are dominated by early-type WNs, with some late-type WNs found predominantly in rich regions of massive stars, although a few are located in the field. 

The WC/WN ratio changes dramatically with galactocentric distance in M33. This can be seen quite obviously in Figure~\ref{fig:WNratio} and in Table~\ref{tab:final}. We believe this is due to a metallicity gradient, consistent with the two-component model observed by Magrini et al.\ (2007).  This is in accord with the basic premise of massive star evolutionary theory (the ``Conti Scenario", Conti 1976): in a high metallicity environment, mass-loss rates will be higher for a given luminosity and mass, and thus more stars will be found in the WC stage at a given time. Massey \& Conti (1983) suggest the change in the WC/WN ratio with galactocentric distance might be due to a change in the IMF slope, but current studies in regions of star formation all suggest that the IMF slope is invariant at the upper end (Bastian et al.\ 2010). Additionally, WR stars can also be formed as a result of Roche-lobe mass transfer in a binary system, but we do not expect binary WRs to significantly influence M33's overall WC/WN ratio.

While the Geneva models (Meynet \& Maeder 2005) do an good job at predicting the WC/WN ratio at lower metallicities, they clearly fail at solar metallicities (i.e., $\log(\rm O/H) + 12 \sim 8.7$) by not predicting a sufficiently high WC/WN ratio. Eldridge \& Vink (2006) suggest that the problem with the evolutionary models is solved by adding a metallicity-dependence mass-loss rate to the stellar winds during the WR stage. However, Meynet et al.\ (2008) argue that such models severely underestimate the number of WRs relative to O-type stars, and instead proposes that the problem is alleviated if an LBV phase is included for the very massive stars that enter a WN phase while still H-burning.  Including an LBV stage would then significantly shorten the predicted WN lifetime, increasing the predicted WC/WN ratio at higher metallicities. Meynet et al.\ (2008) report that tests of the Geneva models at solar metallicity confirm that this helps.

In the future, we plan on conducting similarly complete surveys of M31 ($\log(\rm O/H)+12 \sim 9.0$), IC10 ($\log(\rm O/H)+12 = 8.25$) and NGC6822 ($\log(\rm O/H)+12 = 8.25$). This will allow us to further understand the relationship between metallicity and the WC to WN ratio and it will allow the theorists to improve heavily-relied-upon evolutionary models.

\acknowledgements
We would like to thank Shay Holmes Strong for completing the photometry described in Section~\ref{Phot} and Susan Tokarz for reducing the spectra as described in Section~\ref{SpecCon}. Additionally, we thank Carl Akerlof and Fang Yuan for their helpful suggestions on compiling and running their Astronomical Image Subtraction by Cross-Convolution program. We are grateful for useful comments on this manuscript by Georges Meynet, Laura Magrini, Carol Kehrig, and Peter S.\ Conti. Howard Bond, Steve Howell, Bill Keel, and Scott Kenyon all made useful comments on assorted spectra we sent their way. An anonymous referee made many valuable suggestions which improved the paper. This work was supported by the National Science Foundation under AST-1008020.

\begin{figure}
\epsscale{0.33}
\plotone{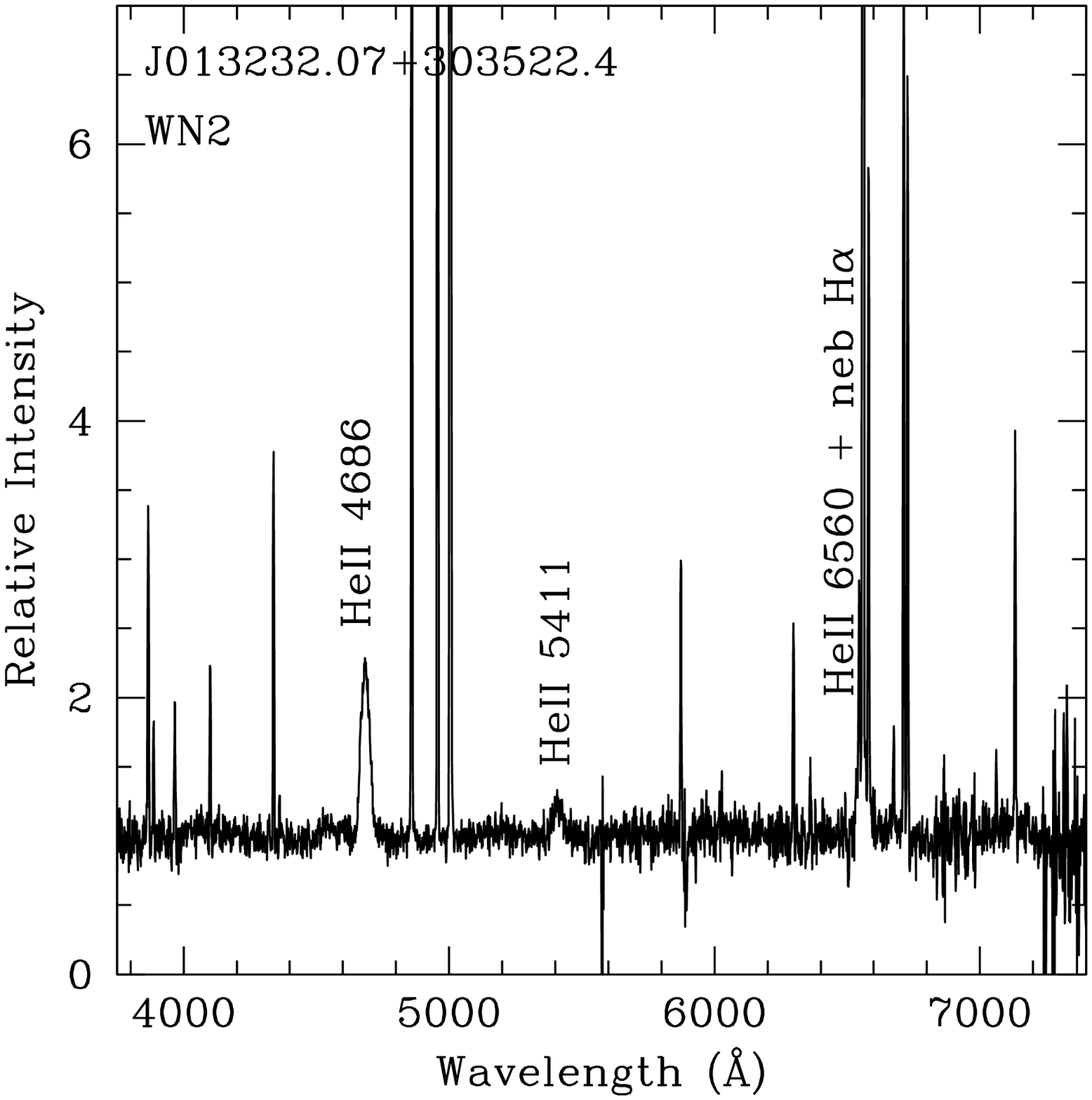}
\plotone{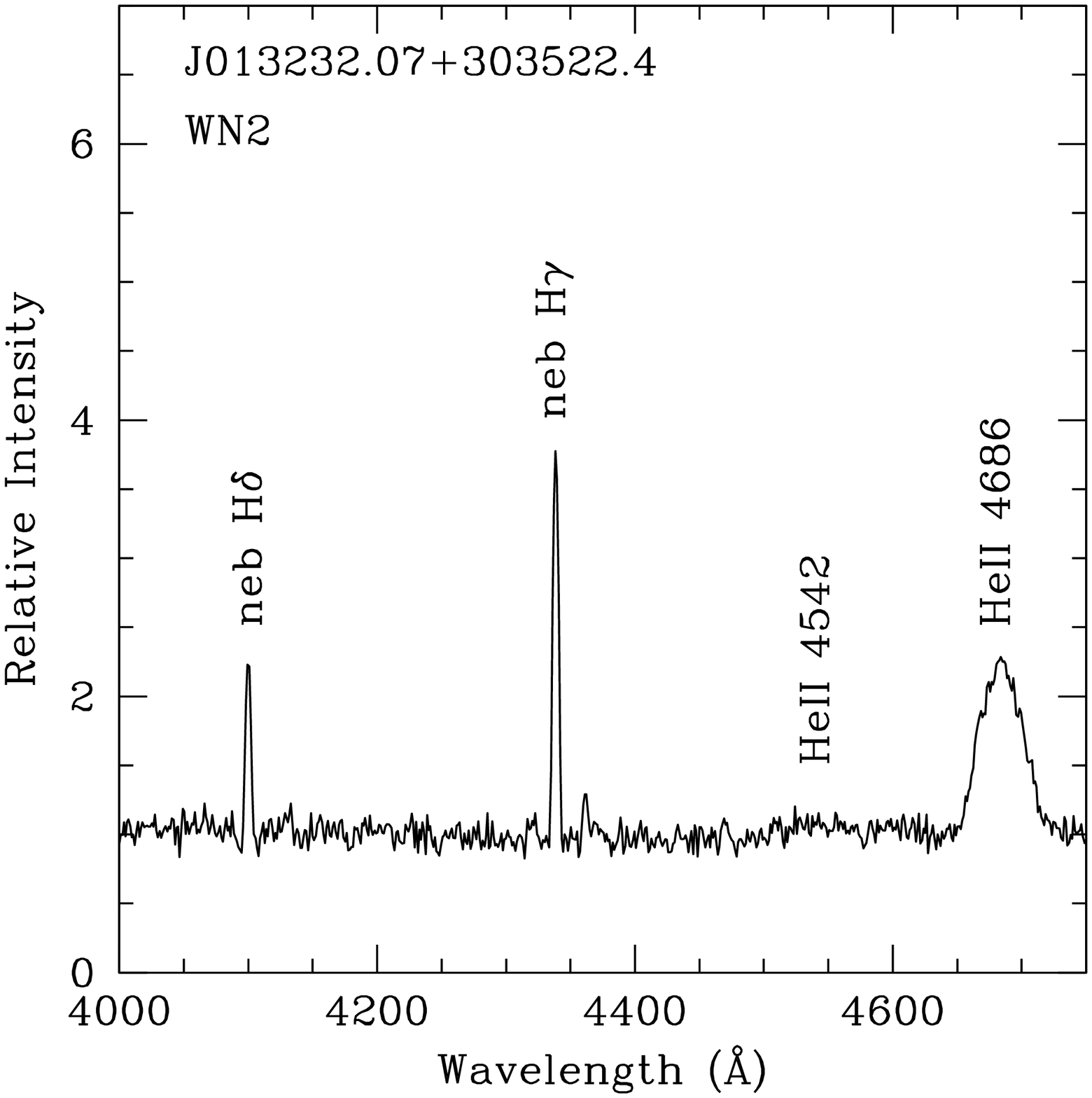}
\plotone{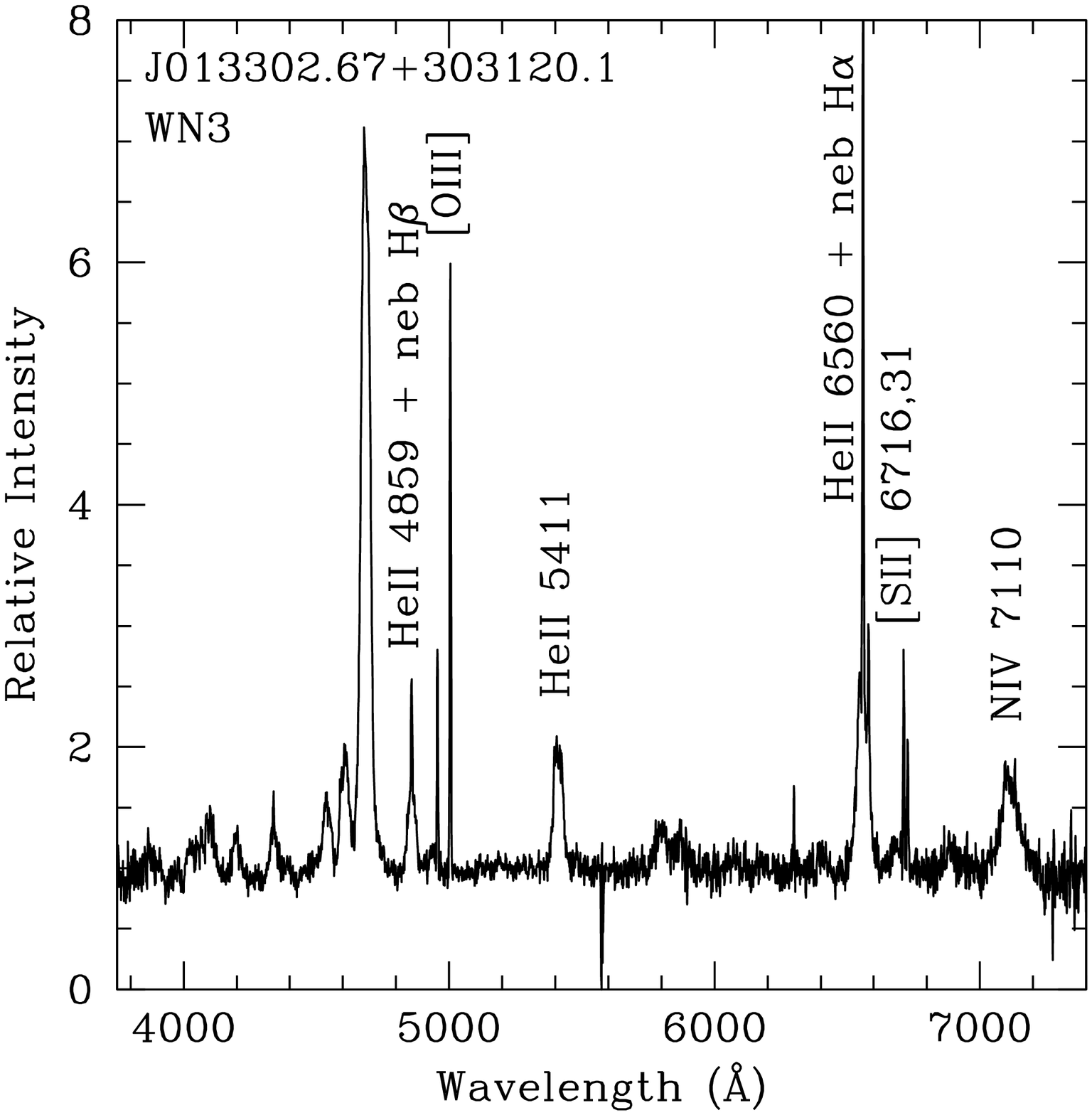}
\plotone{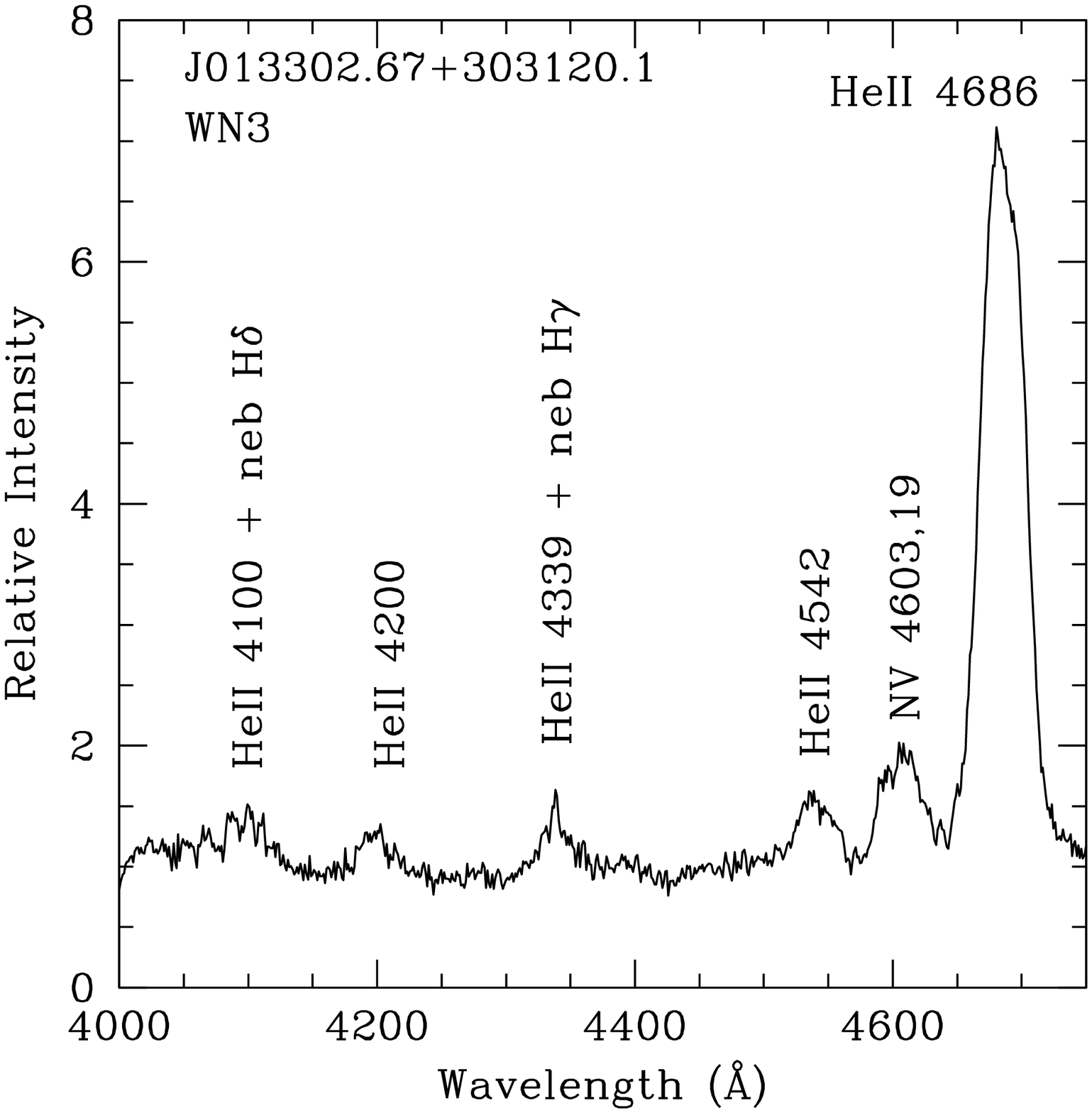}
\plotone{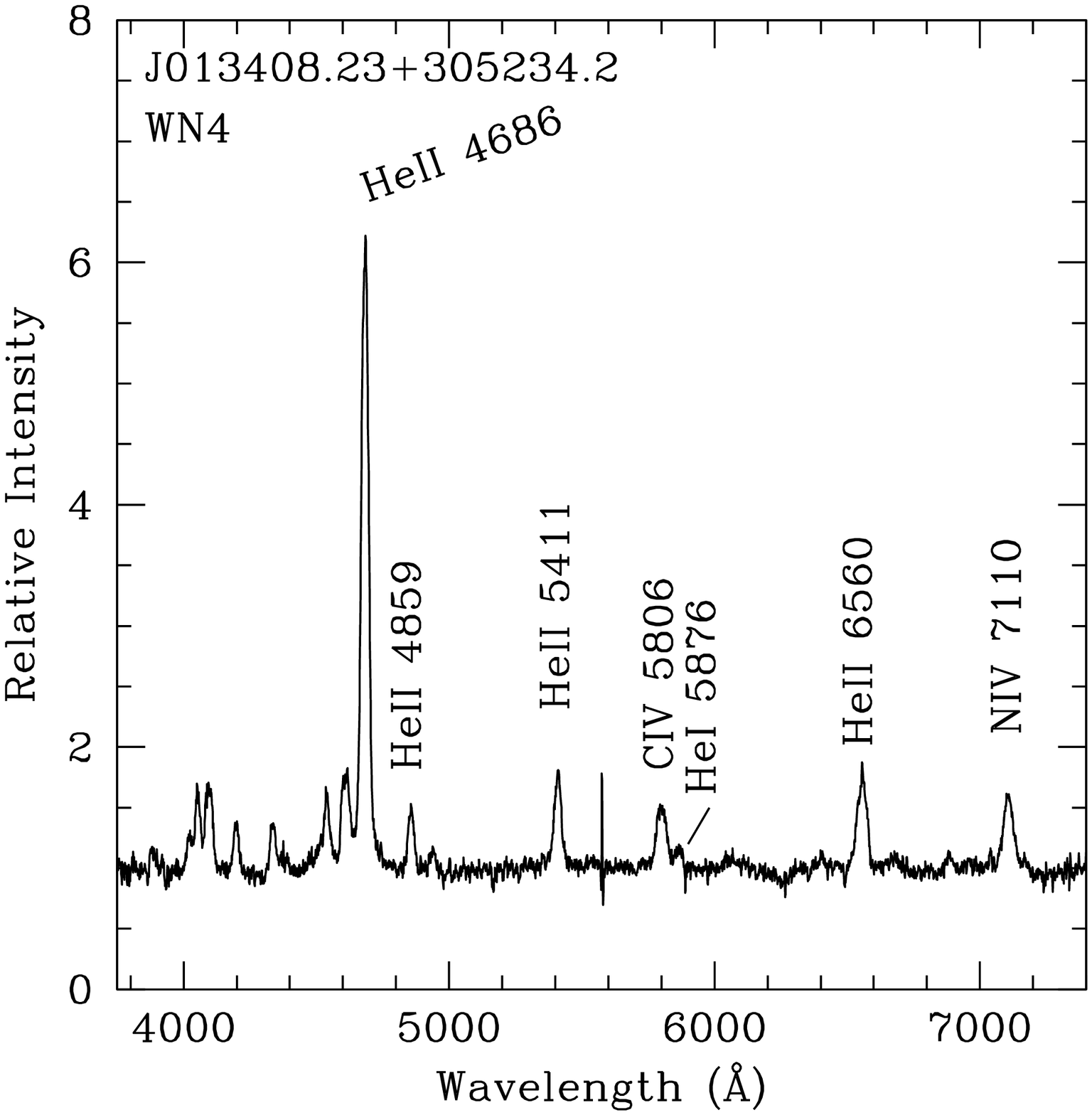}
\plotone{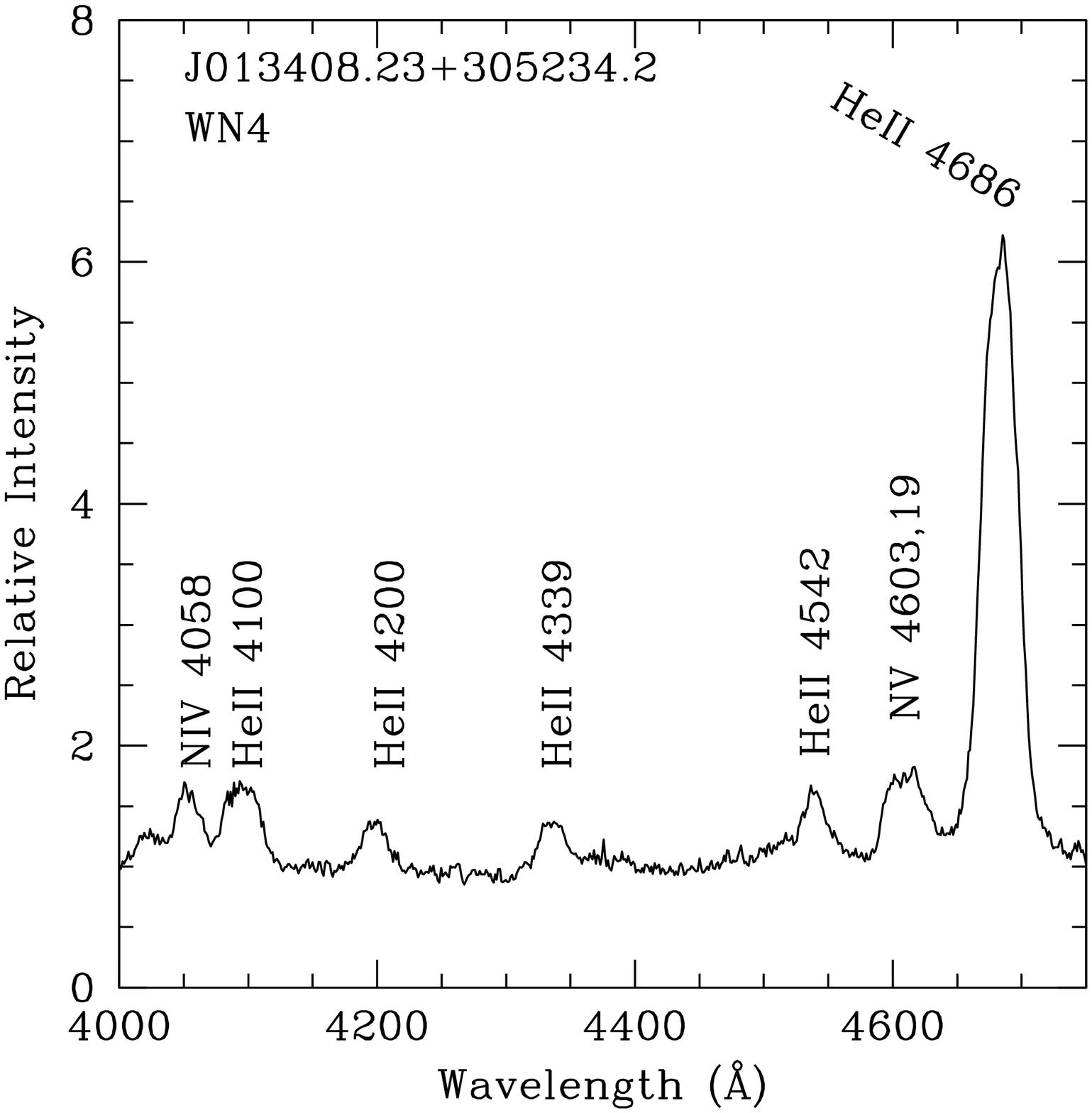}
\caption{\label{fig:wns} The spectra of representative WN stars.  On the left we show the 3750-7400 \AA\ region of the spectrum, while on the right we show an expansion of just the blue region. The latter contains the N III $\lambda 4634, 42$, N IV $\lambda 4058$, and NV $\lambda 4603, 19$ lines which form the primary basis for the classification. We display one example for each of the WN subtypes found in M33.}
\end{figure}

\addtocounter{figure}{-1}
\begin{figure}
\epsscale{0.33}
\plotone{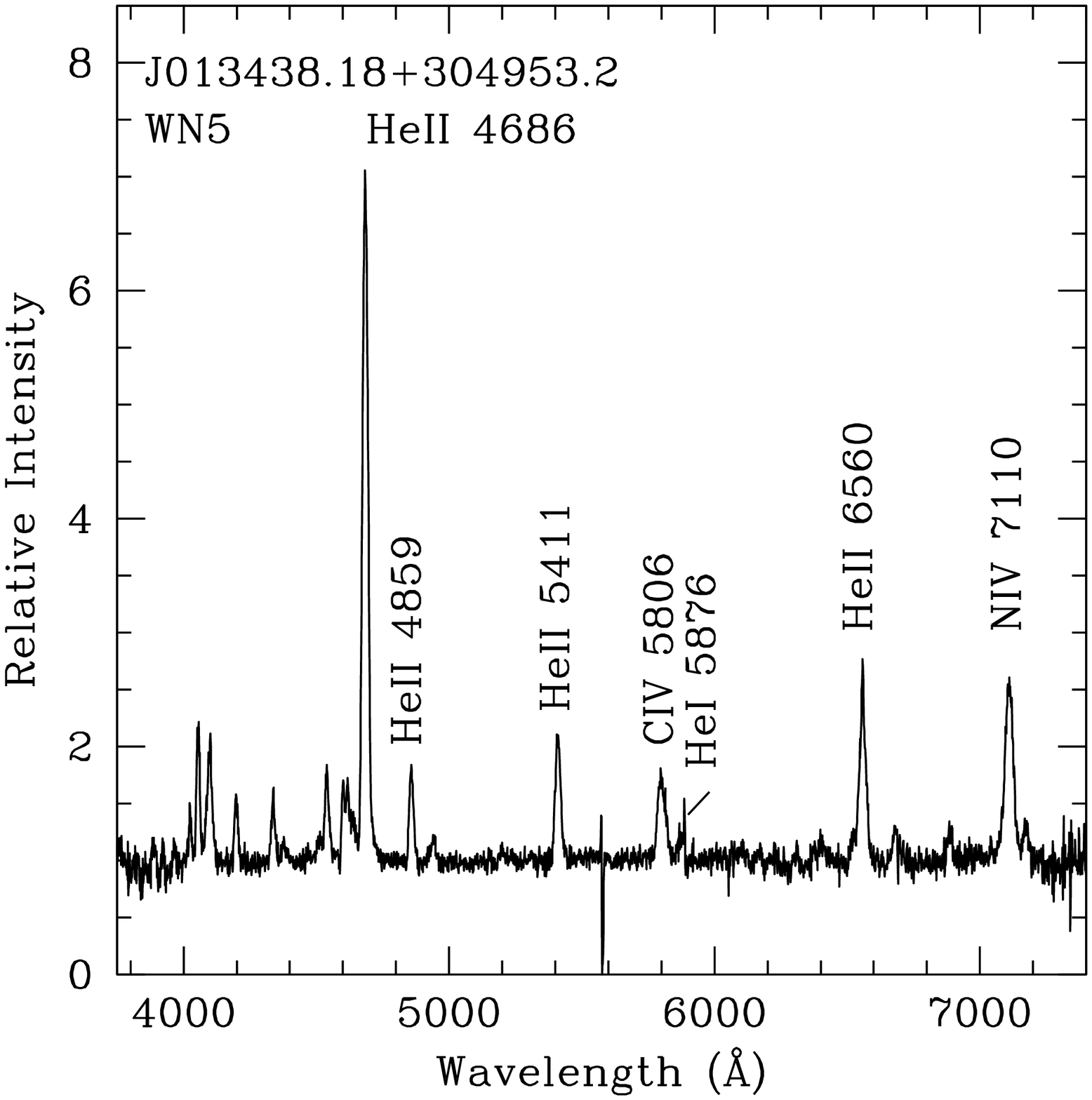}
\plotone{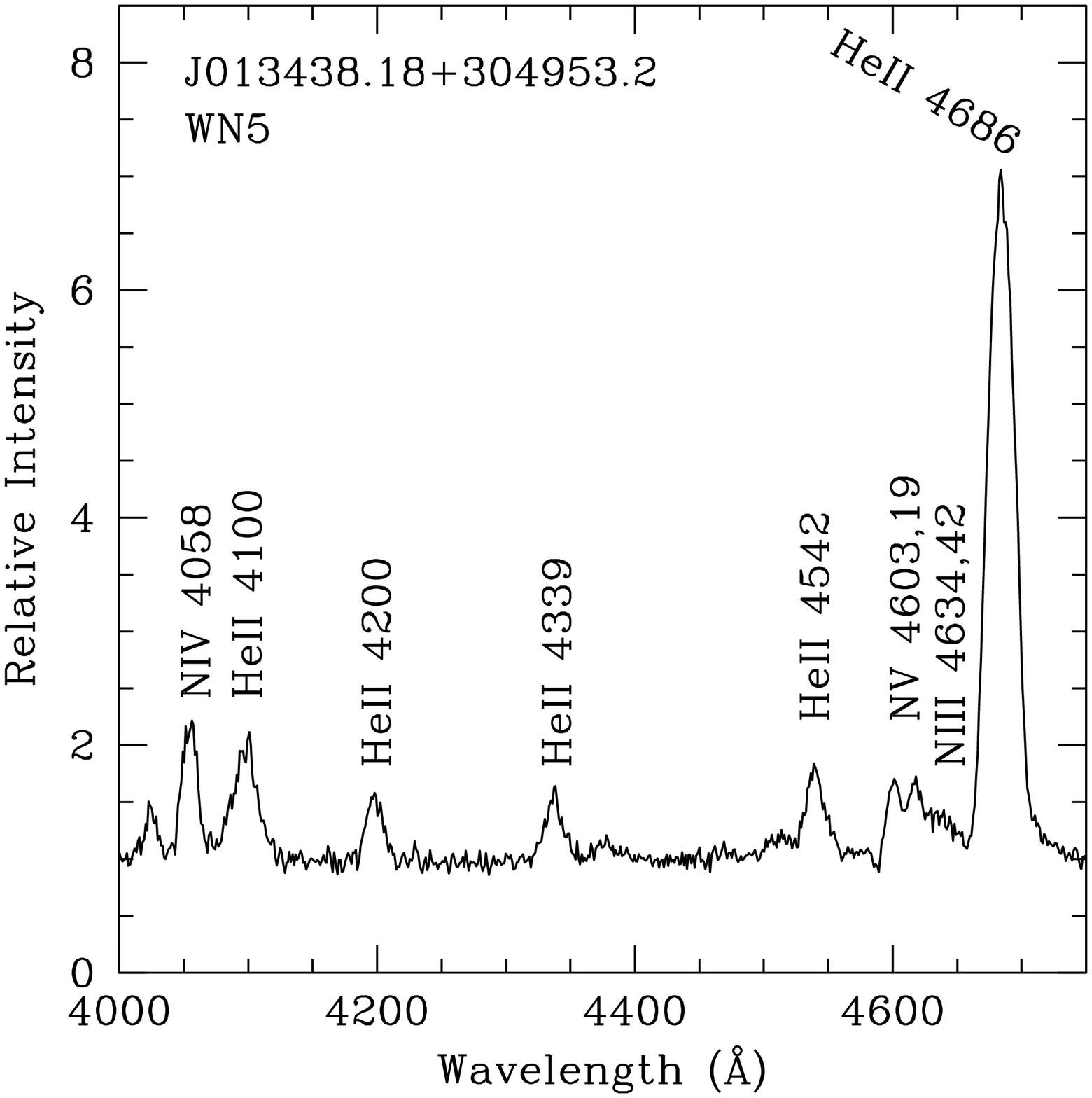}
\plotone{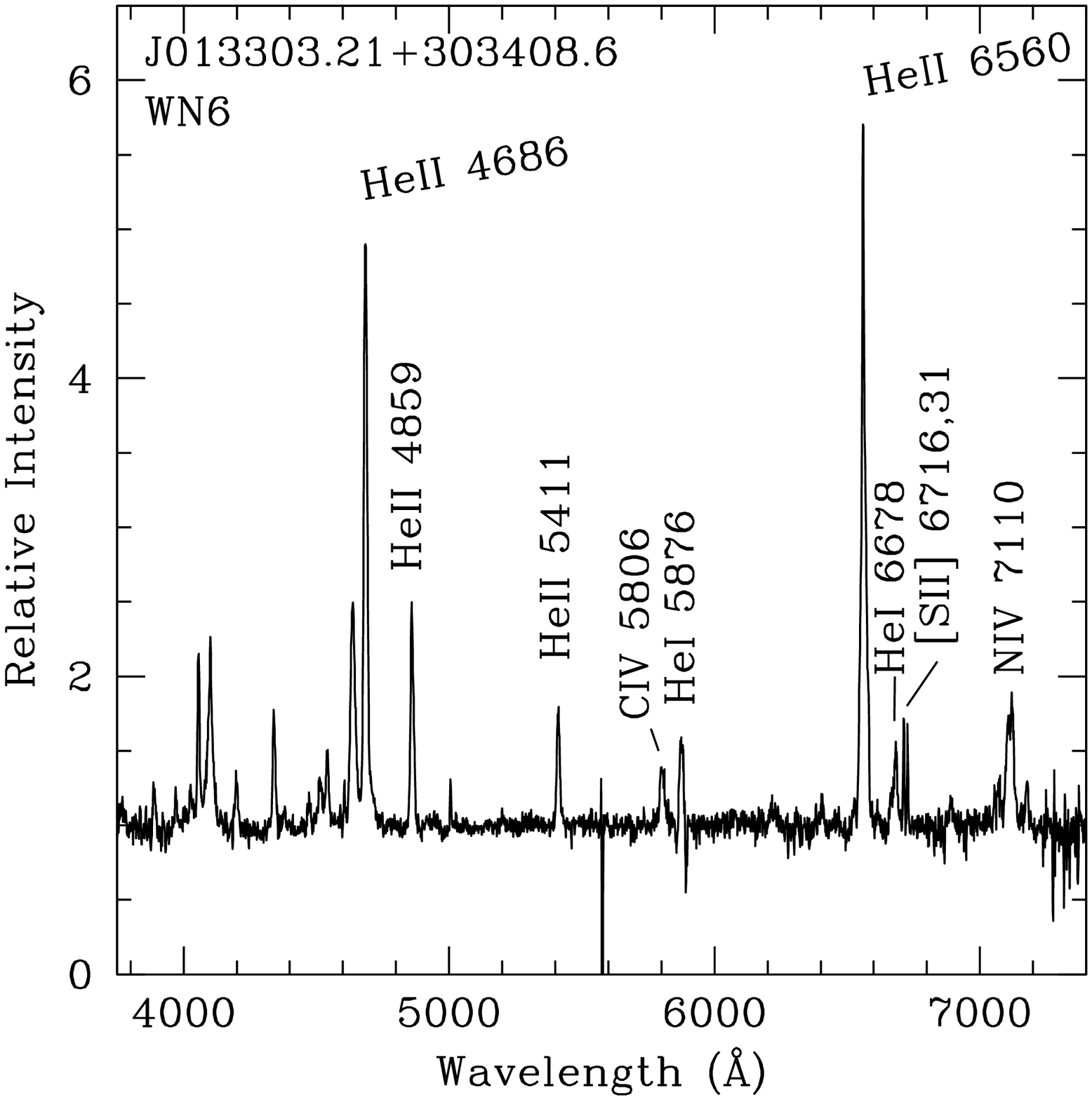}
\plotone{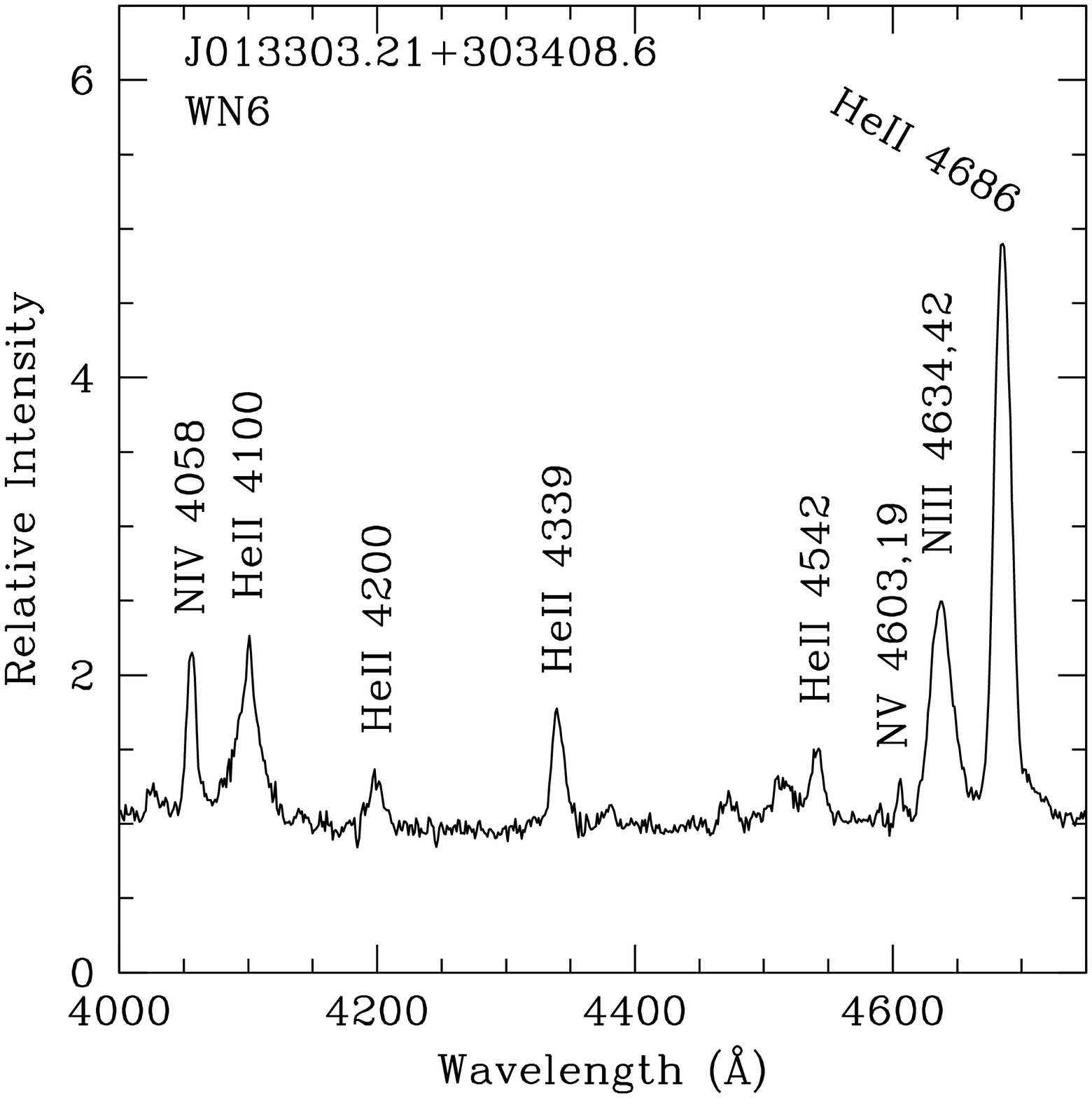}
\plotone{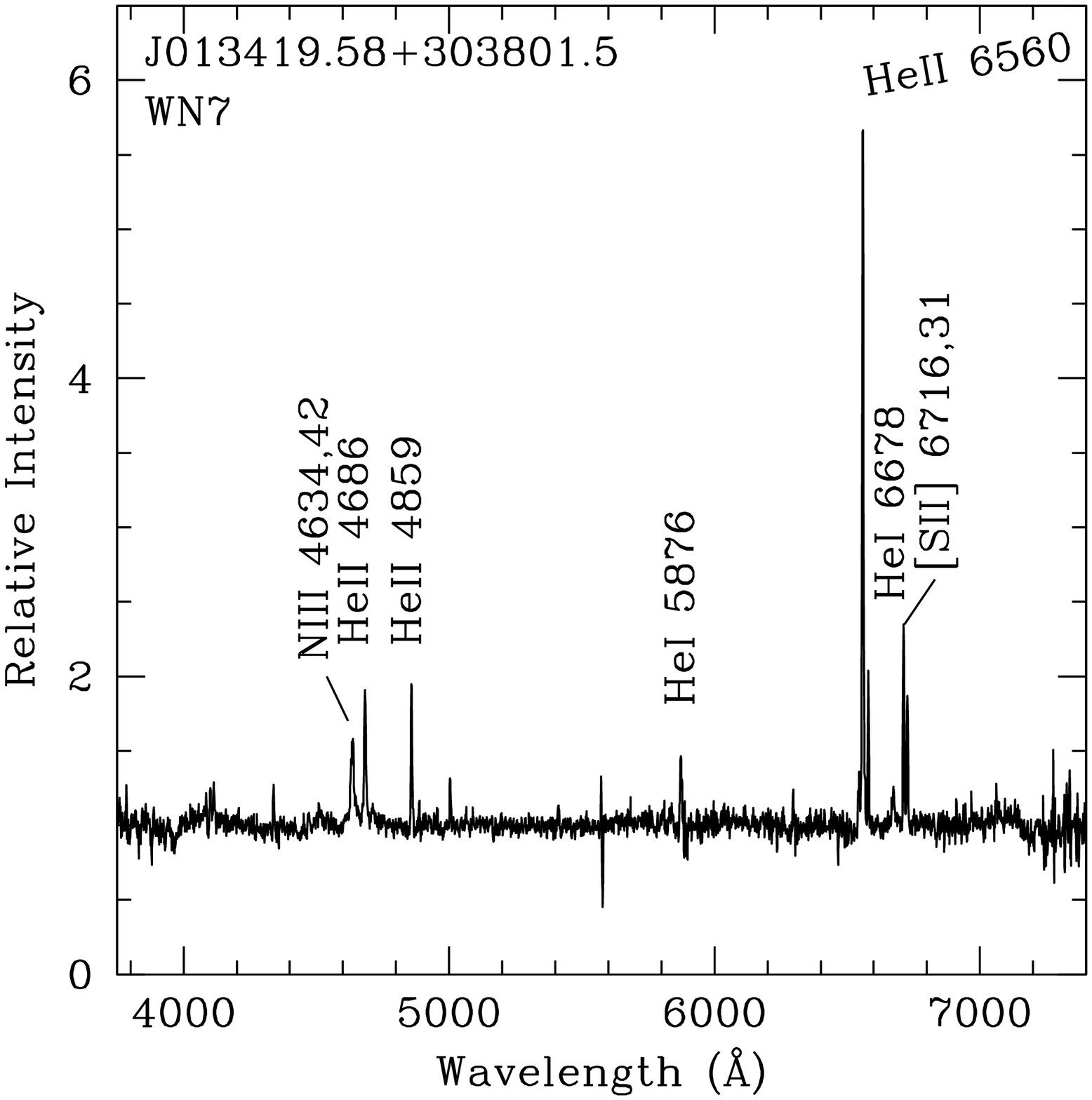}
\plotone{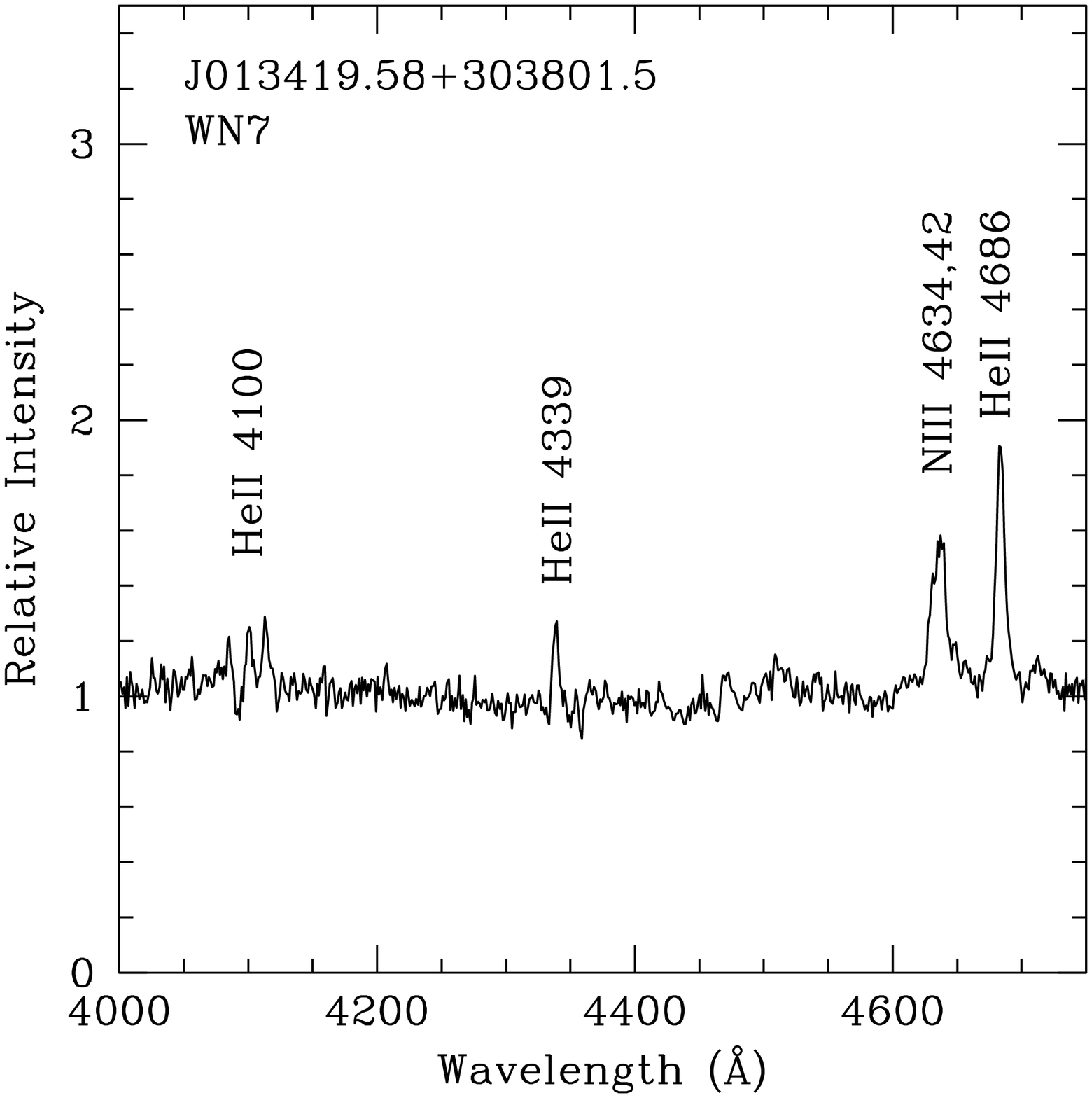}
\caption{Continued}
\end{figure}

\addtocounter{figure}{-1}
\begin{figure}
\epsscale{0.33}
\plotone{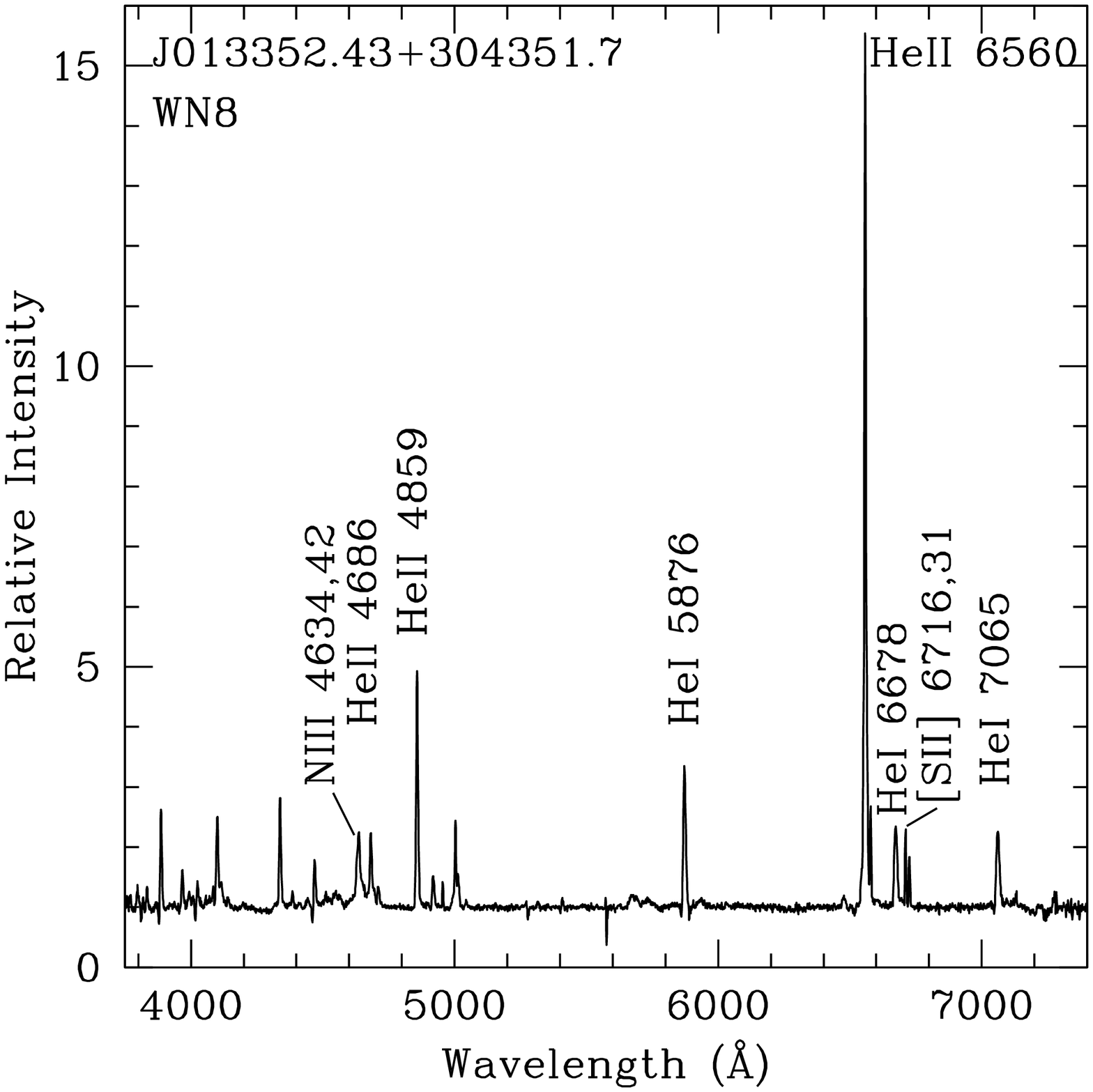}
\plotone{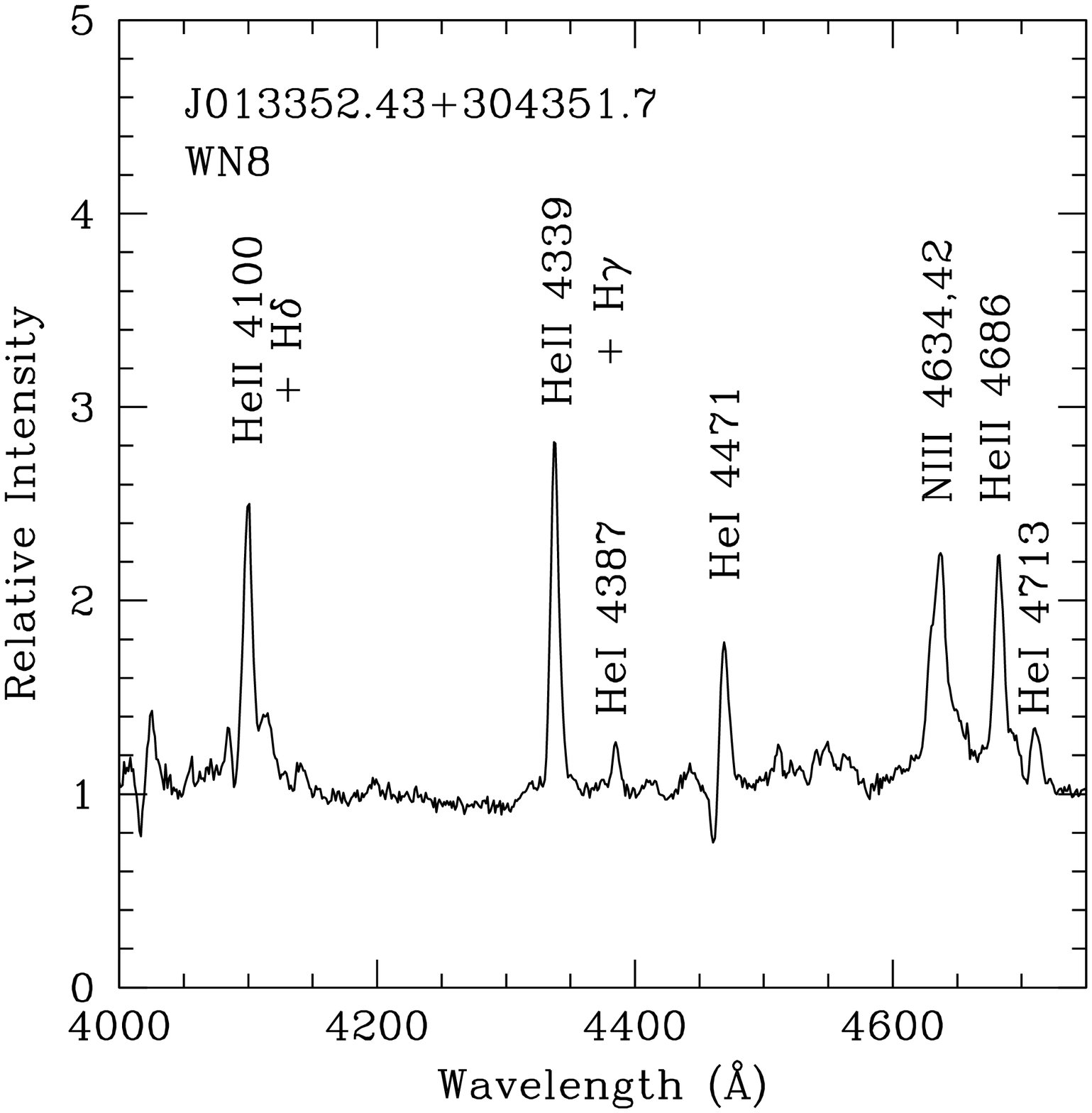}
\caption{Continued}
\end{figure}

\begin{figure}
\epsscale{0.3}
\plotone{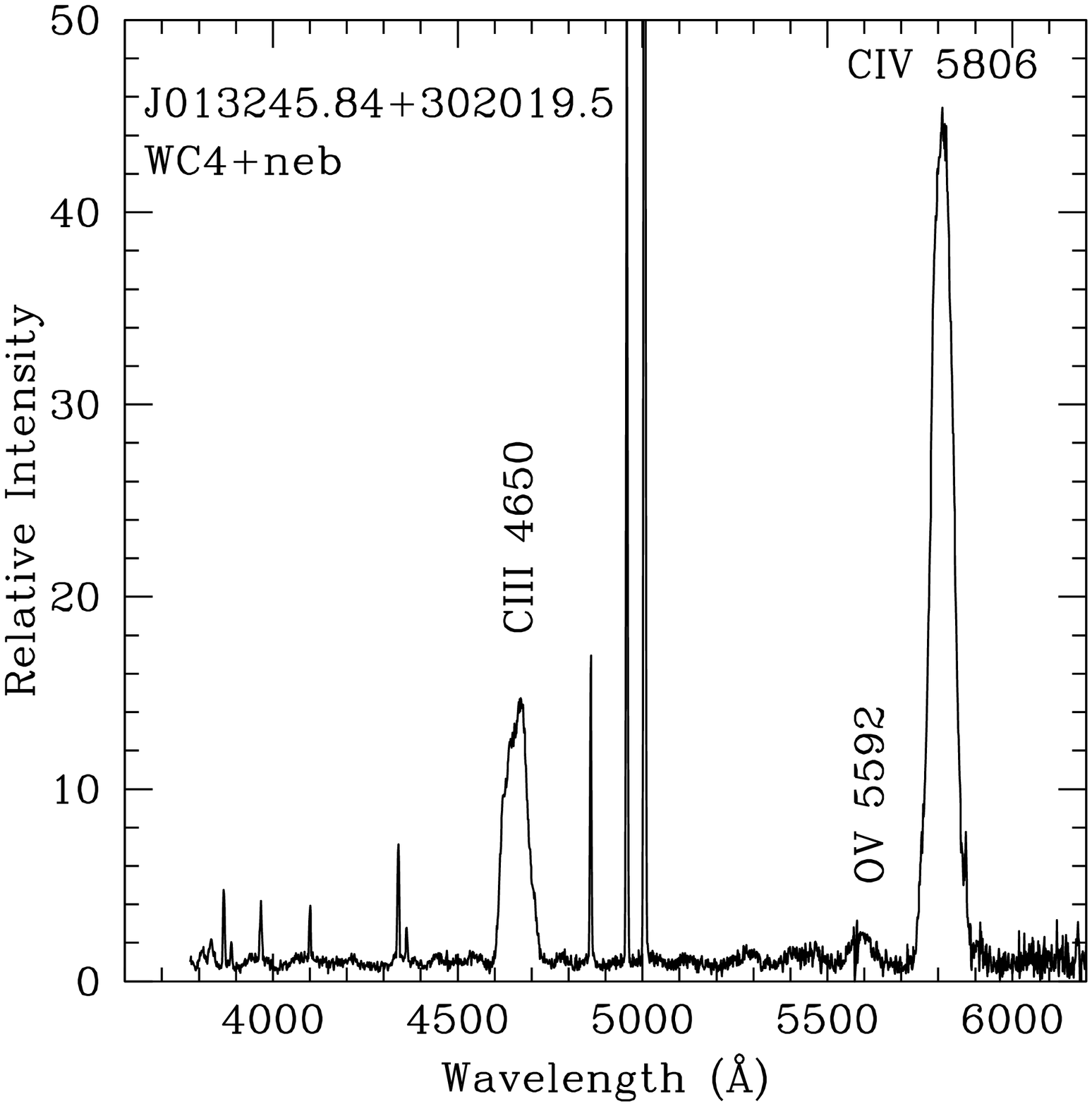}
\plotone{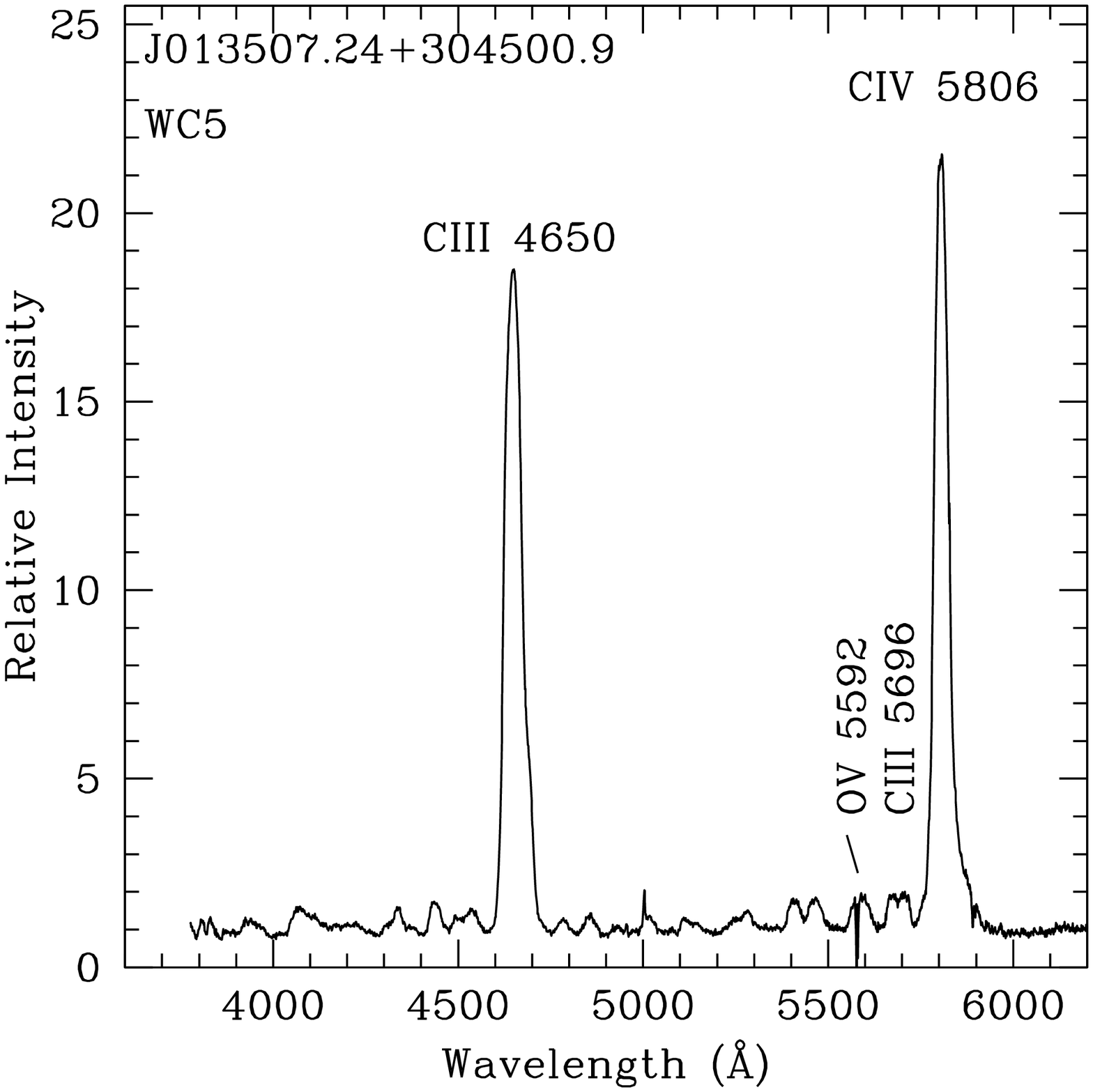}
\plotone{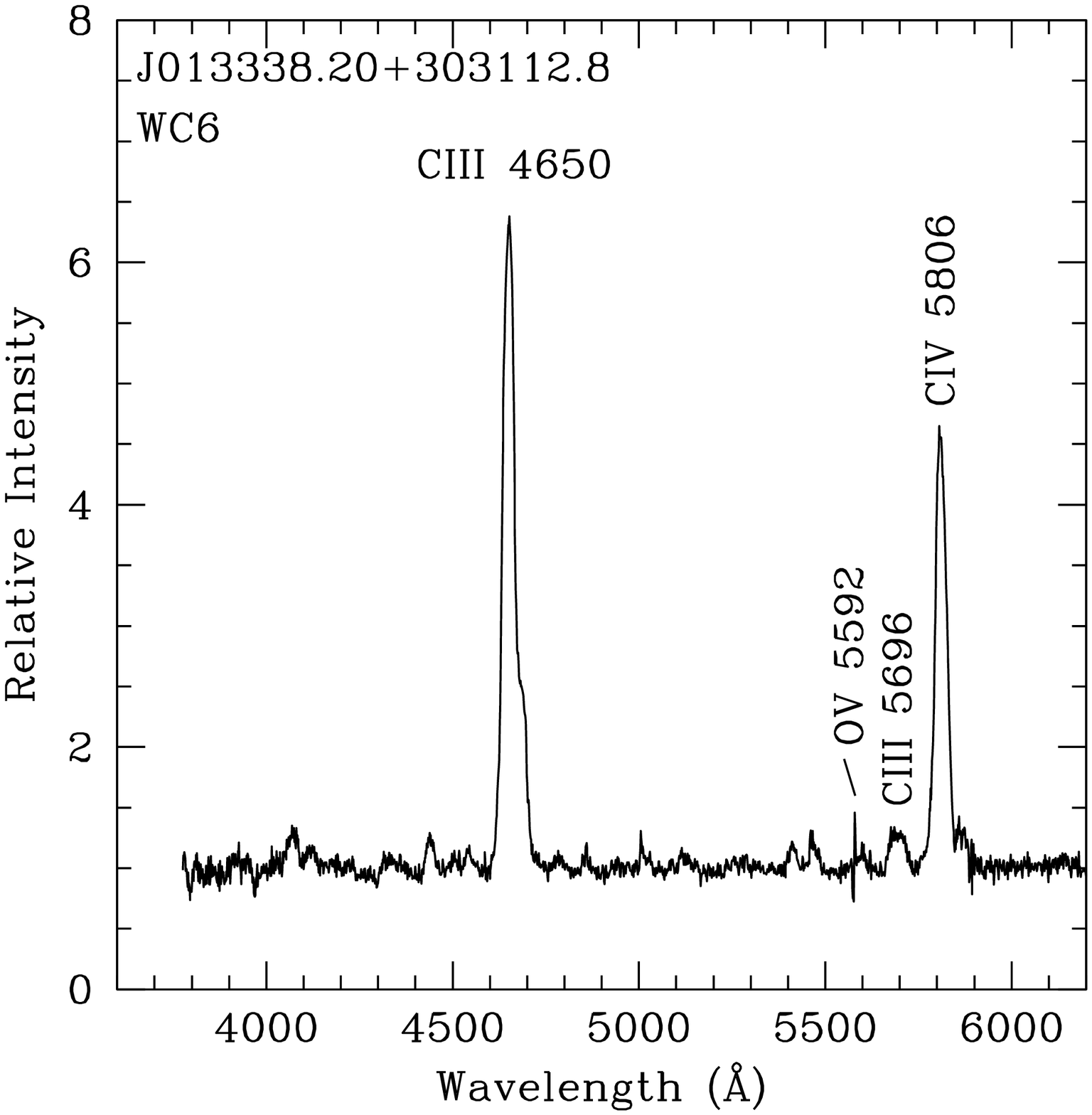}
\caption{\label{fig:wcs} Three WC stars.  The classification of WC stars is based upon the relative strengths  of  C III $\lambda 5696$, O V $\lambda 5592$, and C IV $\lambda 5806$. Of these three, J013338.20+303112.8 (WC6) was previously known (MC 35).  The unmarked narrow lines in the spectrum of J013245.84+302019.5 are nebular. We display examples of each WC subtype found.}
\end{figure}

\begin{figure}
\epsscale{0.48}
\plotone{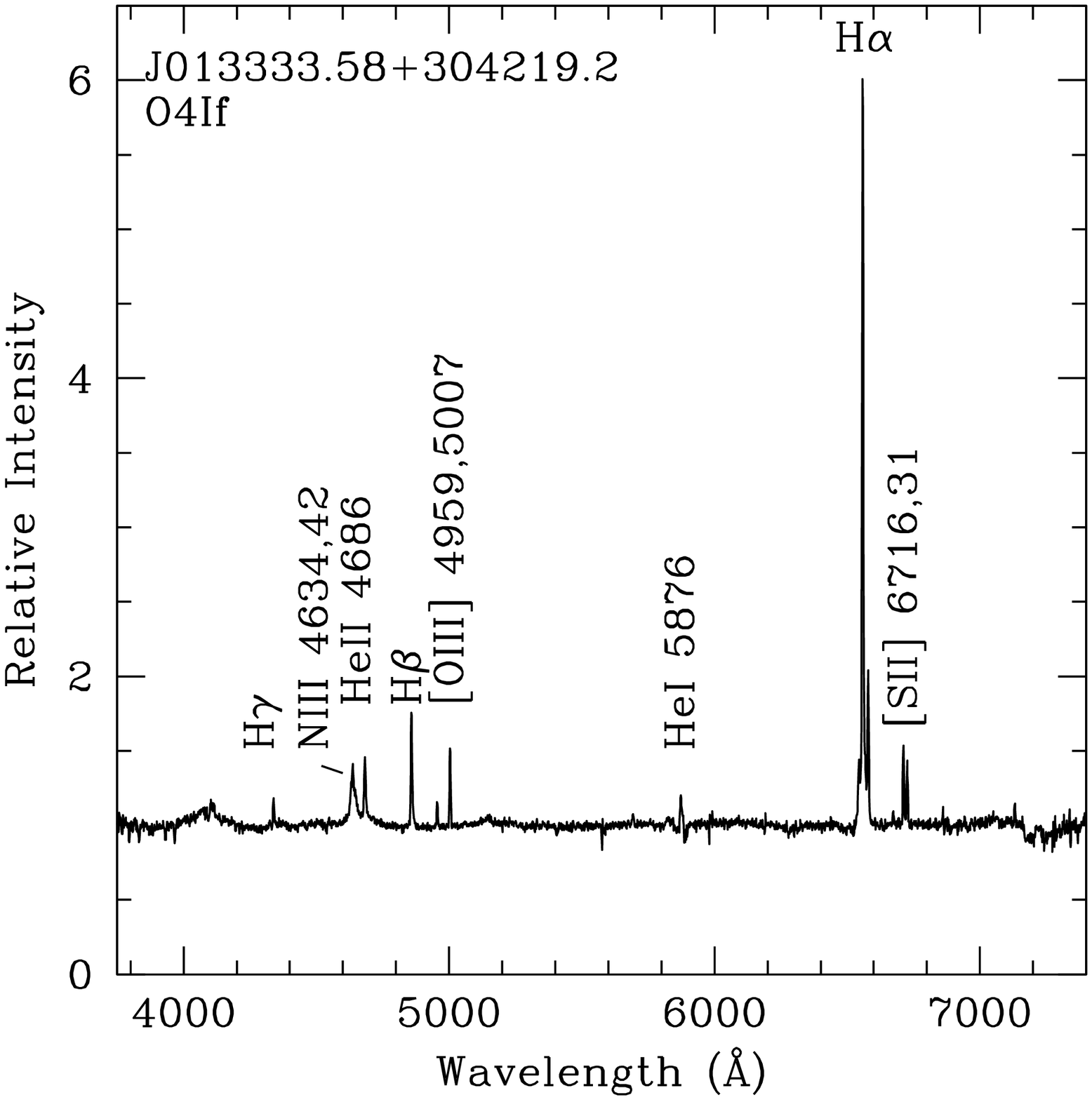}
\plotone{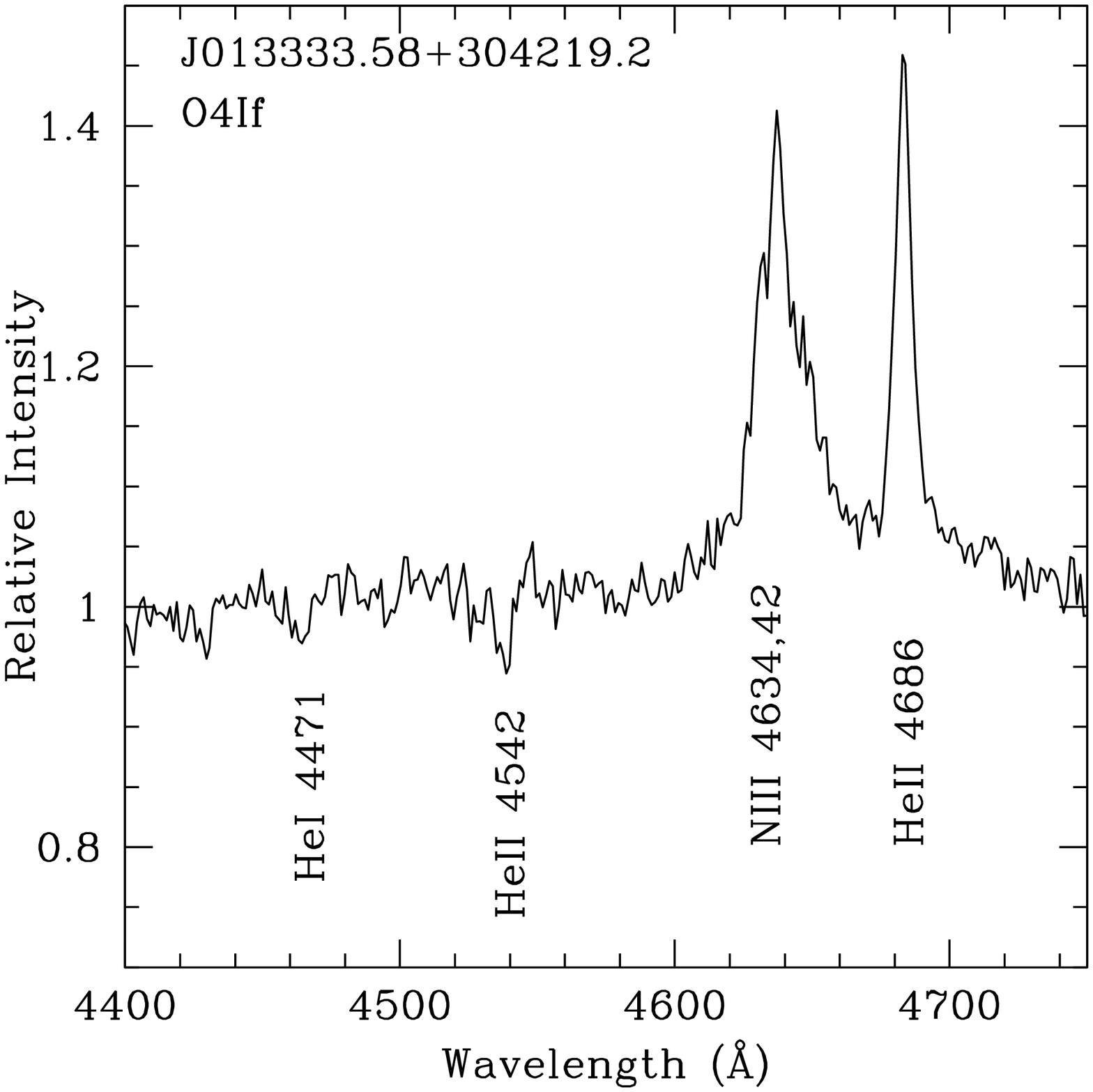}
\caption{\label{fig:o4} The spectrum of the O4 If star J013333.58+304219.2, previously (mis)classified as a WN star by Massey \& Johnson (1998).  Although superficially one might be inclined to describe the spectrum (left) as that of a late-type WN star, the presence of He I and He II absorption features (right) shows that this is actually an O-type star, which we classify here as O4 If.  This is the earliest type supergiant classified to date in M33.}
\end{figure}

\begin{figure}
\epsscale{0.35}
\plotone{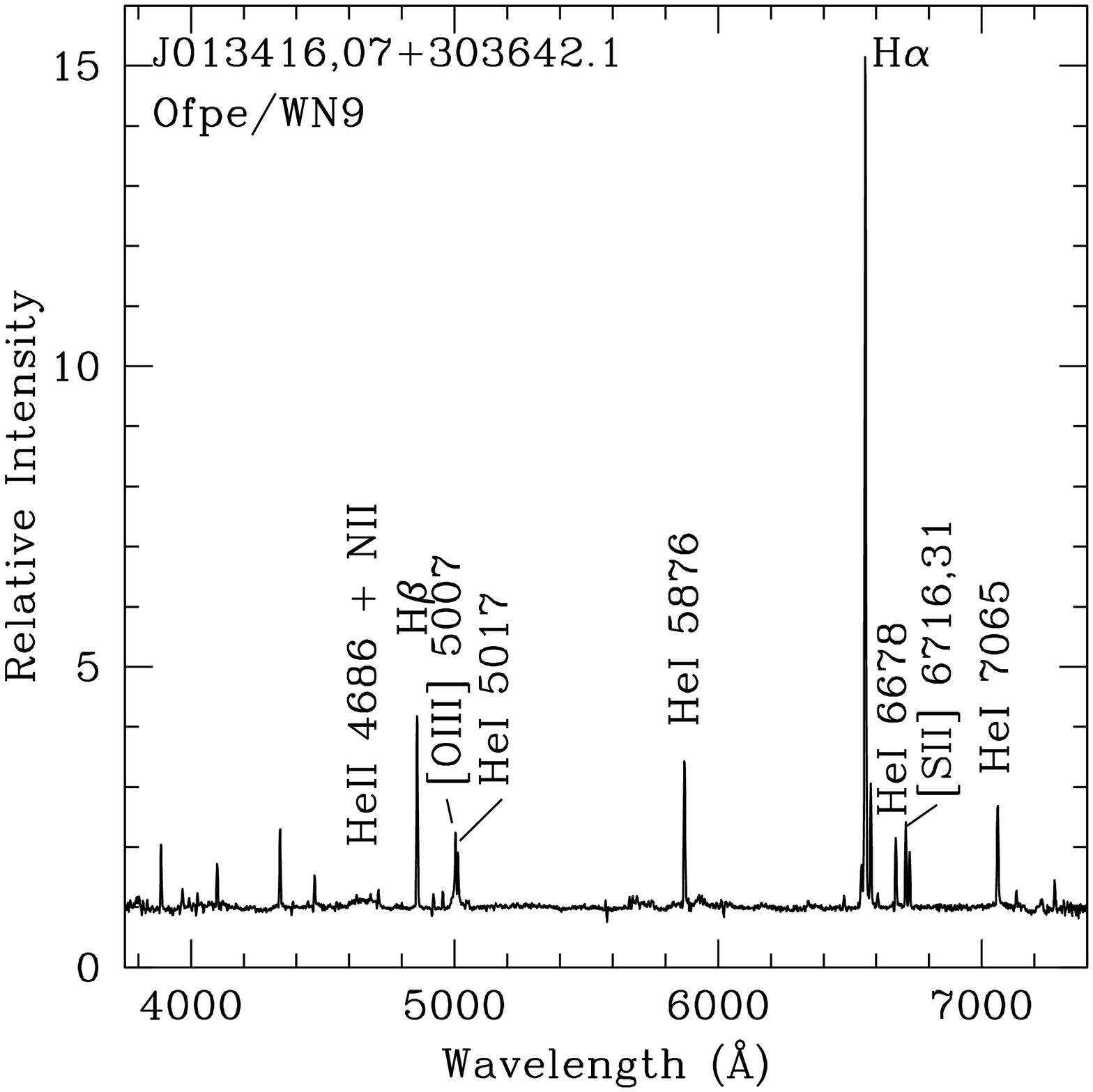}
\plotone{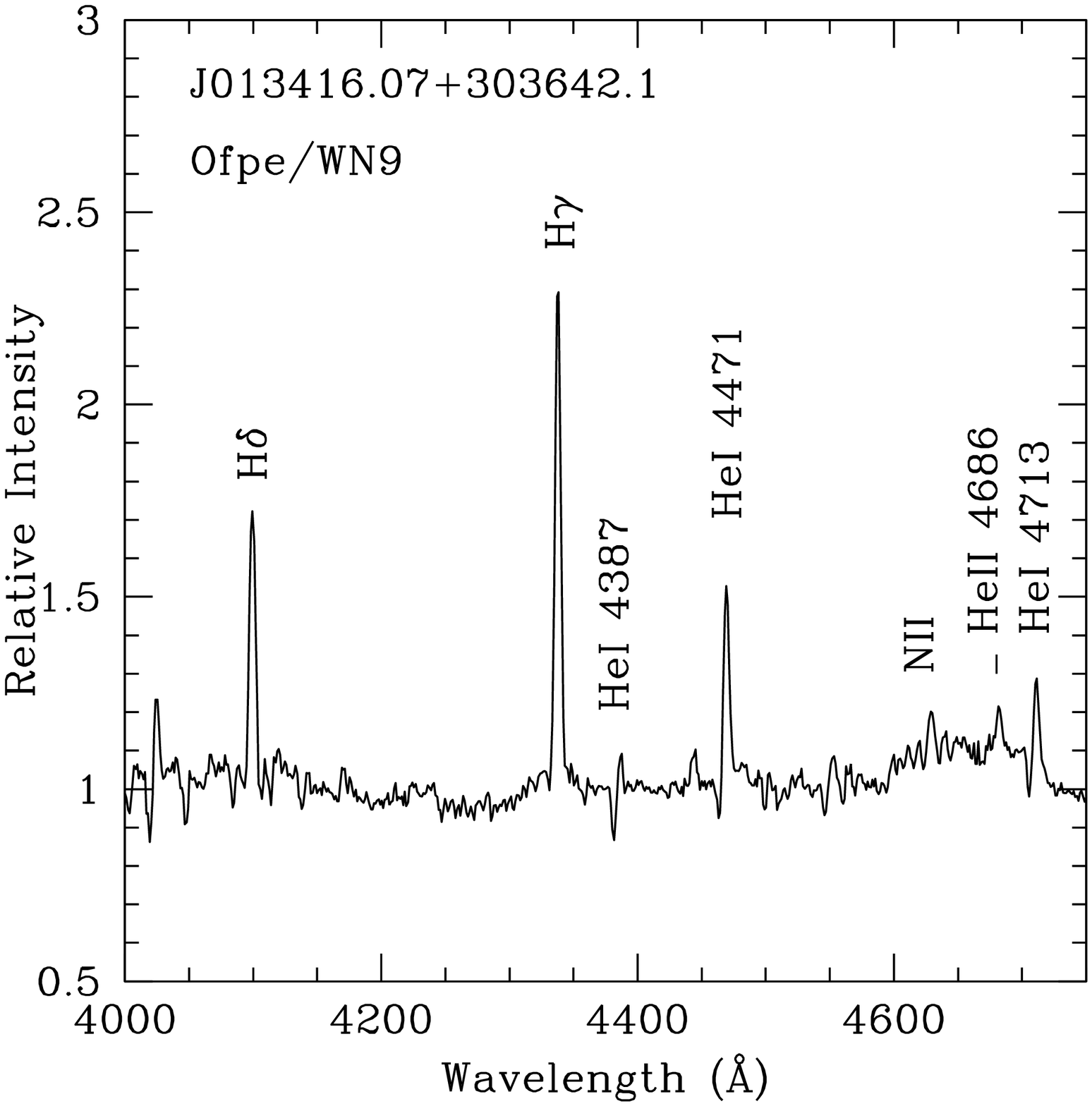}
\plotone{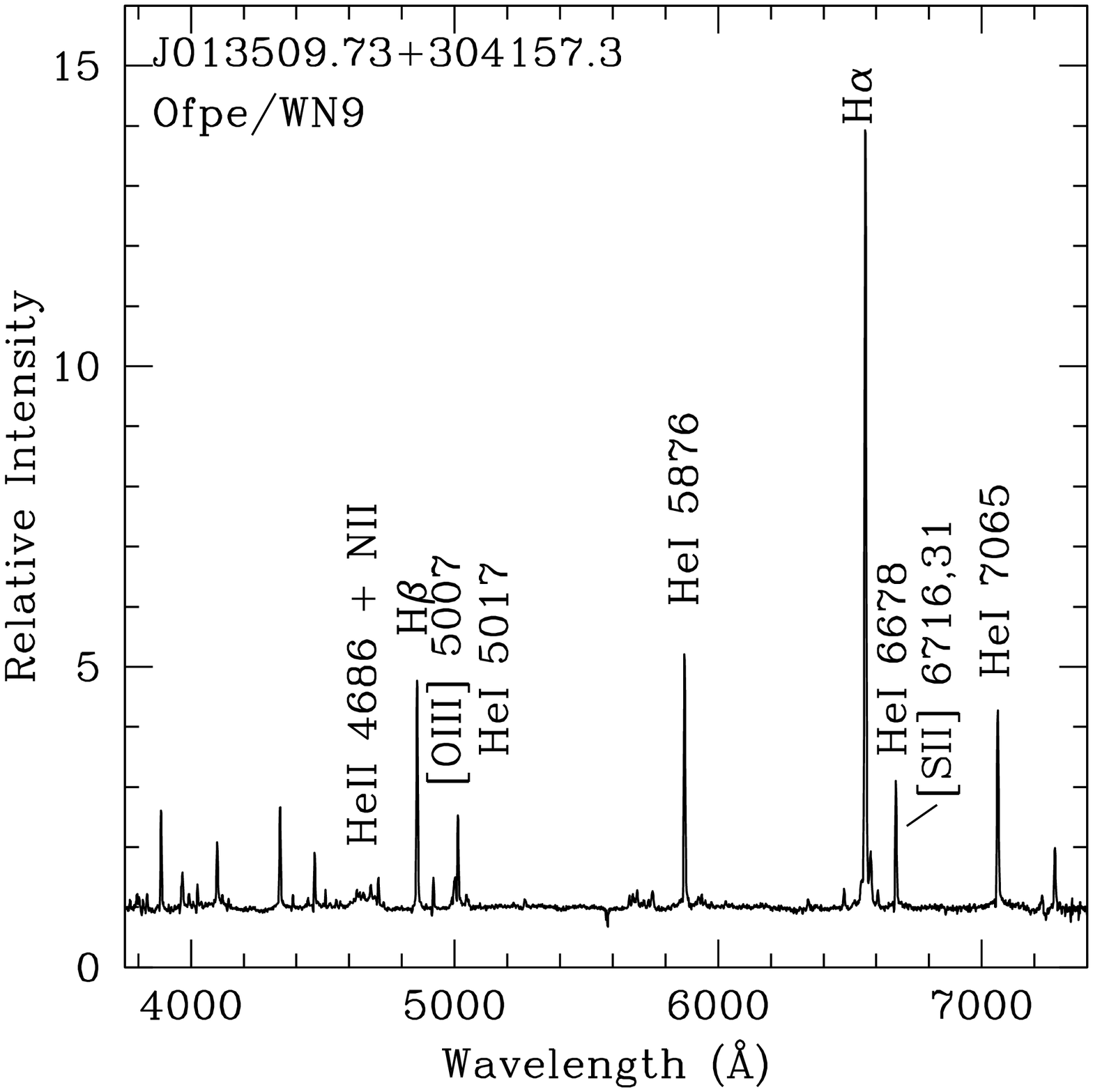}
\plotone{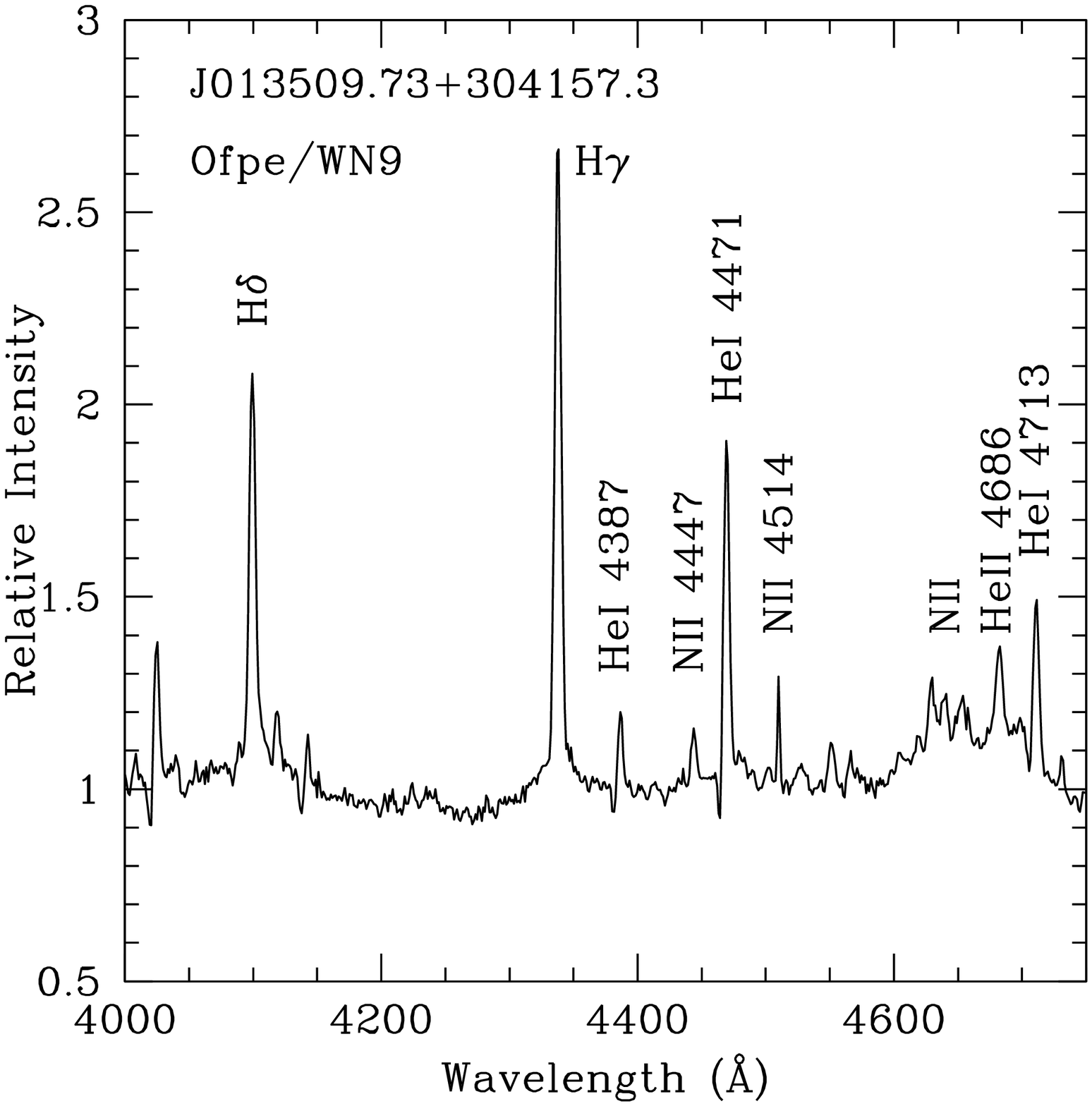}
\plotone{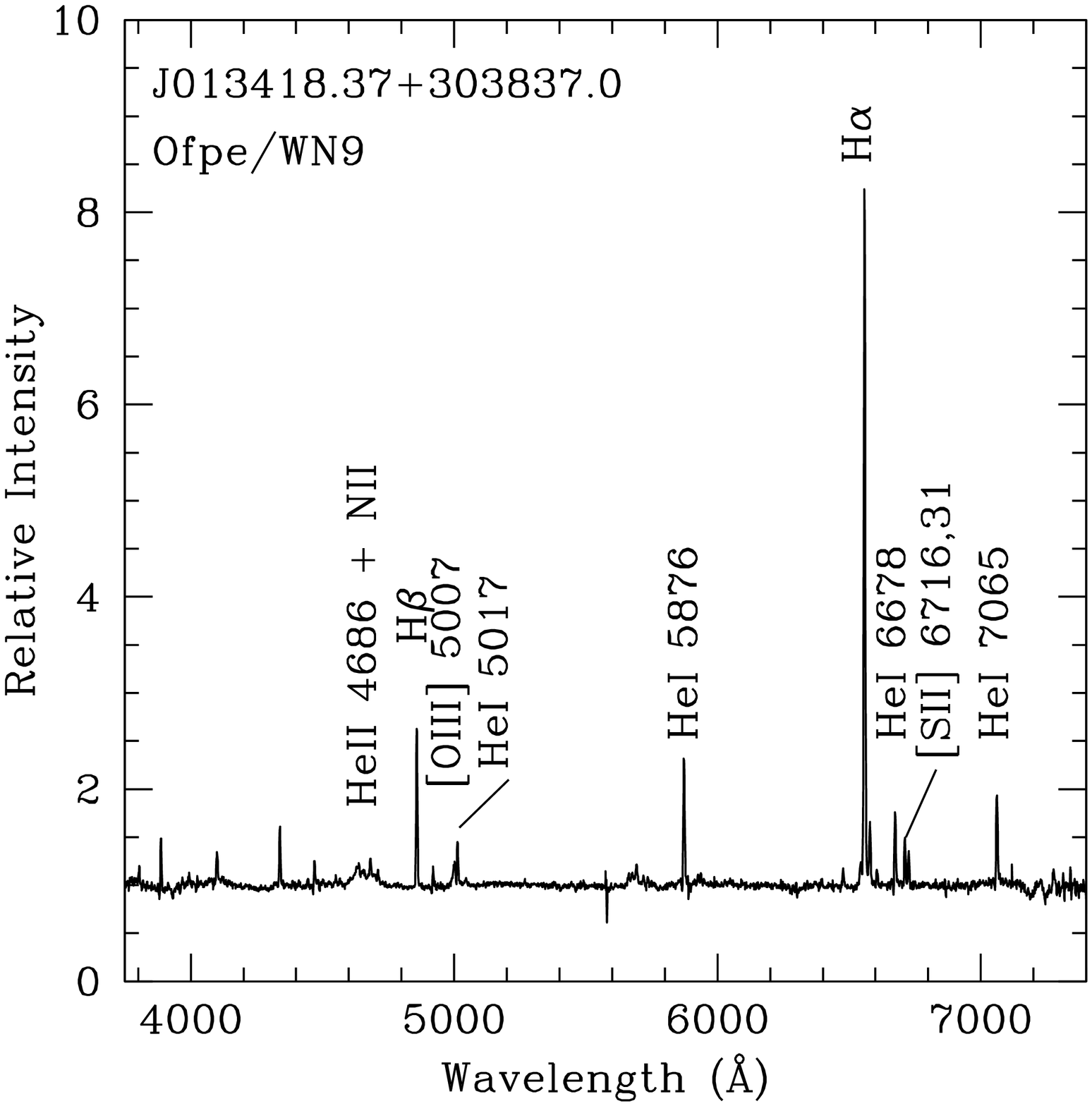}
\plotone{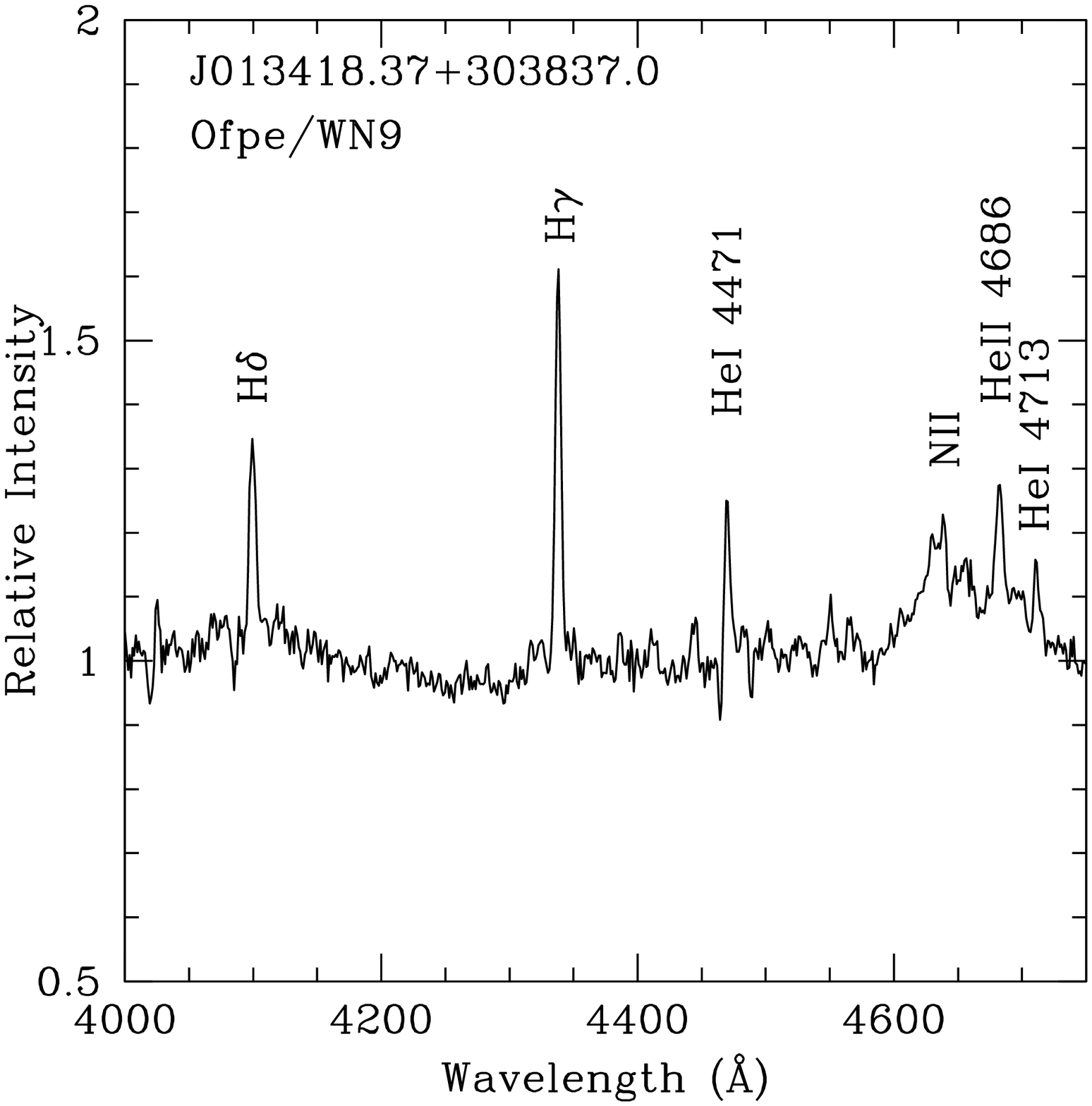}
\caption{\label{fig:Ofpe} Three Ofpe/WN9 stars.  The Ofpe/WN9 class is marked by properties intermediate between that of Of-type stars and very late-type WN type.  J013509.73+304157.3 was previously classified by Massey et al.\ (2007b) as Ofpe/WN9; the object is sometimes referred to as ``Romano's star" based on the discovery spectrum described by Romano (1978).  The other two stars whose spectra are shown here are newly discovered here.}
\end{figure}

\begin{figure}
\epsscale{0.48}
\plotone{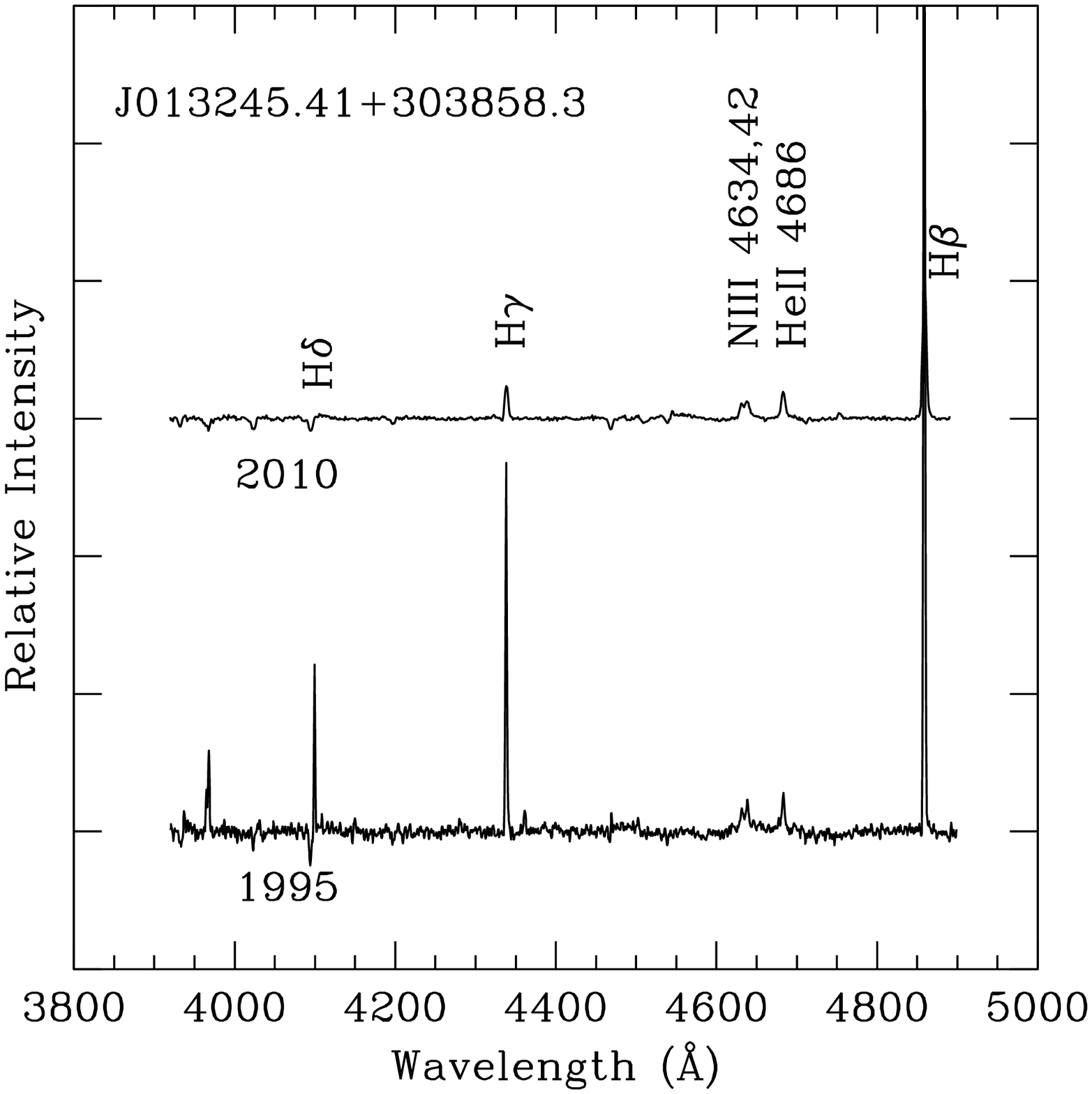}
\plotone{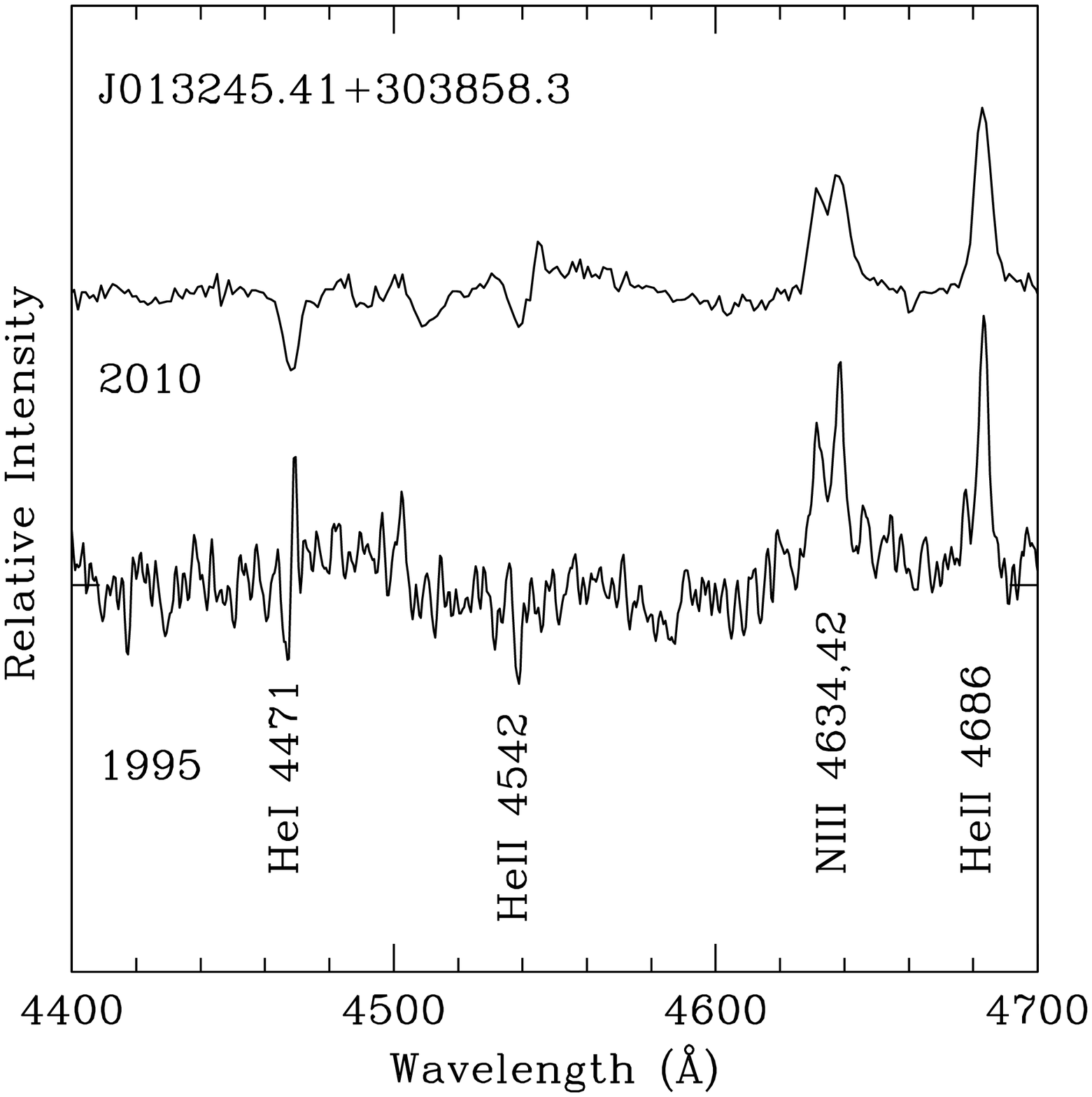}
\caption{\label{fig:uit} Changes in the spectrum of J013245.41+303858.3.  At the top our new (2010) spectrum is shown, while at the bottom the old (1995) spectrum of Massey et al.\ (1996) is shown. The fact that Balmer hydrogen lines are weaker in the new spectrum than in the old may be just an artifact of sky subtraction being non-local with fiber instruments, and the $2\times$ larger diameter fibers used in the 1995 observation. But as the figure on the right shows, the He I $\lambda 4471$ line has also gone from being P Cygni to pure absorption.}
\end{figure}

\begin{figure}
\epsscale{1}
\plotone{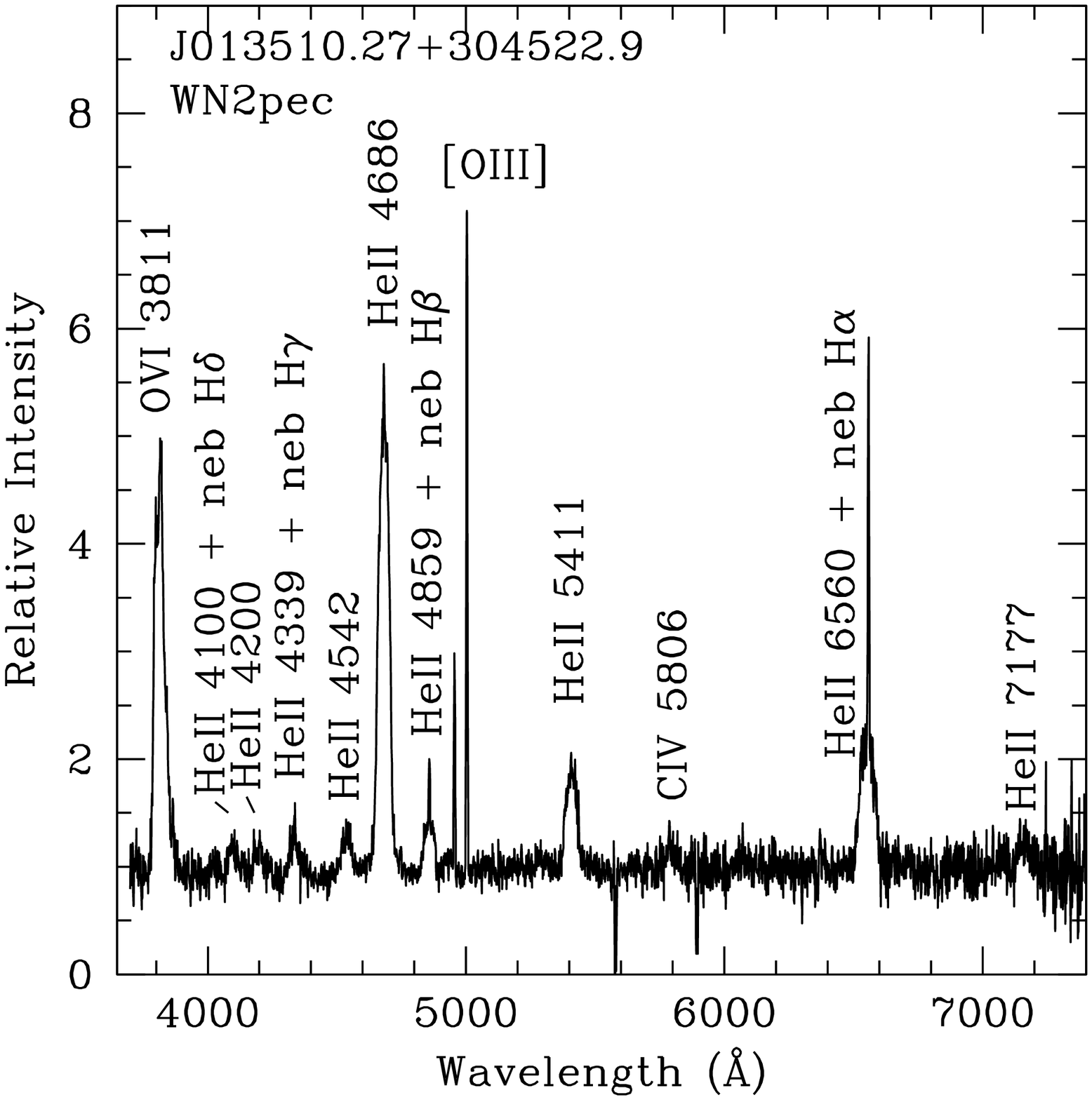}
\caption{\label{fig:ovia} The WN2pec star J013510.27+304522.9.  This star is very unusual in having strong O VI $\lambda 3811$, usually an indication of very high temperature and oxygen abundance.}
\end{figure}

\begin{figure}
\epsscale{0.9}
\plotone{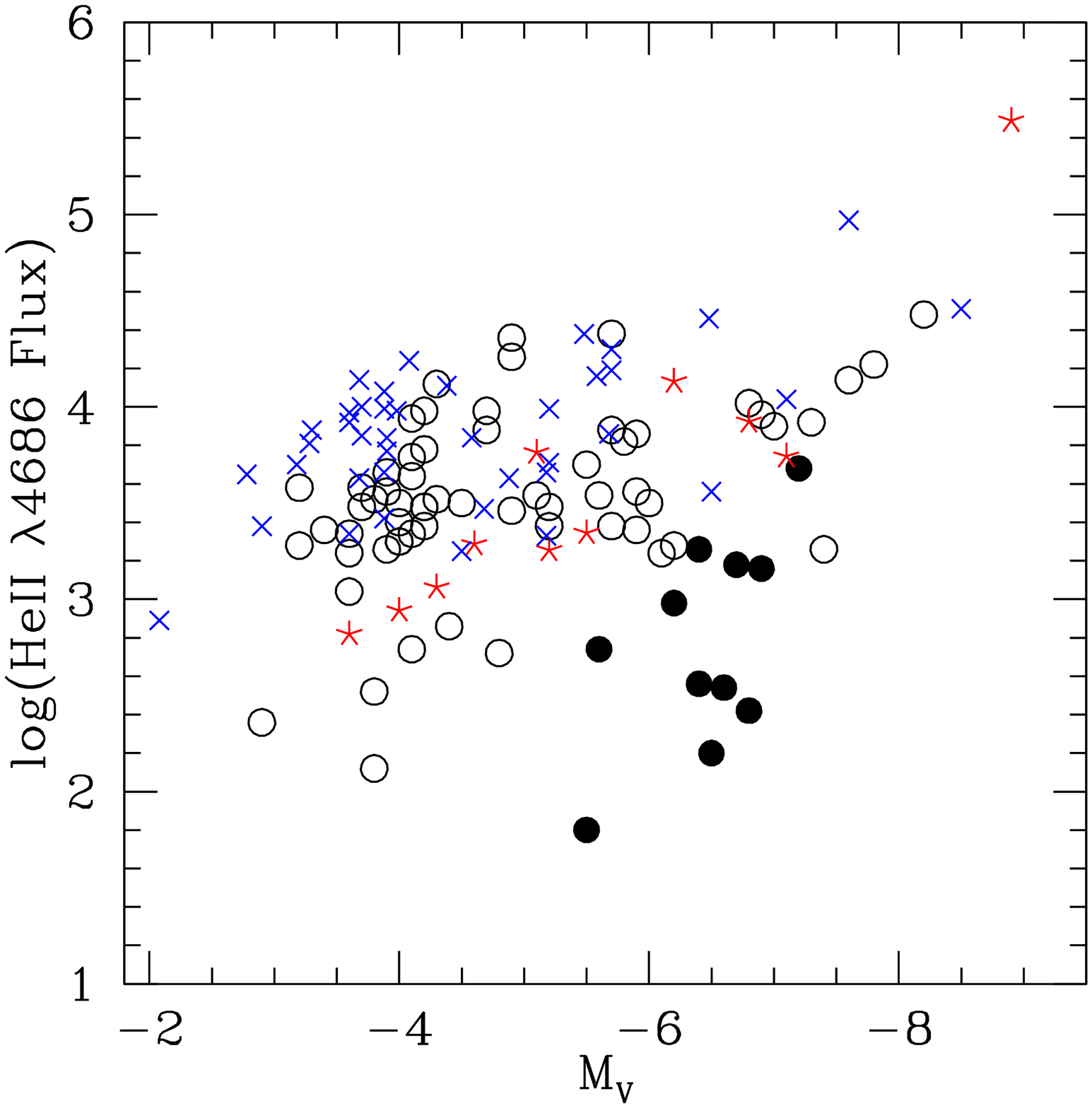}
\caption{\label{fig:ofwn} Sensitivity of Current Survey.  The log of the flux in the He II $\lambda 4686$ emission line is plotted against absolute visual magnitude $M_V$. The line flux has been approximated by $\log(\rm -EW)-M_V/2.5$.  The open circles denote the WNs detected in our survey (both new and previously known). The solid circles are the Of-type stars.  The data for the SMC (red asterisks) and LMC (blue $\times$'s) are shown for comparison, where the data comes from Massey et al.\ (2003) and Conti \& Massey (1989), respectively.  The current survey has detected WNs with line fluxes as weak or weaker than WNs known in the Magellanic Clouds, arguing that the survey has sufficient sensitivity to allow a meaningful determination of the relative
number of WC and WN stars.}
\end{figure}

\begin{figure}
\epsscale{0.75}
\plotone{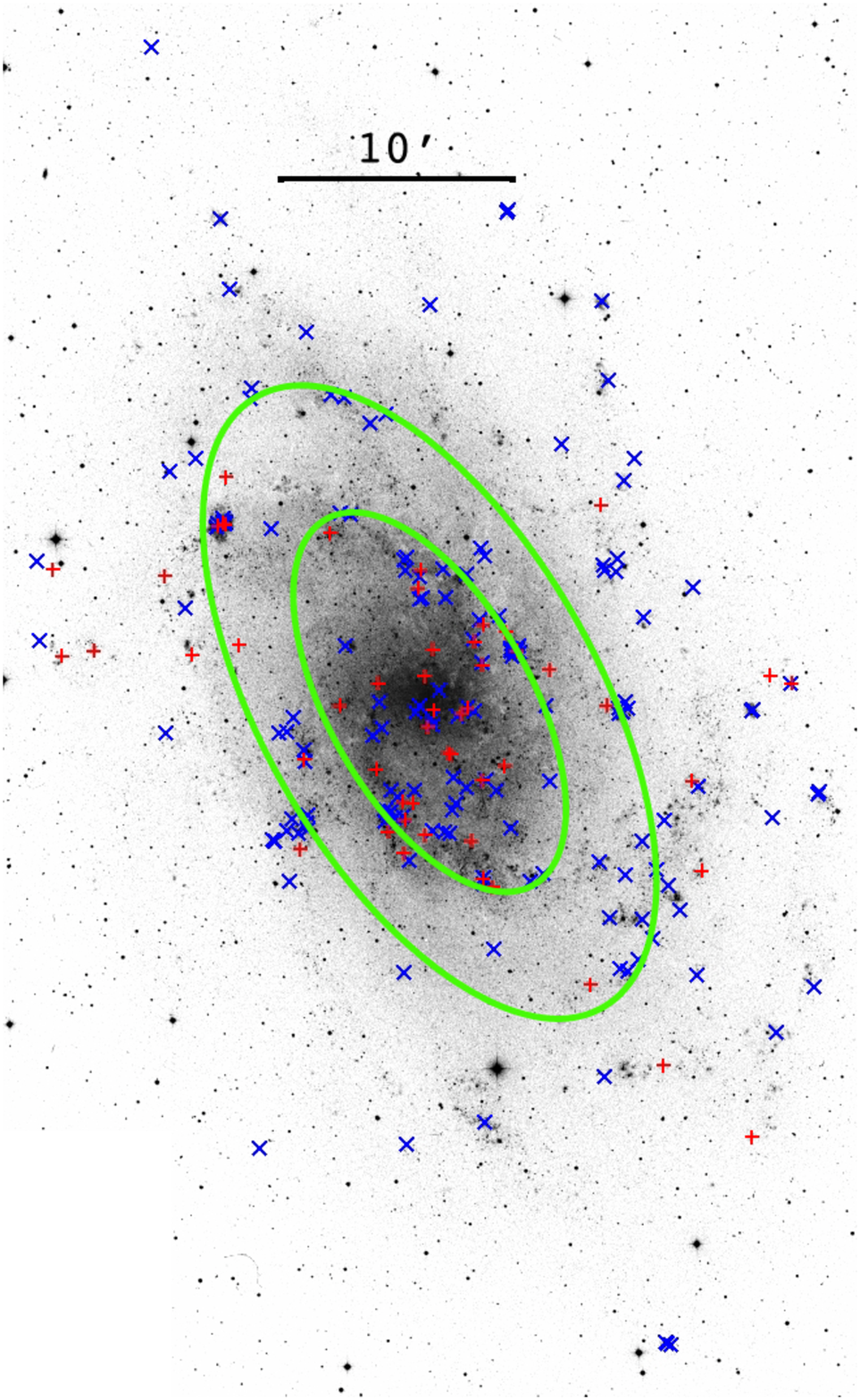}
\caption{\label{fig:WNratio} Location of known WC and WN stars in M33. WN stars are represented as blue $\times$s while WC stars are represented as red $+$s. The green ovals represent distances of $\rho=0.25$ (1.9 kpc) and $\rho=0.50$ (3.8 kpc) within the plane of M33.  Our survey extended well outside the bounds of this figure, except for the small missing region in the lower left corner.}
\end{figure}

\begin{figure}
\epsscale{0.9}
\plotone{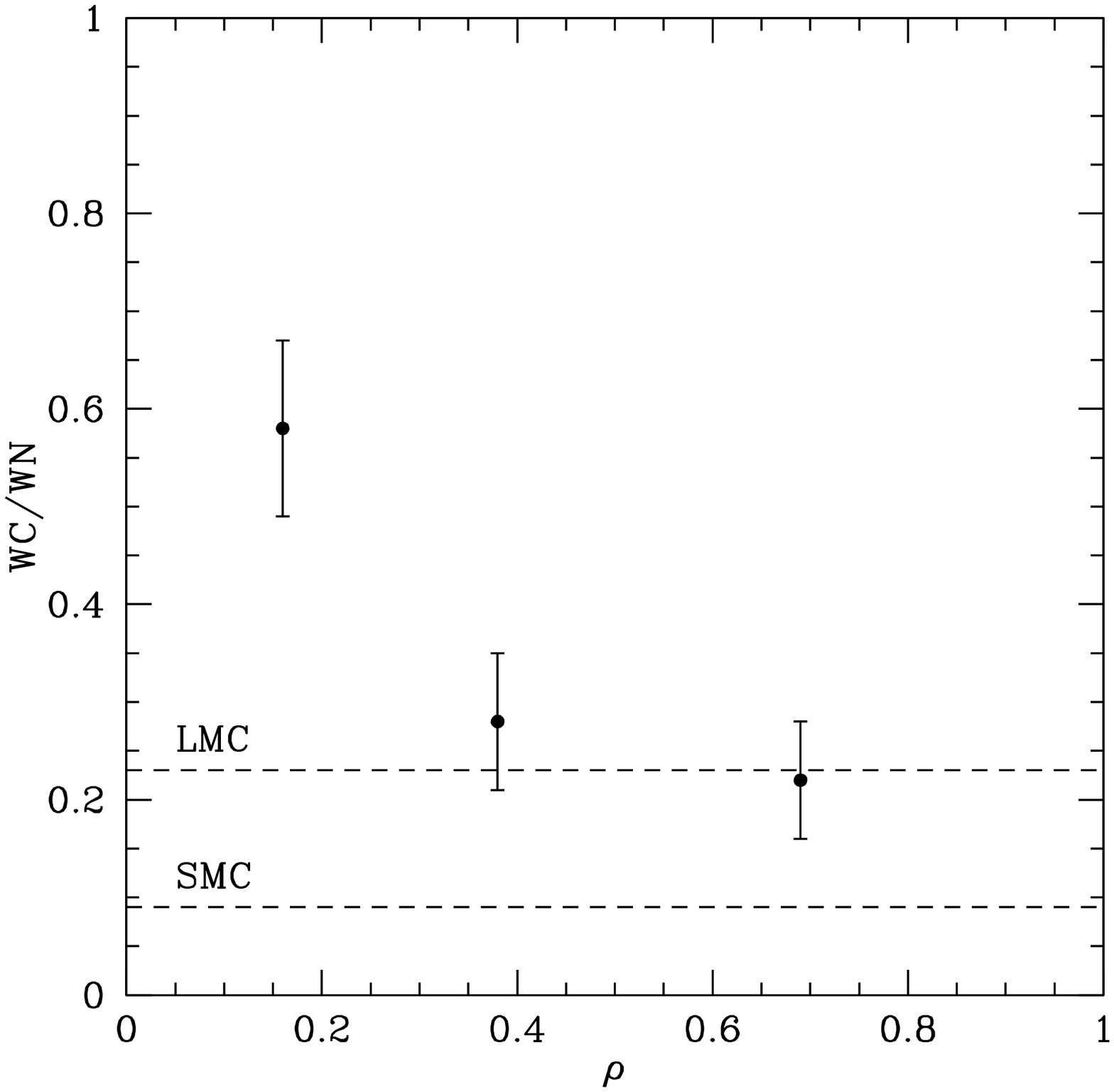}
\caption{\label{fig:wcwnrho} WC/WN ratio vs galactocentric distance.  The ratio of the number of WC-type and WN-type WRs is shown plotted against the distance from the center within the plane of M33. A value of $\rho=1$ corresponds to a distance of 7.53 kpc.  The error bars are based on a 5\% incompleteness.  The values of the WC/WN ratios for the LMC and SMC are also shown.  Note that the LMC and SMC have $\log (\rm O/H)+12=8.37$ and 8.13, respectively, and thus this plot implies that the metallicity of the two outmost bins must be similar to that of the LMC, while that of the inner-most bin is significantly ($\sim 0.5$ dex) greater than that of the LMC.} 
\end{figure}

\begin{figure}
\epsscale{0.48}
\plotone{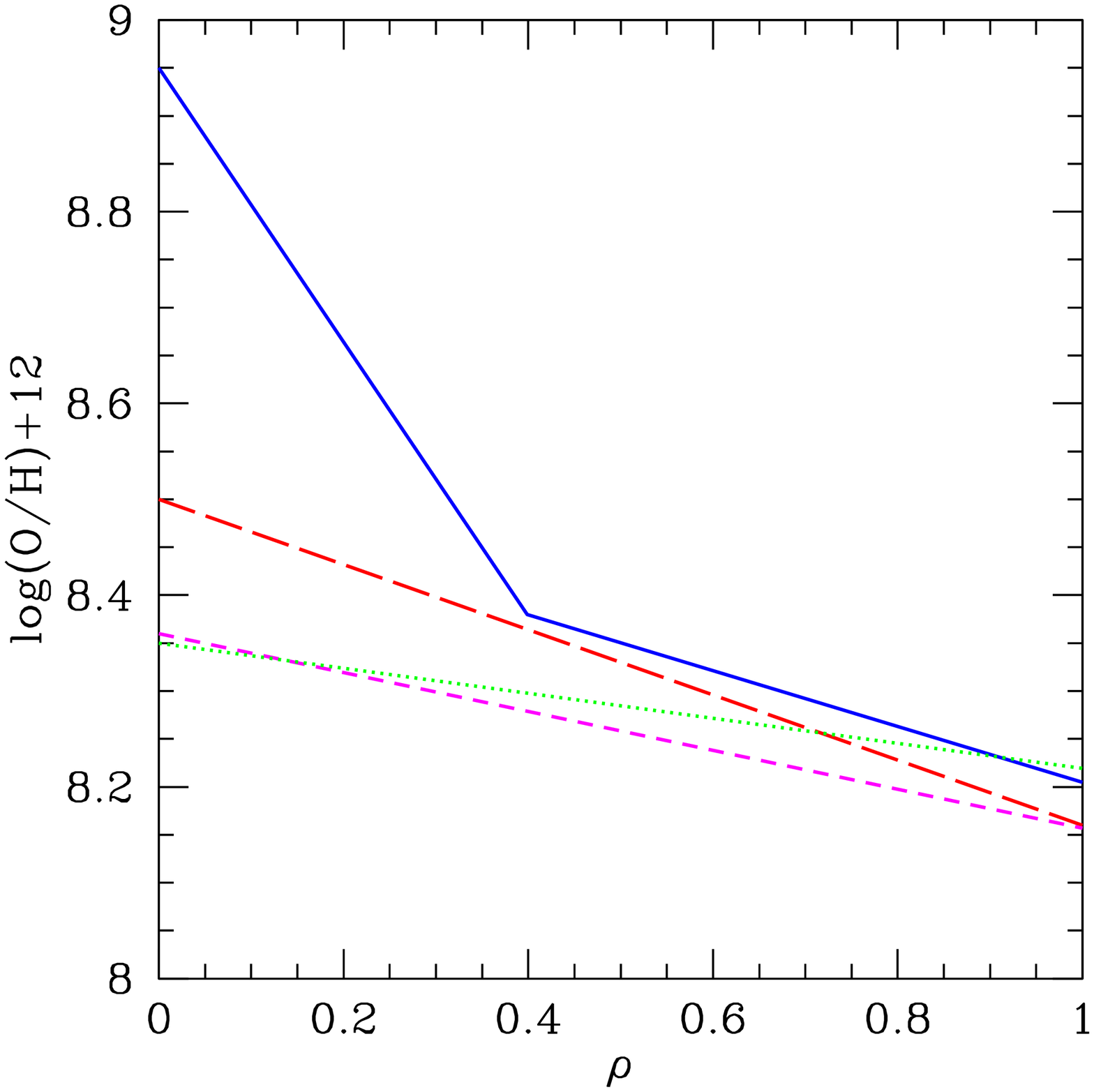}
\plotone{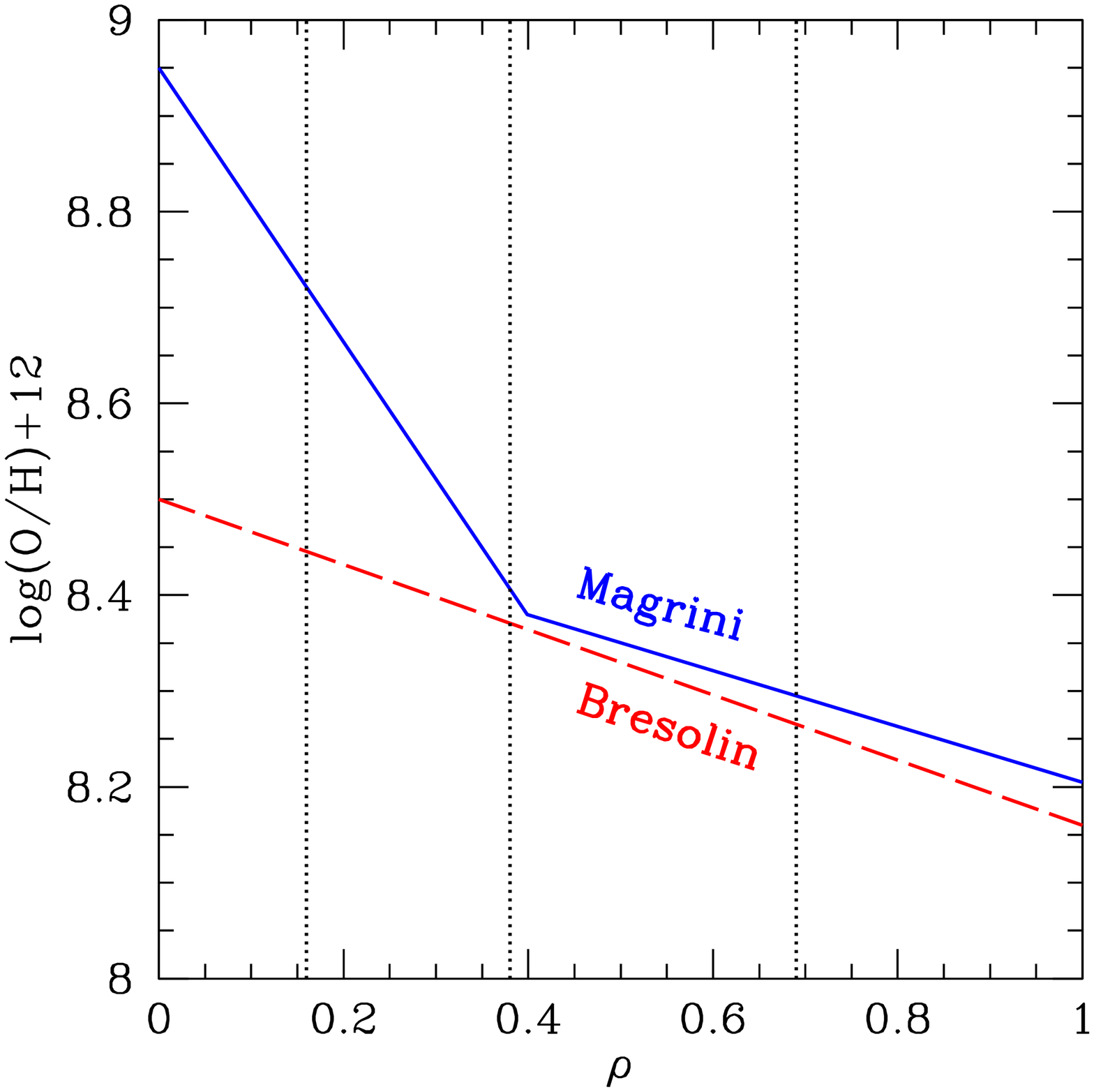}
\caption{\label{fig:metal} Oxygen abundances vs galactocentric distance.  Left: The oxygen abundance gradient from Crockett et al.\ (2006) is shown by a green dotted line, that of Rosolowsky \& Simon (2008) is shown by a short-dashed purple line, that of Magrini et al.\ (2007) with a solid blue line, and that of Bresolin (2011) by a long-dashed red line. Right: We show the oxygen abundance gradient from Magrini et al.\ (2007) and Bresolin (2011), with the vertical dash lines indicating the average $\rho$ values of the three regions denoted in Figure~\ref{fig:wcwnrho} and Table~\ref{tab:WCWN}.}
\end{figure}

\begin{figure}
\epsscale{0.30}
\plotone{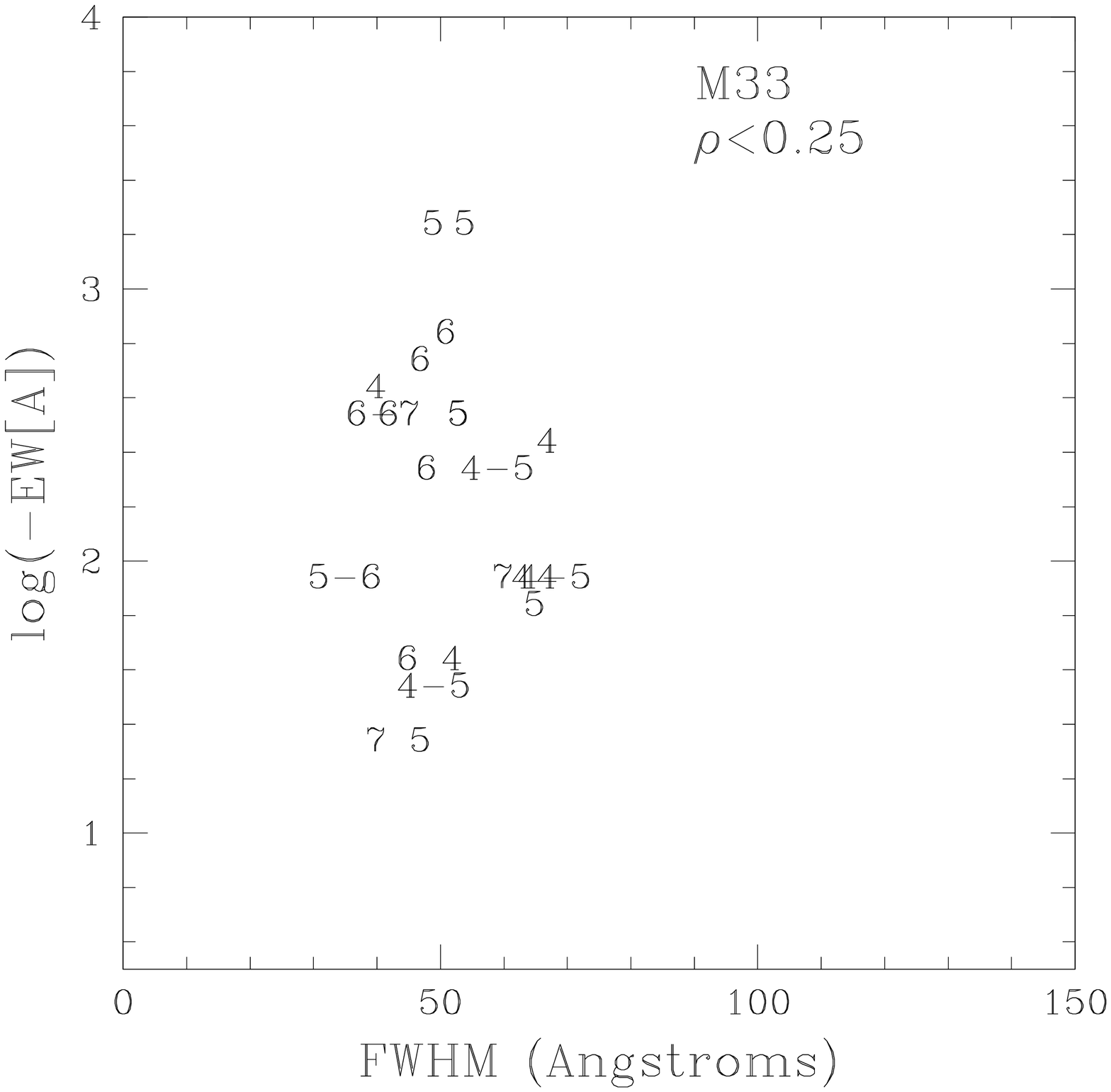}
\plotone{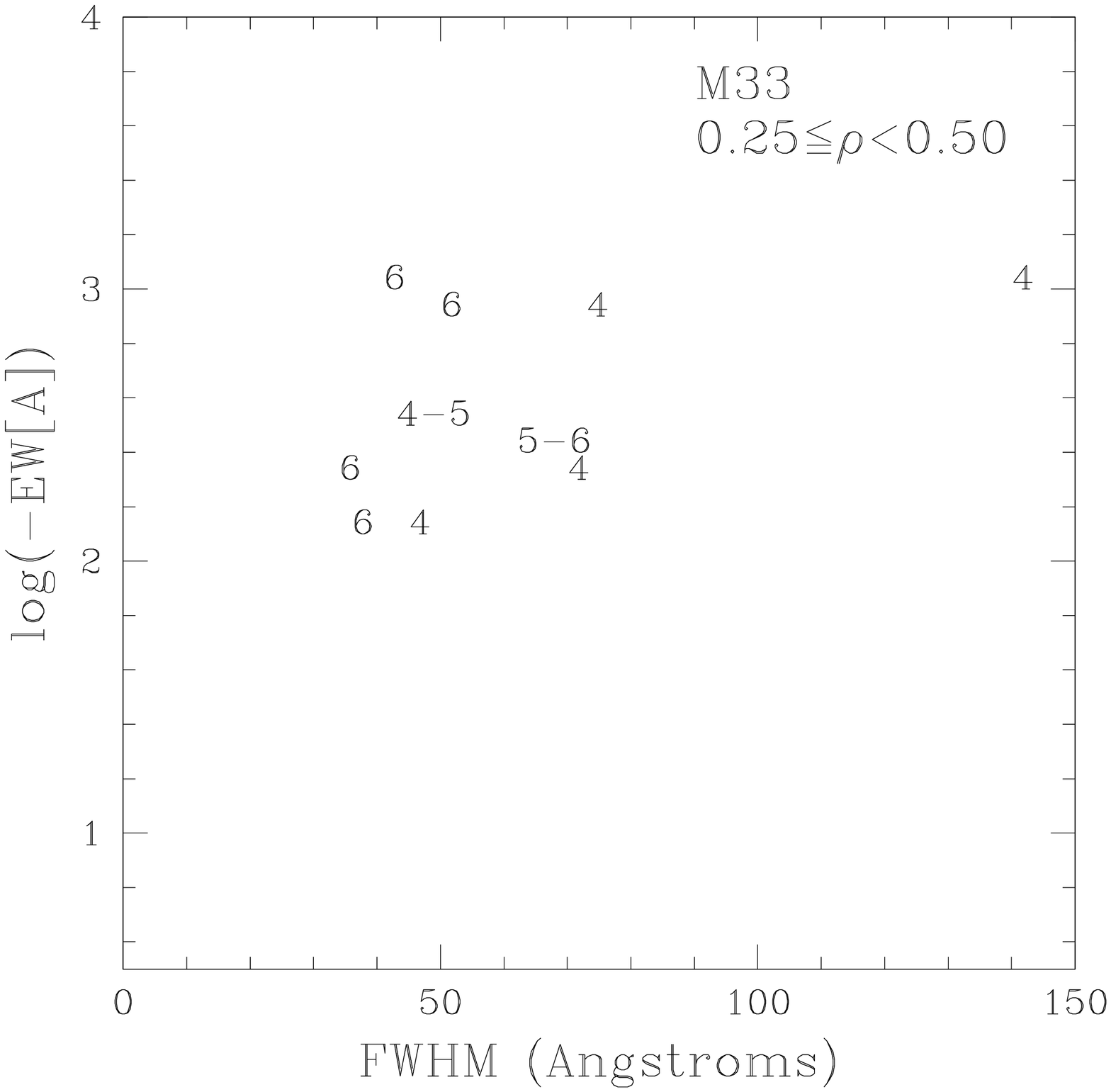}
\plotone{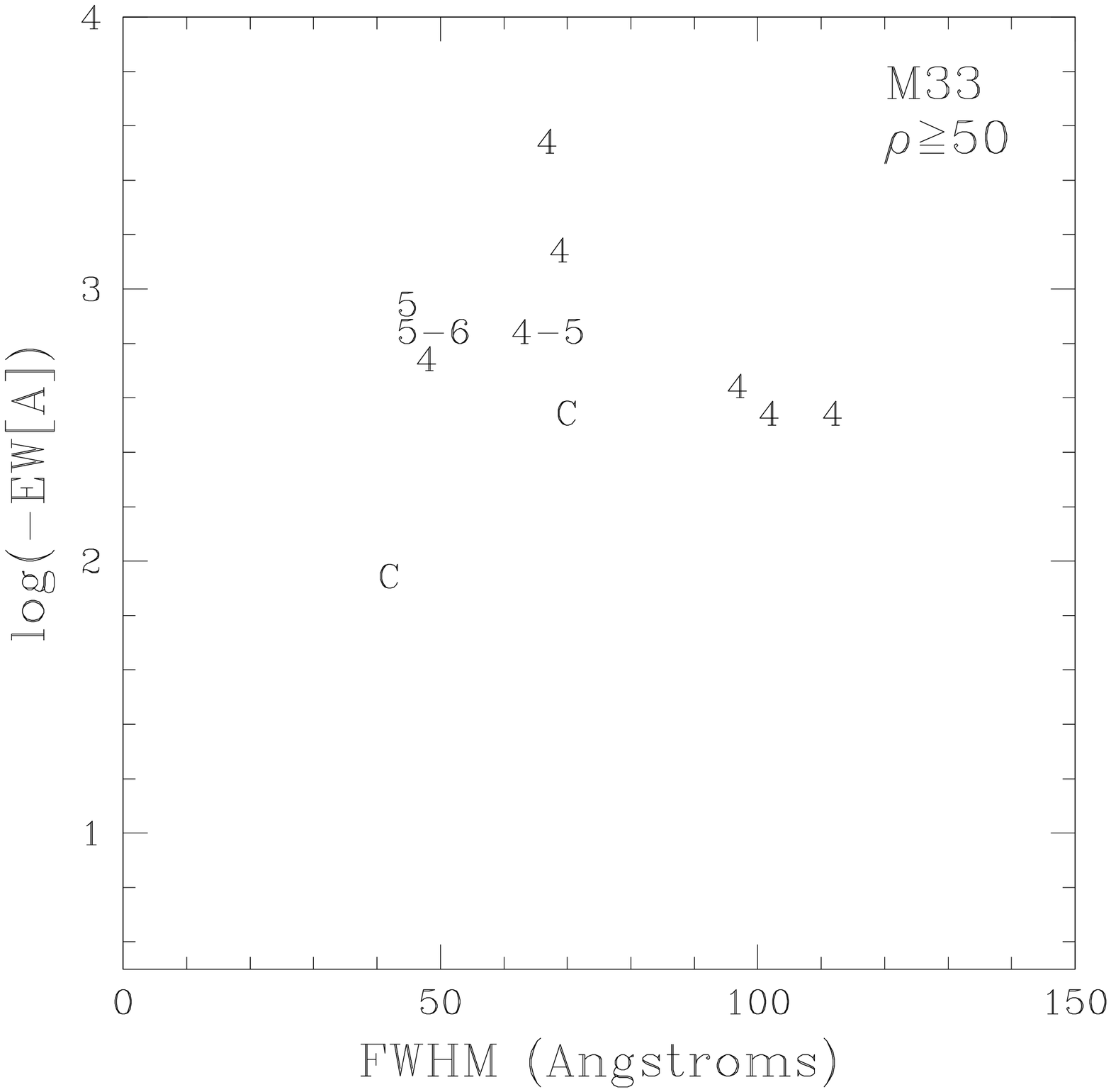}
\plotone{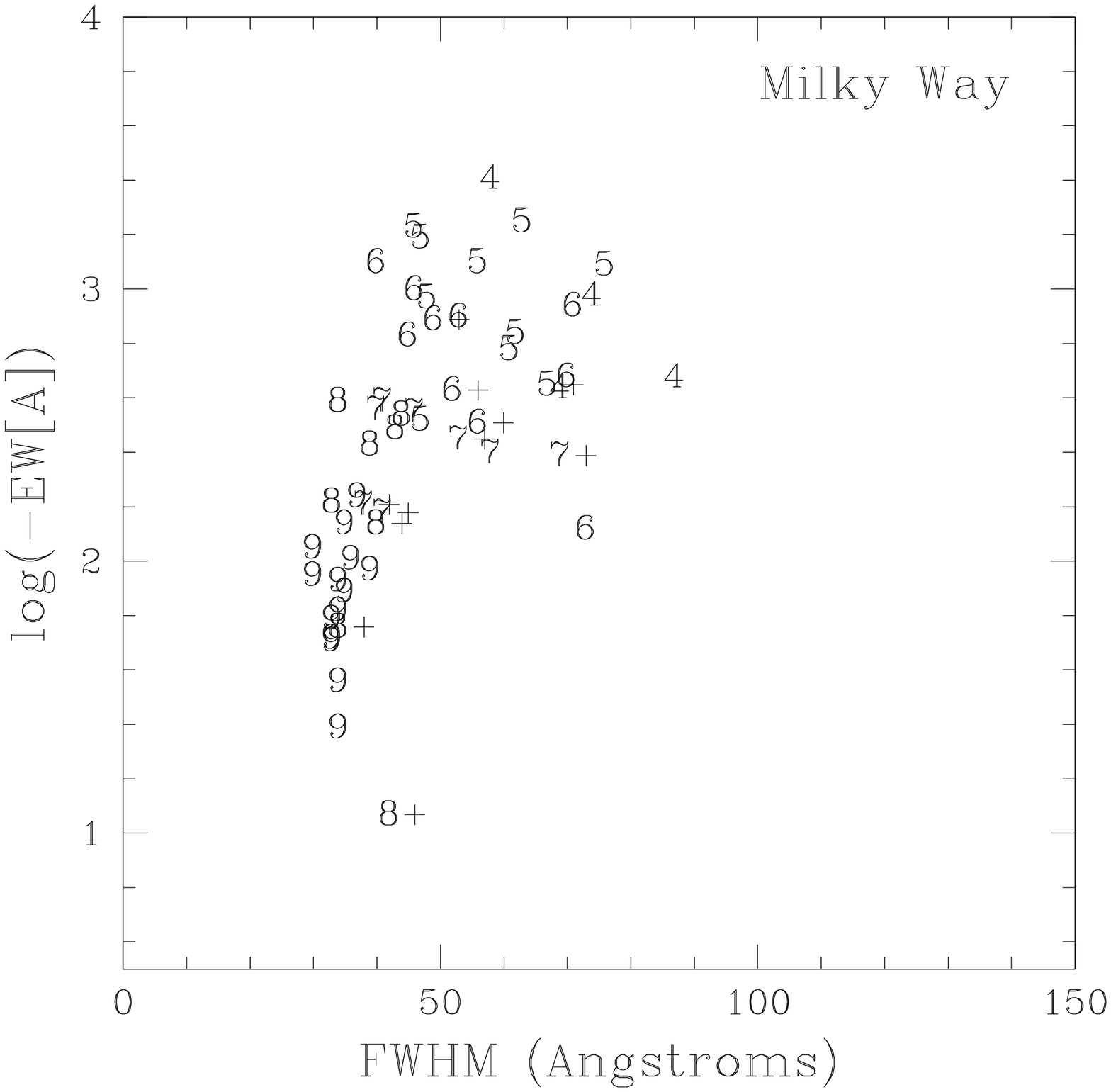}
\plotone{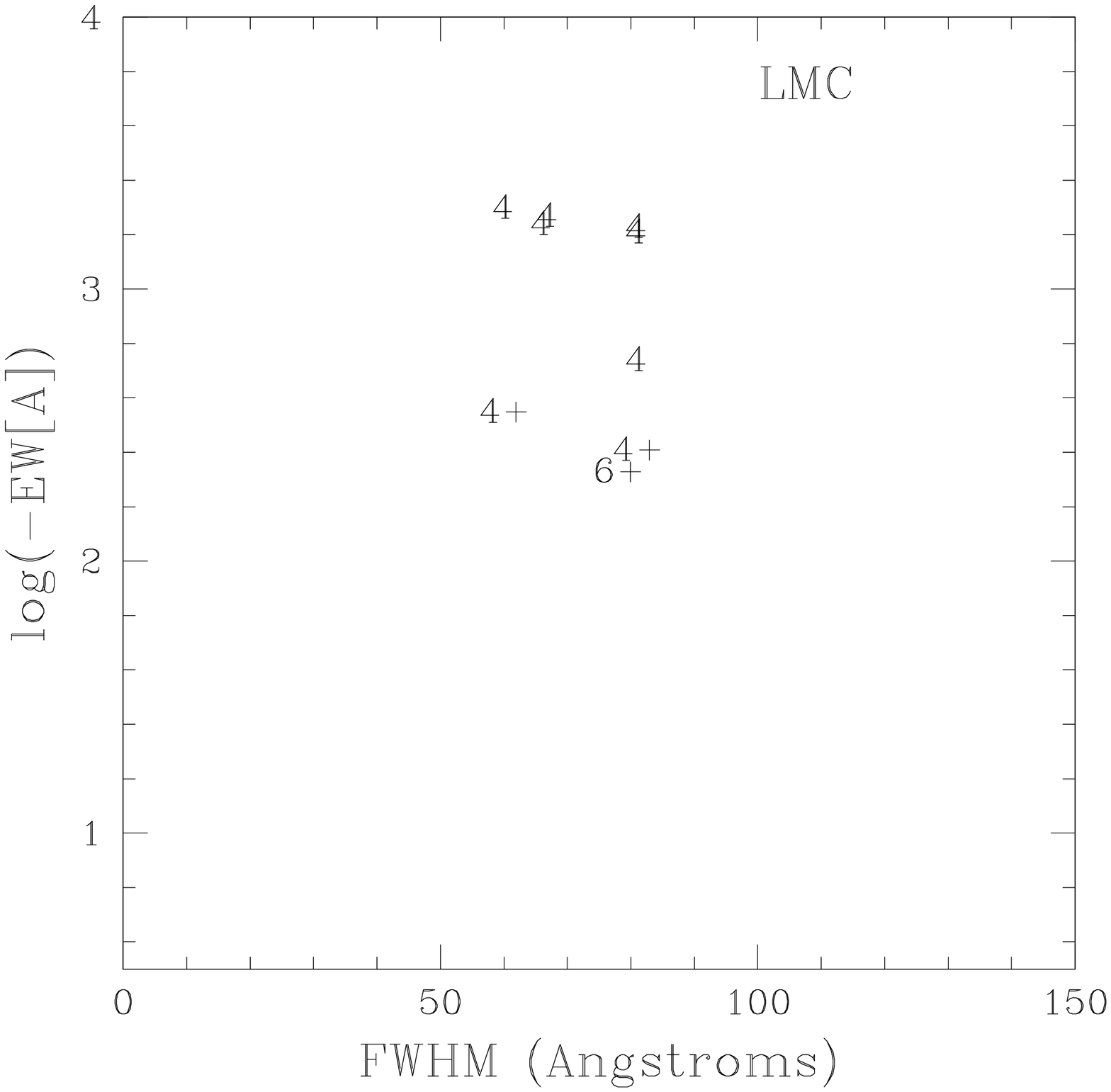}
\caption{\label{fig:WCls} WC line strength vs line width.  The line strength of the CIV $\lambda 5806$ line is plotted against its line width with the spectral subtypes indicated.  The upper panels show the data for the three M33 regions (either based upon measurements here or from the references indicated in Table~\ref{tab:final}), while the while two panels show the data for the Milky Way and LMC (data from Conti \& Massey 1989, tables 3 and 4).  The distribution for the inner region of M33 ($\rho<0.25$, upper left) is more similar to that of the Milky Way (lower left) than the LMC (lower right), while the outer region (upper right) are more like that of the LMC.}
\end{figure}

\newpage
\begin{figure}
\epsscale{0.9}
\plotone{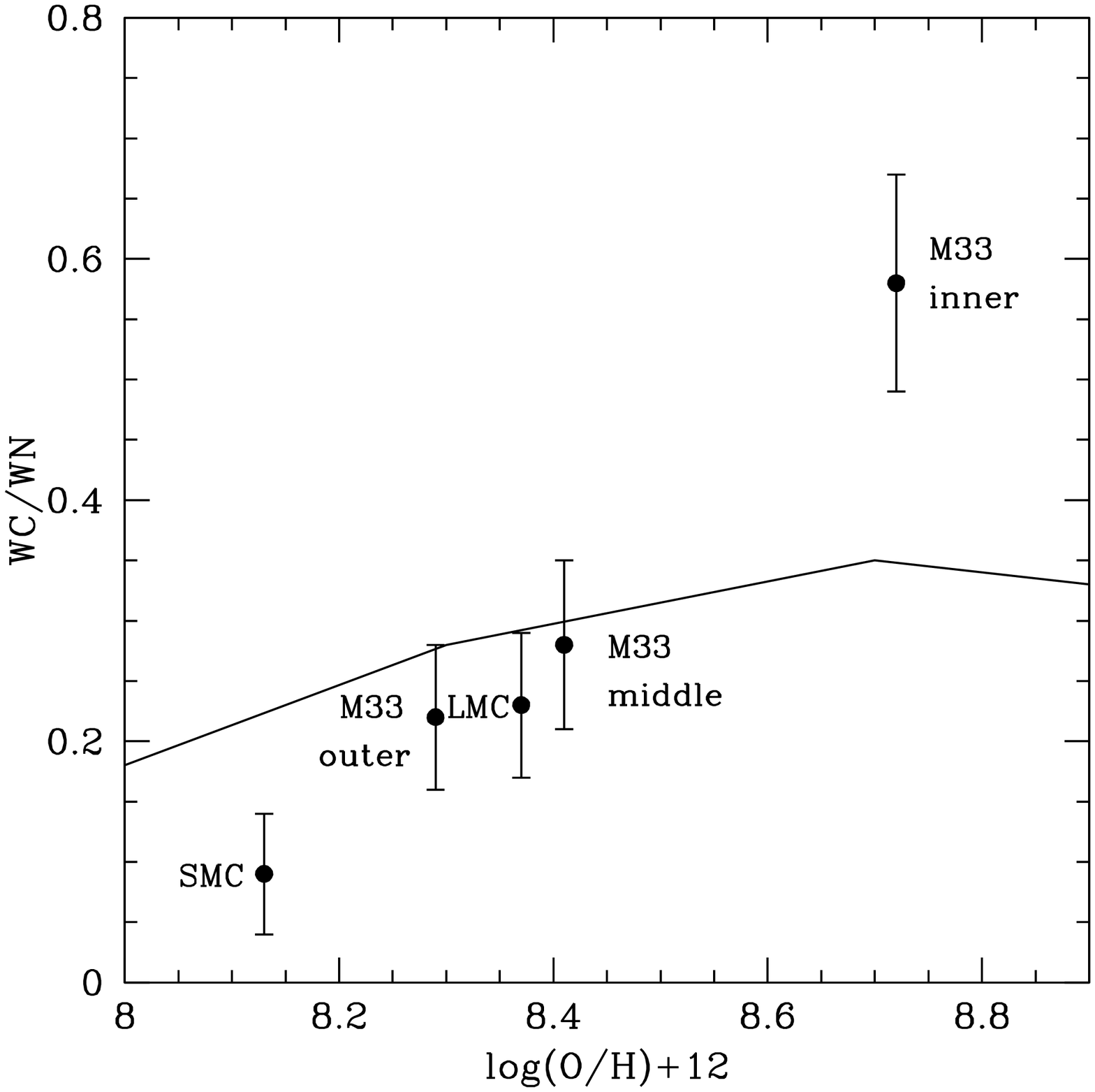}
\caption{\label{fig:modelComp}The WC/WN ratio versus metallicity.
The points are from Table~\ref{tab:WCWN}, and the line shows the evolutionary model results from Meynet \& Maeder (2005) for an initial rotation of $v\sin{i} = 300$ km s$^{-1}$.}
\end{figure}

\begin{deluxetable}{r r r r r r r r r r r}
\rotate
\tablecaption{\label{tab:priorities} Observed Stars and Success Rates}
\tablewidth{0pt}
\tablehead{
&\multicolumn{2}{c}{All}
&
&\multicolumn{7}{c}{$B-V<1$}  \\ \cline{2-3} \cline{5-11} \\
\colhead{Priority\tablenotemark{a}}
&\colhead{Total \#}
&\colhead{Observed}
&
&\colhead{Total \#}
&\colhead{Observed}
&\colhead{Winners\tablenotemark{b}}
&\colhead{Losers}
&\colhead{\% Winners}
&\colhead{\# WRs}
&\colhead{\% WRs}
}
\startdata                             
1 &   92&  83  && 83 &  77 &  68 &  9 & 88.3 & 55 & 68.8\\
2 & 104&  66  && 93 &  56 &    6 & 49 & 10.7 &  2 &   3.6\\
3 & 105&  64  && 82 &  45 &    1 & 44 &   2.2 &  0 &   0.0\\
4 &  75 &  25  && 67 &  19 &    2 & 17 &  10.5 & 0 &   0.0\\
5 & 132 & 30  & & 132 & 30 & 30 & 0 & 100.0 & 29 & 96.7\\
\enddata
\tablenotetext{a}{1--found both by image subtraction and by photometry. 2--High significance
($\ge5\sigma$) photometry candidate.  3---Found by image subtraction and not by photometry.
4---Lower significance (5-6$\sigma$) photometry candidate.  5---Known WRs.}
\tablenotetext{b}{``Winner" means only that there is emission within the emission-line bandpass, not
that the star is a WR.}
\end{deluxetable}

\begin{deluxetable}{l r r r c l}
\tablecaption{\label{tab:nonWRs} Non-Of and Non-WR stars with Emission}
\tablehead{
\colhead{Star}
&\colhead{$V$}
&\colhead{$B-V$}
&\colhead{$U-B$}
&\colhead{Type}
&\colhead{Comment}
}
\startdata
J013322.09+301651.4&19.42&0.32& 0.03&QSO           &z=2.84  MC25\\
J013331.23+303334.2&20.25&0.40&-1.87&HII region   &Nebular He II $\lambda$4686 \\
J013351.46+304057.0&17.73&0.08&-0.93&Unknown       &See text\\
J013423.00+304536.5&20.47&0.27&-1.06&Symbiotic?    &See text\\
J013445.02+304928.0&20.11&0.21&-1.04&QSO           &z=1.99 \\
J013459.47+303701.9&18.37&0.21&-1.23&Hot LBV cand  & Massey et al.\ (2007b)\\
\enddata
\end{deluxetable}

\clearpage

\begin{deluxetable}{l l l l c c r r r r r l}
\rotate
\setlength{\tabcolsep}{0.06in}
\tabletypesize{\small}
\tablecaption{\label{tab:WRs} New WR and Of-type Spectroscopic Results}
\tablewidth{0pt}
\tablehead{
\colhead{Star}
&\colhead{Type}
&\multicolumn{3}{c}{Previous}
&
&\multicolumn{2}{c}{CIII and/or He II}
&
&\multicolumn{2}{c}{CIV $\lambda 5806$}
&\colhead{Comments}  \\    \cline{3-5} \cline{7-8} \cline{10-11}
&&\colhead{Type}
&\colhead{ID}
&\colhead{Ref\tablenotemark{a}}
&&\colhead{log(-EW)}
&\colhead{FWHM}
&
&\colhead{log(-EW)}
&\colhead{FWHM}
}
\startdata
J013232.07+303522.4   & WN2      & \nodata   &  \nodata   & \nodata&&     1.7 &      37 && \nodata & \nodata &                                                    \\ 
J013232.13+303514.3   & WN3+neb  & \nodata   &  \nodata   & \nodata&&     1.8 &      36 &&     1.5 &      50 &                                                    \\ 
J013233.24+302652.3   & WN2+abs  & \nodata   &  \nodata   & \nodata&&     1.0 &      30 && \nodata & \nodata &                                                    \\ 
J013241.39+303416.2   & WN3h     & \nodata   &  \nodata   & \nodata&&     2.1 &      26 &&     0.4 &      25 & H-rich                                             \\ 
J013245.41+303858.3   & Ofpe     &Ofpe/WN9   & UIT 008    & \nodata&&     0.3 &      14 && \nodata & \nodata &                                                    \\ 
J013245.74+303854.4   & WN2+O7   &WN         & MC 3       & \nodata&&     1.1 &      35 && \nodata & \nodata &                                                    \\ 
J013245.84+302019.5   & WC4+neb -& \nodata   &  \nodata   & \nodata&&     3.0 &      76 &&     3.5 &      65 & OV present                                         \\ 
J013251.04+303535.1   & OIf      & \nodata   &  \nodata   & \nodata&&     0.5 &       7 && \nodata & \nodata & Neb subtraction poor                               \\
\enddata
\tablenotetext{a}{References for old spectral types are from Massey \& Johnson 1998 unless otherwise noted.  1---Kehrig et al.\ 2011; 2---Massey et al.\ 2007b; 3---Drissen et al.\ 2008; 4---Abbott et al.\ 2004.}
\end{deluxetable}

\begin{deluxetable}{l c c c c c l}
\tablecaption{\label{tab:NoLonger} Stars No Longer Considered WRs}
\tablehead{
\colhead{Star}
&\multicolumn{2}{c}{Old}
&
&\multicolumn{2}{c}{New}
&\colhead{Other ID} \\  \cline{2-3} \cline{5-6} 
&\colhead{Type}
&\colhead{Ref.\tablenotemark{a}}
&
&\colhead{Type}
&\colhead{Ref.\tablenotemark{a}}
}
\startdata
J013322.09+301651.4  &WN  & 1 &&      QSO & 2 & [MJ98] 37, MC25  \\
J013333.58+304219.2  &WN  &  3  &&      O4 If & 2   &  [MJ98] 46,  C22  \\
J013344.68+304436.7  &WN8 &  4     &&      O If   & 5 & [MJ98] 77, OB66-25  \\
J013416.07+303642.1  &WN11& 6    &&     P Cyg LBV&  7 & B517, H 108 \\
J013432.68+304655.4  & WN6  & 8    &&        OIf   & 2        &   N604-WR10 \\
\enddata
\tablenotetext{a}{1---Massey \& Conti 1983; 2---This work; 3---Massey \& Johnson 1998; 4---Massey et al.\ 1996; Abbott et al.\ 2004; 6---Crowther et al.\ 1997; 7---Massey et al.\ 2007b; 8---Drissen et al.\ 2008.}
\end{deluxetable}

\begin{deluxetable}{l l l c c c r r l l l l}
\rotate
\setlength{\tabcolsep}{0.06in}
\tabletypesize{\small}
\tablecaption{\label{tab:final} WR Stars in M33}
\tablewidth{0pt}
\tablehead{
\colhead{Star}
&\colhead{$\rho$\tablenotemark{a}}
&\colhead{Type}
&\colhead{Ref\tablenotemark{b}}
&\colhead{$V$\tablenotemark{c}}
&\colhead{$m_{\lambda 4750}$\tablenotemark{d}}
&\colhead{$M_V$\tablenotemark{e}}
&\colhead{$M_{\lambda 4750}$\tablenotemark{e}}
&\colhead{OB\tablenotemark{f}}
&\colhead{MJ\tablenotemark{g}}
&\colhead{ID\tablenotemark{h}}
&\colhead{Comment}
}
\startdata
J013232.07+303522.4 & 0.88 & WN2        & 1 & 20.99 & 20.76 & -4.0 & -4.3 & 131 & \nodata &  \nodata    &             \\
J013232.13+303514.3 & 0.88 & WN3+neb    & 1 & 20.74 & 20.43 & -4.2 & -4.6 & 131 & \nodata &  \nodata    &             \\
J013233.24+302652.3 & 0.85 & WN2+abs    & 1 & 21.22 & 22.23 & -3.8 & -2.8 & 123 & \nodata &  \nodata    &             \\
J013237.62+304004.3 & 0.87 & WC         & 2 & 19.77 & 20.00 & -5.2 & -5.0 & Fld & WR1     & MCA 1       &             \\
J013237.72+304005.7 & 0.87 & Ofpe/WN9   & 2 & 17.75 & 20.00 & -7.2 & -5.0 & 132 & WR2     & UIT 003     &             \\
J013240.82+302454.3 & 0.79 & WNE        & 2 & 20.72 & 20.51 & -4.3 & -4.5 & 121 & WR3     & MC 1        &             \\
J013241.39+303416.2 & 0.76 & WN3h       & 1 & 20.75 & 20.41 & -4.2 & -4.6 & Fld & \nodata &  \nodata    &             \\
J013241.95+304024.6 & 0.83 & WCE        & 2 & 20.55 & 21.09 & -4.4 & -4.0 & 132 & WR4     & MC 2        &             \\
\enddata
\tablenotetext{a}{Distance from the center within the plane of M33, normalized to the $D_25$ isophotoal radius of 30\farcm8 assuming $\alpha_{2000}=01^h33^m50\fs89$, $\delta_{2000}=30^\circ 39\arcmin36\farcs8$, an inclination of 56$^\circ$ and a position angle of the major axis of 23$^\circ$, following Kwitter \& Aller 1981 and Zaritsky et al.\ 1989.}
\tablenotetext{b}{References for spectral types: 1---This work; 2---Massey \& Johnson 1998; 3---Kehrig et al.\ 2011;
4---Massey et al.\ 2007b; 5---Drissen et al.\ 2008;  6---Abbott et al.\ 2004}
\tablenotetext{c}{From Massey et al.\  2005.}
\tablenotetext{d}{AB magnitude through CT filter centered at 4750 \AA\ and calibrated using the values from
Massey \& Johnson 1998.}
\tablenotetext{e}{Absolute magnitudes computed assuming a true distance modulus of 24.60 (830 Mpc) and
adopting an average reddening of $E(B-V)=0.12$ based on Massey et al.\ 2007a.  Adopting $R_V=3.1$ leads
to an $A_V=0.37$ mag and $A_{m_{4750}}=0.45$ mag.}
\tablenotetext{f}{OB association as defined by Humphreys \& Sandage 1980.  Parenthesis means the star is just outside
the boundaries of the association, while ``Fld" implies it is a field star, not in a cataloged OB association.}
\tablenotetext{g}{Massey \& Johnson 1998 WR designation.}
\tablenotetext{h}{IDs are from Massey \& Johnson 1998 and references therein, other than the HII regions
designations N592 and N604 from Drissen et al.\ 2008 and the ``K10" designations from Kehrig et al.\ 2011).}
\tablenotetext{i}{J013332.64+304127.2 was shown to be a late-type WN star by Massey \& Conti 1983 (MC28), and
later identified by HeII $\lambda 4686$ imaging by Armandroff \& Massey 1985, where it is listed as AM 2.
However, Massey et al.\ 2006 found that the spectrum has since changed dramatically, resembling a hot LBV.}
\end{deluxetable}

\begin{deluxetable}{l c l}
\tablecaption{\label{tab:isolated} Isolated WR Stars in M33}
\tablewidth{0pt}
\tablehead{
\colhead{Star}
&\colhead{$\rho$}
&\colhead{Type}
}
\startdata
J013425.11+301950.3 & 0.98 & WN3 \\
J013438.18+304953.2 & 0.53 & WN5\\
J013507.24+304500.9 & 0.84 & WC5\\
J013510.27+304522.9 & 0.87 & WN2pec\\
\enddata
\end{deluxetable}

\begin{deluxetable}{l r r r r r r r r r r r}
\tablecaption{\label{tab:Mvs}Absolute Visual Magnitudes\tablenotemark{a}}
\tablewidth{0pt}
\tablehead{
\colhead{Subtype}
&\multicolumn{3}{c}{M33}
&
&\multicolumn{3}{c}{LMC}
&
&\multicolumn{3}{c}{SMC}  \\ \cline{2-4} \cline{6-8} \cline {10-12}
&\colhead{$M_V$}
&\colhead{$\sigma$}
&\colhead{$N$}
&
&\colhead{$M_V$}
&\colhead{$\sigma$}
&\colhead{$N$}
&
&\colhead{$M_V$}
&\colhead{$\sigma$}
&\colhead{$N$}
}
\startdata
WN2       & $-6.0$ & 0.9&  3 && $-2.6$ &\nodata & 1 && $-6.8$ & \nodata &1 \\
WN3       & $-4.8$ & 1.2& 25 && $-4.0$ & 0.6    &11 && $-5.0$ & 1.2 & 6 \\
WN4-4.5   & $-5.2$ & 1.3& 19 && $-4.2$ & 0.6    &33 &&$-5.2$ & \nodata &1 \\
WN5       & $-5.4$ & 1.5&  5 && $-5.1$ & 1.4    & 3 && \nodata & \nodata &\nodata\\
WN6       & $-4.9$ & 0.9&  5 && $-4.8$ & 1.5    & 3 && $-6.2$ & \nodata & 1 \\
WN7       & $-5.6$ & 1.4& 10 && $-5.9$ & 1.0    & 4 && \nodata & \nodata &\nodata \\
WN8       & $-4.5$ & 0.9&  6 && $-5.5$ & 0.0    & 2 && \nodata &  \nodata &\nodata  \\
WN9       & $-4.4$ & 0.9& 11 && $-5.2$ & 0.7    & 2 && \nodata &  \nodata & \nodata \\
WC4       & $-5.9$ & 1.3& 14 && $-5.0$ & 0.9    &11 &&\nodata &  \nodata & \nodata  \\
WC5       & $-4.9$ & 0.8&  4 && \nodata & \nodata &\nodata && \nodata & \nodata & \nodata\\
WC6       & $-4.5$ & 0.9& 11 && \nodata & \nodata &\nodata && \nodata & \nodata &\nodata \\
WC7       & $-4.2$ & 0.3&  2 && \nodata & \nodata &\nodata && \nodata & \nodata &\nodata\\
\enddata
\tablenotetext{a}{The values are the means of the WRs of a given subtype
excluding ones with uncertain types and/or with companions.}
\end{deluxetable}

\begin{deluxetable}{l l l}
\tablecaption{\label{tab:more} Additional WR Candidates}
\tablewidth{0pt}
\tablehead{
\colhead{Star}
&\colhead{$V$}
&\colhead{$m_{\lambda 4650}$}
}
\startdata
J013302.73+301131.6 & 20.00 & 19.65 \\
J013404.07+304658.3 & 19.53 & 19.40 \\
J013411.45+303637.3 & 21.34 & 21.24 \\
J013432.60+304211.3 & 19.34 & 19.10 \\
J013434.26+304637.8 & 20.48 & 20.71 \\
J013442.41+305019.0 & 20.95 & 21.19 \\
\enddata
\end{deluxetable}

\begin{deluxetable}{c c c r r c}
\tablecaption{\label{tab:WCWN}Numbers of WC and WN Stars}
\tablewidth{0pt}
\tablehead{
\colhead{Region}
&\colhead{$\overline{\rho}$}
&\colhead{$\log(\rm O/H)+12$}
&\colhead{\#WCs}
&\colhead{\#WNs}
&\colhead{WC/WN}}
\startdata
$\rho<0.25$ & 0.16 & 8.72 & 26 & 45 & 0.58$\pm$0.09 \\
         $0.25\ge\rho<0.50$    & 0.38 & 8.41 & 15 & 54 & 0.28$\pm$0.07 \\
$\rho\ge0.50$ & 0.69 & 8.29 & 12 & 54 & 0.22$\pm$0.06 \\
LMC                           &\nodata& 8.37 & 25 &109 & 0.23$\pm$0.06 \\
SMC                           &\nodata& 8.13 &  1 & 11 & 0.09$\pm$0.05 \\
\enddata
\end{deluxetable}
\end{document}